\documentclass[11pt]{article}
\usepackage{amsmath,amsfonts,amssymb,graphicx,epsfig,pdflscape}
\usepackage[usenames]{color}
\usepackage{pstricks}
\numberwithin{equation}{section}

\newcommand{\nc}{\newcommand}
\nc{\bib}{\bibitem}
\nc{\al}{\alpha}
\nc{\D}{\Delta}
\nc{\la}{\lambda}
\nc{\var}{\varphi}
\nc{\pa}{\partial}
\nc{\nn}{\nonumber \\ }
\nc{\be}{\begin{equation}}
\nc{\ee}{\end{equation}}
\nc{\bea}{\begin{eqnarray}}
\nc{\eea}{\end{eqnarray}}
\nc{\bra}[1]{\langle {#1}|}
\nc{\ket}[1]{|{#1}\rangle}
\nc{\chit}{\raisebox{0.25ex}{$\chi$}}
\nc{\chih}{\raisebox{0.25ex}{$\hat\chi$}}
\nc{\ch}{{\rm ch}}
\nc{\g}{\mathfrak{g}}
\nc{\h}{\mathfrak{h}}
\nc{\ph}{\hat{p}}
\nc{\qh}{\hat{q}}
\nc{\rh}{\hat{r}}
\nc{\sh}{\hat{s}}
\nc{\Mh}{\hat{M}}
\nc{\Vh}{\hat{V}}
\nc{\Ob}{\bar{O}}
\nc{\Ac}{\mathcal{A}}
\nc{\Cc}{\mathcal{C}}
\nc{\Ic}{\mathcal{I}}
\nc{\Kc}{\mathcal{K}}
\nc{\Mc}{\mathcal{M}}
\nc{\Qc}{\mathcal{Q}}
\nc{\Sc}{\mathcal{S}}
\nc{\Vc}{\mathcal{V}}
\nc{\Vch}{\hat{\mathcal{V}}}
\nc{\Wc}{\mathcal{W}}
\def\vvdots{\mathinner{\mkern1mu\raise1pt\vbox{\kern7pt\hbox{.}}\mkern2mu
  \raise4pt\hbox{.}\mkern2mu\raise7pt\hbox{.}\mkern1mu}}
\definecolor{lightblue}{rgb}{.61,.61,1}
\definecolor{midblue}{rgb}{.7,.7,1}
\definecolor{lightlightblue}{rgb}{.85,.85,1}
\definecolor{lightestblue}{rgb}{.96,.96,1}
\definecolor{lightpurple}{rgb}{1,.65,1}

\begin{document}

\topmargin -5mm
\oddsidemargin 5mm

\setcounter{page}{1}

\vspace{8mm}
\begin{center}
{\LARGE {\bf Staggered and affine Kac modules over $A_1^{(1)}$}}

\vspace{8mm}
{\LARGE J{\o}rgen Rasmussen}
\\[.4cm]
{\em School of Mathematics and Physics, University of Queensland}\\
{\em St Lucia, Brisbane, Queensland 4072, Australia}
\\[.4cm]
{\tt j.rasmussen\,@\,uq.edu.au}
\end{center}

\vspace{12mm}
\centerline{{\bf{Abstract}}}
\vskip.4cm
\noindent
This work concerns the representation theory of the affine Lie algebra $A_1^{(1)}$\! at fractional level
and its links to the representation theory of the Virasoro algebra.
We introduce affine Kac modules as certain finitely generated submodules of Wakimoto modules.
We conjecture the existence of several classes of staggered $A_1^{(1)}$-modules and provide evidence in the form of
detailed examples.
We extend the applicability of the Goddard-Kent-Olive coset construction to include the affine Kac and staggered modules.
We introduce an exact functor between the associated category of $A_1^{(1)}$-modules and the corresponding 
category of Virasoro modules.
At the level of characters, its action generalises the Mukhi-Panda residue formula.
We also obtain explicit expressions for all irreducible
$A_1^{(1)}$-characters appearing in the decomposition of Verma modules, re-examine the
construction of Malikov-Feigin-Fuchs vectors, and extend the Fuchs-Astashkevich theorem from the Virasoro algebra to $A_1^{(1)}$\!.
\renewcommand{\thefootnote}{\arabic{footnote}}
\setcounter{footnote}{0}

\newpage
\tableofcontents

\section{Introduction}

Kac-Moody algebras~\cite{Kac67,Moody67,KacBook} were introduced half a century ago
and include the (infinite-dimensional) affine Lie algebras frequently appearing in theoretical and mathematical physics,
including conformal field theory~\cite{BPZ84,KZ84,DiFMS} and string theory~\cite{GW86,PolBook}.
Focus in this paper is on the so-called non-twisted affine $s\ell(2)$ algebra, also known as $A_1^{(1)}$\!.
We are particularly interested in its representation theory and how this is linked to the representation theory of
the Virasoro algebra~\cite{Vir70,IK11}.

A Wess-Zumino-Witten model \cite{WZ71,Wit83,Wit84} is an example of a conformal field theory whose symmetry algebra
contains an affine Lie algebra at non-negative integer level. 
Affine $s\ell(2)$ algebras at such levels also appear in the Goddard-Kent-Olive coset construction~\cite{GKO85,GKO86} of the
unitary minimal models \cite{BPZ84,FQS84}. 
In fact, the entire family of Virasoro minimal models (including the non-unitary ones)
can be described by extending the construction to fractional level \cite{Kent86,MW90}.
Concretely, one then considers the diagonal coset
\be
 \frac{(A_1^{(1)})_k\oplus(A_1^{(1)})_n}{(A_1^{(1)})_{k+n}},
 \qquad k=\frac{p}{p'}-2,\qquad (p,p')=1,\qquad n,p,p'\in\mathbb{N},
\label{cosetAAA}
\ee
where the indices of $(A_1^{(1)})$ denote the various levels, 
and where $\mathbb{N}$ is the set of positive integers.

Further supporting the relevance of non-integer levels, 
for each admissible level $k$ ($p\geq2$), there exists a finite set of 
irreducible $A_1^{(1)}$-representations, known as admissible representations~\cite{KW88,KW88b}, 
whose characters form a (finite-dimensional) representation of the modular group.
Following (\ref{cosetAAA}), these characters are related through branching rules with branching functions
given by characters of the coset Virasoro algebra. As discussed below, these relations admit extensions to certain
classes of reducible but not completely reducible representations and characters.

A hallmark of a {\em logarithmic} conformal field theory is the presence of such Virasoro representations.
For example, on an indecomposable staggered module~\cite{Roh96,GK96,KR09}, the Virasoro generator $L_0$ acts non-diagonalisably, 
rendering the module reducible.
Moreover, it has been realised~\cite{Ras1012,BGT1102,MDRR1503} that finitely generated submodules
of Feigin-Fuchs modules~\cite{FF82,FF89}, rather than quotients of Verma modules, 
are the fundamental building blocks in the logarithmic minimal models $\mathcal{LM}(p,p')$~\cite{PRZ0607,RP0707}. 
Due to their innate characterisation in terms of Kac-table labels, these submodules are known as (Virasoro) Kac modules.

In a logarithmic minimal model with an {\em extended}\, symmetry algebra, it seems natural to expect that there is a similar
class of modules. A main objective of the present work is thus the introduction of 
{\em affine Kac modules} as certain indecomposable submodules of Wakimoto modules \cite{Wak86,BF90}, 
anticipating a central role for them in logarithmic minimal models with $A_1^{(1)}$-extended symmetry algebra. 
In further preparation for the study of logarithmic $A_1^{(1)}$\! models, see also \cite{Gab01,LMRS02,LMRS04,Ras0508,Rid09},
we conjecture the existence of several classes of staggered $A_1^{(1)}$-modules whose structures closely resemble
those of staggered Virasoro modules.

Returning to the coset construction (\ref{cosetAAA}), we find that its pertinence extends
to the affine Kac modules and the proposed family of staggered $A_1^{(1)}$-modules. 
The ensuing branching functions agree with and supplement those found in~\cite{PR13} 
by computing the so-called logarithmic limit~\cite{Ras0405,Ras0406,PS1207} of the branching rules for admissible characters.
Our results can therefore be seen as reaffirming the applicability of the logarithmic limit
in the exploration of logarithmic conformal field theories.

The $A_1^{(1)}$\! and Virasoro representation theories are evidently very similar. 
For fractional level $k>-2$, we find that this is made manifest by a dense
and exact functor between the category of irreducible, affine Kac and staggered $A_1^{(1)}$-modules and the corresponding 
category of modules over the Virasoro algebra with central charge $1-6(k+1)/(k+2)$.
Loosely speaking, on the Loewy diagram of such an $A_1^{(1)}$-module, the functor acts by first removing all irreducible 
subquotients with only finite-dimensional $L_0$-eigenspaces and then reinterpreting the remaining nodes as subsingular vectors 
in the Loewy diagram of a Virasoro module.
At the level of characters, the functor generalises the Mukhi-Panda residue formula~\cite{MP90} for admissible characters,
see also \cite{KY92}.

The remainder of this paper is organised as follows.
In Section~\ref{Sec:Rep}, we review some basic aspects of $A_1^{(1)}$\! and its representation theory.
In Section~\ref{Sec:Frac}, we specialise our considerations to fractional levels $k>-2$, including a detailed discussion of
the associated Verma modules and the Malikov-Feigin-Fuchs singular vectors \cite{MFF86} generating the submodules.
We extend the Fuchs-Astashkevich theorem \cite{Ast97} from the Virasoro algebra to $A_1^{(1)}$\! and 
obtain explicit expressions for all irreducible characters appearing in the decomposition of Verma modules over $A_1^{(1)}$\!.
These affine characters are expressed in terms of the {\em reduced theta functions} defined in Appendix~\ref{Sec:GenTheta}.
We also recall the $A_1^{(1)}$\! string functions.
In Section~\ref{Sec:Kac}, we define the affine Kac modules and discuss the corresponding affine Kac characters. 
We classify the irreducible affine Kac modules and identify the larger set of affine Kac modules 
which happen to be isomorphic to highest-weight quotient modules.
In Section~\ref{Sec:Stag}, we discuss the structure of staggered $A_1^{(1)}$-modules and conjecture the existence
of three infinite classes of such modules. Two examples of staggered modules are discussed in detail, illustrating that the Cartan generator
$J_0^3$ may or may not act diagonalisably. The Virasoro generator $L_0$, on the other hand, acts non-diagonalisably in both examples.
In Section~\ref{Sec:Coset}, we apply the coset construction (\ref{cosetAAA}) to affine Kac characters and derive
an expression for the corresponding branching functions in terms of $A_1^{(1)}$\! string functions, 
confirming a result of~\cite{PR13}. From this, we obtain the branching rules and functions for staggered characters.
Using details of the reciprocal of the Jacobi triple product, discussed in Appendix~\ref{Sec:rec}, alternative expressions
for the branching functions are presented in Appendix~\ref{Sec:Alt}.
In Section~\ref{Sec:Fun}, we recall the Mukhi-Panda residue formula and discuss how it generalises from admissible
characters to general irreducible, affine Kac and staggered characters. These results are then elevated to the level of modules
by the introduction of the aforementioned functor. Its action on affine Kac and staggered modules is described in detail.
Section~\ref{Sec:Concl} contains some concluding remarks.

\section{Basics on $A_1^{(1)}$}
\label{Sec:Rep}

\subsection{Algebraic aspects}

The affine Lie algebra
\be
 A_1^{(1)}=\big\langle J_n^a,\,K,\,D;\,a\in\{+,-,3\},\,n\in\mathbb{Z}\big\rangle
\ee
is generated by the modes $J_n^a$, the central element $K$ and the derivation $D$,
having Lie products
\be
 [J_n^3,J_m^\pm]=\pm J_{n+m}^\pm,\quad
 [J_n^+,J_m^-]=2J_{n+m}^3+nK\delta_{n+m,0},\quad
 [J_n^3,J_m^3]=\tfrac{1}{2}nK\delta_{n+m,0},\quad
 [D,J_n^a]=nJ_n^a.
\ee
All other Lie products vanish.
Following a widespread tradition in the literature, we are using the notation $J_n^+=E_n$, $J_n^3=\frac{1}{2}H_n$,
and $J_n^-=F_n$, where $\{E_0,H_0,F_0\}$ is the standard basis for $A_1$.
The derived algebra,
\be
 [A_1^{(1)},A_1^{(1)}]=\big\langle J_n^a,\,K;\,a\in\{+,-,3\},\,n\in\mathbb{Z}\big\rangle,
\ee
enjoys the Cartan decomposition
\be
 [A_1^{(1)},A_1^{(1)}]=\mathfrak{n}_{-}\oplus\mathfrak{n}_0\oplus\mathfrak{n}_+,
\ee
where
\be
 \mathfrak{n}_-:=\langle J_0^-,J_{-1}^+\rangle,\qquad
 \mathfrak{n}_0:=\langle J_0^3,K\rangle,\qquad
 \mathfrak{n}_+:=\langle J_{1}^-,J_0^+\rangle
\ee
are subalgebras of $[A_1^{(1)},A_1^{(1)}]$.
Associated with this,
\be
 \mathfrak{b}_+:=\langle J_0^3,K,D,J_{1}^-,J_0^+\rangle,\qquad
 \mathfrak{b}_+':=\langle J_0^3,K,J_{1}^-,J_0^+\rangle
\ee
are Borel subalgebras of $A_1^{(1)}$ and $[A_1^{(1)},A_1^{(1)}]$, respectively.

On $A_1^{(1)}$, the linear dagger operation defined by
\be
 K^\dagger:=K,\qquad 
 D^\dagger:=D,\qquad
 (J^\pm_n)^\dagger:=J^\mp_{-n},\qquad 
 (J_n^3)^\dagger:=J_{-n}^3,\qquad n\in\mathbb{Z},
\label{dagger}
\ee
is an anti-homomorphism in the sense that
\be
 [A,B]^\dagger=[B^\dagger,A^\dagger],\qquad \forall A,B\in A_1^{(1)}.
\ee
The operation is thus an anti-automorphism.
It is extended to the universal enveloping algebra $U(A_1^{(1)})$ by setting $I^\dagger=I$ 
and $(AB)^\dagger=B^\dagger A^\dagger$ for $A,B\in U(A_1^{(1)})$.

On the modules of our interest, the central element $K$ acts as a scalar multiple of the identity operator, $K=kI$,
where $k\in\mathbb{C}$ is called the level. In fact, we often work over the quotient of $U(A_1^{(1)})$ by the
two-sided ideal generated by $K-kI$.
The root system of the underlying Lie algebra $A_1$ is normalised
by $\theta^2=2$, where $\theta$ is the highest root, such that $k^\vee:=\frac{2k}{\theta^2}=k$. 
It is furthermore convenient to introduce the {\em shifted level}
\be
 t:=k+2,
\ee
where $h^\vee=t-k^\vee=2$ is the dual Coxeter number of $A_1$. Throughout, we will assume that
\be
 t\neq0,
\ee
thereby steering clear of the so-called {\em critical level}, $k=-2$. In fact, we are primarily interested in $t\in\mathbb{Q}_+$.

The Segal-Sugawara (or affine Sugawara) construction
\be
 L_n=\frac{1}{2t}\sum_{m\in\mathbb{Z}}\Big(\!:\!J_{n-m}^-J_m^+\!:+:\!J_{n-m}^+J_m^-\!:
   +\,2\!:\!J_{n-m}^3J_m^3\!:\!\Big),\qquad n\in\mathbb{Z},
\ee
with $:...:$ denoting the usual normal ordering, yields a realisation of the Virasoro algebra
\be
 [L_n,L_m]=(n-m)L_{n+m}+\frac{c}{12}(n^3-n)\delta_{n+m,0},\qquad n,m\in\mathbb{Z},
\ee
with central charge
\be
 c=c_k:=\frac{3k}{k+2}.
\ee
It follows that
\be
 [L_n,J_m^a]=-mJ_{n+m}^a,
\ee
so we may represent $D$ by the Virasoro mode $-L_0$.
Thus combining the Segal-Sugawara construction with the affine Lie algebra itself, we obtain
what we here refer to as the {\em affine current algebra} $A_1^{(1)}$\!.
Except in Section~\ref{Sec:Coset}, our focus will be on the affine Lie algebra $A_1^{(1)}$ itself,
although we shall mostly use $-L_0$ as the derivation. We also note that
\be
 L_0=\frac{1}{t}\Big(J_0^3(J_0^3+1)+\sum_{m=0}^\infty J_{-m}^-J_m^+
  +\sum_{m=1}^\infty(J_{-m}^+J_m^-+2J_{-m}^3J_m^3)\Big).
\label{L0}
\ee

The universal enveloping algebra of $\mathfrak{n}_+$ decomposes as
\be
 U(\mathfrak{n}_+)=\big\langle I\big\rangle\oplus\big(U(\mathfrak{n}_+)J_1^-+U(\mathfrak{n}_+)J_0^+\big),
\label{Unp}
\ee
where only the first summation is direct. Indeed,
\be
 U(\mathfrak{n}_+)J_1^-\cap U(\mathfrak{n}_+)J_0^+\neq\emptyset,
\ee
as illustrated by
\be
 (J_0^+)^3J_1^-=\big(J_1^-(J_0^+)^2+6J_1^3J_0^+-6J_1^+\big)J_0^+.
\label{JJ}
\ee
It follows from (\ref{L0}) and (\ref{Unp}) that
\be
 L_0-\tfrac{1}{t}J_0^3(J_0^3+1)\in U(\mathfrak{n}_-)\big(U(\mathfrak{n}_+)J_1^-+U(\mathfrak{n}_+)J_0^+\big).
\ee
For later convenience, especially when evaluating the action of $L_0-\tfrac{1}{t}J_0^3(J_0^3+1)$ 
on specific vectors in Sections~\ref{Sec:StagEx} and~\ref{Sec:j10}, we note the decompositions
\begin{align}
 J_1^+&=\big[\!-\tfrac{1}{2}(J_0^+)^2\big]J_1^-+\big[J_1^3+\tfrac{1}{2}J_0^+J_1^-\big]J_0^+,
\label{J1+}
 \\[.2cm]
 J_2^+&=\big[\!-\tfrac{1}{2}(J_1^3J_0^++J_1^+)J_0^+\big]J_1^-
   +\big[(J_1^3)^2+\tfrac{1}{2}(J_1^3J_0^++J_1^+)J_1^-\big]J_0^+,
\label{J2+}
\\[.2cm]
 J_2^-&=\big[\tfrac{1}{2}J_1^-J_0^+-J_1^3\big]J_1^-+\big[\!-\tfrac{1}{2}(J_1^-)^2\big]J_0^+,
\label{J2-}
\\[.2cm]
 J_1^3&=\big[\tfrac{1}{2}J_0^+\big]J_1^-+\big[\!-\tfrac{1}{2}J_1^-\big]J_0^+,
\label{J13}
\\[.2cm]
 J_2^3&=\big[\tfrac{1}{2}J_0^+(\tfrac{1}{2}J_1^-J_0^+-J_1^3)\big]J_1^-
   +\big[\!-\tfrac{1}{2}(\tfrac{1}{2}J_0^+(J_1^-)^2+J_2^-)\big]J_0^+.
\label{J23}
\end{align}

\subsection{Verma modules}

For $j,h\in\mathbb{C}$, let $B_{j,h}$ be a one-dimensional $\mathfrak{b}_+$-module such that
\be
 J_0^3v=jv,\qquad Kv=kv,\qquad Dv=-hv,\qquad \mathfrak{n}_+v=0,\qquad \forall v\in B_{j,h}.
\ee
Setting $D=-L_0$, the corresponding Verma module over $A_1^{(1)}$ is defined as
\be
 \Vc_j:=\mathrm{Ind}_{\mathfrak{b}_+}^{A_1^{(1)}}\!B_{j,h_j},
\ee
where, in accordance with (\ref{L0}),
\be
 h_j=\frac{j(j+1)}{t}.
\ee
This module enjoys a vector-space decomposition of the form
\be
 \Vc_j=\bigoplus_{N\in\mathbb{N}_0}\bigoplus_{Q\in\mathbb{Z}}\,[\Vc_j]_{Q,N},
\label{VjDec}
\ee
where the weight spaces are defined as
\be
 [\Vc_j]_{Q,N}:=\{v\in\Vc_j\,|\,J_0^3v=(j+Q)v,\,L_0v=(h_j+N)v\}.
\ee
A nonzero vector in $[\Vc_j]_{Q,N}$ is said to be of {\em charge} $Q$ and at {\em grade} $N$.
It is convenient to view the Verma module $\Vc_j$ as freely generated from the highest-weight vector $\ket{j}$ satisfying
\be
 J_0^3\ket{j}=j\ket{j},\qquad K\ket{j}=k\ket{j},\qquad L_0\ket{j}=h_j\ket{j},\qquad
 \mathfrak{n}_+\ket{j}=0,
\label{hwv}
\ee
by the action of the universal enveloping algebra $U(\mathfrak{n}_-)$:
\be
 \Vc_j=U(\mathfrak{n}_-)\ket{j}.
\ee
As indicated, in the ket notation, we generally suppress the dependence on $k$.

For a module $\Mc$ with finite-dimensional generalised weight spaces with respect to the pair $(L_0,J_0^3)$, we define
the corresponding {\em affine character} as
\be
 \chit[\Mc](q,z):=\mathrm{Tr}_{\Mc}^{\phantom{V}}q^{L_0-\frac{c}{24}}z^{-J_0^3}.
\label{chiM}
\ee
With $\chit_j(q,z)=\chit[\Vc_j](q,z)$ denoting the affine character of the Verma module $\Vc_j$,
it follows from (\ref{VjDec}) that
\be
 \chit_j(q,z)=\sum_{N\in\mathbb{N}_0}\sum_{Q\in\mathbb{Z}}(\dim[\Vc_j]_{Q,N})q^{h_j-\frac{c}{24}+N}z^{-j-Q},
\ee
where $\dim[\Vc_j]_{Q,N}$ is independent of $j$.
The minus sign on $J_0^3$ in (\ref{chiM}) ensures that, for fixed $N\in\mathbb{N}_0$,
\be
 \sum_{Q\in\mathbb{Z}}(\dim[\Vc_j]_{Q,N})z^{-Q}\in\mathbb{N}((z)),
\label{Laurent}
\ee
meaning that the sum is a formal Laurent series in $z$ with positive integer coefficients. 
Likewise, for $Q$ fixed, we have the power series
\be
 \sum_{N\in\mathbb{N}_0}(\dim[\Vc_j]_{Q,N})\,q^N\!\in\mathbb{N}_0[[q]].
\label{Taylor}
\ee
Explicit expressions for the affine character $\chit_j(q,z)$ include
\be
 \chit_j(q,z)
 =\frac{q^{h_j-\frac{c}{24}}\,z^{-j}}{\varphi(q,z)}
 =\frac{q^{h_j-\frac{c}{24}+\frac{1}{8}}\,z^{-j-\frac{1}{2}}}{\Theta_{1,2}(q,z)-\Theta_{-1,2}(q,z)}
 =\frac{q^{\frac{1}{t}(j+\frac{1}{2})^2}\,z^{-j-\frac{1}{2}}}{\eta(q,z)},
\label{chiVh}
\ee
where our conventions for $\varphi(q,z)$, $\eta(q,z)$ and $\Theta_{n,m}(q,z)$ 
are given in Appendix~\ref{Sec:GenTheta}.

\subsection{Reducibility}

The Verma module $\Vc_j$ is reducible if and only if $j=j_{r,s}$
for some $r,s\in\mathbb{Z}$ \cite{KK79}, where
\be
 2j_{r,s}+1=r-st,\qquad rs>0\ \ \mathrm{or}\ \ r>0,\,s=0.
\label{2j1}
\ee
Correspondingly, in the reducible Verma module $\Vc_{j_{r,s}}$, there exists a singular vector
$\ket{r,s}$ of charge $Q=-r$ and grade $N=rs$, that is,
\be
 \mathfrak{n}_+\ket{r,s}=0,\qquad
 J_0^3\ket{r,s}=\big(j_{r,s}-r\big)\ket{r,s},\qquad
 L_0\ket{r,s}=\big(h_{r,s}+rs\big)\ket{r,s},
\ee
where
\be
 h_{r,s}:=h_{j_{r,s}}=\frac{(r-s t)^2-1}{4t}.
\ee
Up to scaling, this singular vector is unique.
Extending the domain for the indices $r,s$ of $j$ and $h$ to all of $\mathbb{Z}$, we have
\be
 j_{-r,s}=j_{r,s}-r,\qquad h_{-r,s}=h_{r,s}+rs,\qquad r,s\in\mathbb{Z},
\ee
and note that $j_{-r,-s}=-j_{r,s}-1$ and $h_{-r,-s}=h_{r,s}$.
One can thus form the quotient modules
\be
 \Qc_{r,s}:=\Vc_{j_{r,s}}/\Vc_{j_{-r,s}},\qquad rs>0\ \ \mathrm{or}\ \ r>0,\,s=0.
\ee
Likewise, the {\em irreducible} module $\Ic_{r,s}$ is constructed by
quotienting out the maximal proper submodule of $\Vc_{j_{r,s}}$.
The affine character of $\Qc_{r,s}$ is given by
\be
 \chit_{r,s}(q,z)
  =\chit_{j_{r,s}}(q,z)-\chit_{j_{-r,s}}(q,z)
  =\frac{q^{h_{r,s}-\frac{c}{24}+\frac{1}{8}}\,z^{-j_{r,s}-\frac{1}{2}}}{\eta(q,z)}
     \big(1-q^{rs}z^{r}\big)
\label{Kact}
\ee
and is referred to as an {\em affine Kac character}.
Using (\ref{zm1}), we note that
\be
 \chit_{-r,-s}(q,z)=-\chit_{r,s}(q,z^{-1}).
\label{chizm1}
\ee
Affine Kac characters were introduced in~\cite{PR13} and will play an important role in the present work,
but do not otherwise seem to have attracted much attention in the literature.

It is stressed that a {\em reducible} affine Kac character $\chit_{r,s}(q,z)$
is actually the character of several distinct modules. 
By construction, it is the character of the highest-weight quotient module $\Qc_{r,s}$.
However, it is also the character of the {\em direct sum} of the irreducible modules whose characters
appear in the decomposition of $\chit_{r,s}(q,z)$. In addition, as discussed in Section~\ref{Sec:Kac}, 
it is the character of the corresponding {\em affine Kac module} defined as a particular finitely generated submodule of a
Wakimoto module of weight $j_{r,s}$ \cite{Wak86,BF90}.

\section{Fractional levels}
\label{Sec:Frac}

\subsection{Admissibility}

We are interested in the affine Lie algebra $A_1^{(1)}$\! at fractional level such that
the shifted level $t=k+2$ is positive.
We thus introduce the parameterisation
\be
 t=\frac{p}{p'},\qquad \mathrm{gcd}(p,p')=1,\qquad p,p'\in\mathbb{N},
\label{tpp}
\ee
in which case the weights enjoy the periodicity
\be
 j_{r,s}=j_{r+\ell p,s+\ell p'},\qquad \ell\in\mathbb{Z}.
\label{jperiodic}
\ee
Such a fractional level is said to be {\em admissible} if $p\geq2$. 
The associated {\em admissible weights} $j_{r_0,s_0}$ are of the form
\be
 2j_{r_0,s_0}+1=r_0-s_0 t,\qquad 
1\leq r_0\leq p-1,\qquad0\leq s_0\leq p'-1.
\label{jadm}
\ee
In many applications, one works with the corresponding irreducible highest-weight representations,
also known as {\em admissible representations}~\cite{KW88,KW88b}.
Among these, the ones for which $s=0$, such that $2j\in\mathbb{N}_0$, play an important role
in this work and are here referred to as {\em quasi-integrable}. Weights satisfying $2j\in\mathbb{N}_0$
are likewise called quasi-integrable.
The familiar {\em integrable} representations are recognised as the quasi-integrable ones for which $r=1,\ldots,p-1$
and $k\in\mathbb{N}_0$.
As already indicated in (\ref{tpp}), we stress that we shall allow 
\be
 p\geq1,
\label{p1}
\ee
such that all $t\in\mathbb{Q}_+$ are covered.
Accordingly, we refer to levels of the form (\ref{tpp}) simply as {\em fractional}.

For fractional level, following (\ref{jperiodic}),
the Verma module characters are invariant under simultaneous translations of $r,s$ by multiples of $p,p'$
such that $(r+\ell p)(s+\ell p')>0$ or $r+\ell p>0,\,s+\ell p'=0$:
\be
 \chit_{j_{r+\ell p,s+\ell p'}}(q,z)=\chit_{j_{r,s}}(q,z),\qquad \ell\in\mathbb{Z}.
\ee 
Contrarily, the affine Kac characters (\ref{Kact}) are {\em all distinct}.

\subsection{Verma modules}

For $t\in\mathbb{Q}_+$ and $j_{r,s}$ as in (\ref{2j1}), the submodule structure 
of the Verma module $\Vc_{j_{r,s}}$ depends critically on whether or not $r$ is a multiple of $p$.
Indeed, for $r$ {\em not} a multiple of $p$, the submodule structure is described by a Loewy
diagram of the form~\cite{KK79,MFF86,Mal91,KacBook}
\psset{unit=.8cm}
\setlength{\unitlength}{.8cm}
\be
\begin{pspicture}(0,0.6)(9,2.5)
 \rput(-4,0.73){$r\not\in p\mathbb{Z}$\,:}
 \rput(-1.06,.73){$\bullet$}
 \rput(0,1.5){$\bullet$}
 \rput(0.8,1.5){$\to$}
 \rput(1.6,1.5){$\bullet$}
 \rput(2.4,1.5){$\to$}
 \rput(3.2,1.5){$\bullet$}
 \rput(4,1.5){$\to$}
 \rput(4.8,1.5){$\bullet$}
 \rput(5.6,1.5){$\to$}
 \rput(6.4,1.5){$\bullet$}
 \rput(7.2,1.5){$\to$}
 \rput(8,1.5){$\bullet$}
 \rput(8.8,1.5){$\to$}
 \rput(0,0){$\bullet$}
 \rput(0.8,0){$\to$}
 \rput(1.6,0){$\bullet$}
 \rput(2.4,0){$\to$}
 \rput(3.2,0){$\bullet$}
 \rput(4,0){$\to$}
 \rput(4.8,0){$\bullet$}
 \rput(5.6,0){$\to$}
 \rput(6.4,0){$\bullet$}
 \rput(7.2,0){$\to$}
 \rput(8,0){$\bullet$}
 \rput(8.8,0){$\to$}
 \rput(-1.6,0.73){$\al_0$}
 \rput(0,2){$\beta_0$}
 \rput(1.6,2){$\al_{-1}$}
 \rput(3.2,2){$\beta_{-1}$}
 \rput(4.8,2){$\al_{-2}$}
 \rput(6.4,2){$\beta_{-2}$}
 \rput(8,2){$\al_{-3}$}
 \rput(0,-0.5){$\beta_1$}
 \rput(1.6,-0.5){$\al_{1}$}
 \rput(3.2,-0.5){$\beta_{2}$}
 \rput(4.8,-0.5){$\al_{2}$}
 \rput(6.4,-0.5){$\beta_{3}$}
 \rput(8,-0.5){$\al_{3}$}
 \rput(-0.6,1.25){$\nearrow$}
 \rput(-0.6,0.25){$\searrow$}
 \rput(0.8,0.73){$\searrow$}
 \rput(0.8,0.73){$\nearrow$}
 \rput(2.4,0.73){$\searrow$}
 \rput(2.4,0.73){$\nearrow$}
 \rput(4,0.73){$\searrow$}
 \rput(4,0.73){$\nearrow$}
 \rput(5.6,0.73){$\searrow$}
 \rput(5.6,0.73){$\nearrow$}
 \rput(7.2,0.73){$\searrow$}
 \rput(7.2,0.73){$\nearrow$}
 \rput(8.8,0.73){$\searrow$}
 \rput(8.8,0.73){$\nearrow$}
 \rput(9.8,0){$\ldots$}
 \rput(9.8,1.5){$\ldots$}
\end{pspicture} 
\\[1cm]
\label{diagram2}
\ee
whereas for $r$ a nonzero multiple of $p$, it is described by a Loewy diagram of the form
\psset{unit=.8cm}
\setlength{\unitlength}{.8cm}
\be
\begin{pspicture}(0,0)(9,0.8)
 \rput(-4,0.05){$r\in p\mathbb{Z}^\times$:}
 \rput(0,0){$\bullet$}
 \rput(0.8,0){$\to$}
 \rput(1.6,0){$\bullet$}
 \rput(2.4,0){$\to$}
 \rput(3.2,0){$\bullet$}
 \rput(4,0){$\to$}
 \rput(4.8,0){$\bullet$}
 \rput(5.6,0){$\to$}
 \rput(6.4,0){$\bullet$}
 \rput(7.2,0){$\to$}
 \rput(8,0){$\bullet$}
 \rput(8.8,0){$\to$}
 \rput(0,0.57){$\gamma_1$}
 \rput(1.6,0.57){$\gamma_2$}
 \rput(3.2,0.57){$\gamma_3$}
 \rput(4.8,0.57){$\gamma_4$}
 \rput(6.4,0.57){$\gamma_5$}
 \rput(8,0.57){$\gamma_6$}
 \rput(9.8,0){$\ldots$}
\end{pspicture} 
%\\[0.5cm]
\label{diagram1}
\ee
Each node in these diagrams represents a Verma module of highest weight 
$\al_i$, $\beta_i$ or $\gamma_i$, where
an arrow connecting two such modules, $\Vc_\mu\to\Vc_{\mu'}$, indicates that the target module,
$\Vc_{\mu'}$, is a submodule of the initial module, $\Vc_\mu$.
Moreover, the submodules in (\ref{diagram2}) or (\ref{diagram1}) are all generated by singular vectors,
where the maximal dimension of the space of singular vectors of any given weight is one.

Every Verma module of the form $\Vc_{j_{r,s}}$ appears as a submodule of exactly one of the $(p+1)p'$
distinct Verma modules
\be
 \Vc_{j_{\rho,s_0}},\qquad 0\leq \rho\leq p,\qquad 0\leq s_0\leq p'-1,
\label{Vambient}
\ee
where it is noted that $j_{0,s_0}=j_{-p,-(p'-s_0)}=j_{p,p'+s_0}$ and $j_{p,s_0}$ are indeed of the form (\ref{2j1}).
The submodule structure of $\Vc_{j_{r,s}}$ thus follows from the Loewy diagram, (\ref{diagram2})
or (\ref{diagram1}), of the corresponding ambient Verma module (\ref{Vambient});
one merely has to identify the ambient Loewy diagram and locate the position of the 
Verma module $\Vc_{j_{r,s}}$ in that diagram. To see this in action, we first recall the submodule
structures of the Verma modules (\ref{Vambient}). 

For $\rho=r_0$, where as usual $1\leq r_0\leq p-1$, the Loewy diagram is of the form (\ref{diagram2}) with
\be
 \al_i(r_0,s_0)=j_{r_0+2ip,s_0}=j_{r_0,s_0}+ip,\qquad
 \beta_i(r_0,s_0)=j_{-r_0+2ip,s_0}=j_{r_0,s_0}-r_0+ip,\qquad
 i\in\mathbb{Z}.
\label{ab}
\ee
For $\rho=\delta p$, where $\delta=0,1$, the Loewy diagram 
can be obtained from a Loewy diagram of the form (\ref{diagram2}) by imposing
\be
 \al_{i-\delta}\equiv\beta_{i},\qquad i\in\mathbb{Z}.
\label{a=b}
\ee
The double-string diagram thereby collapses to a single-string diagram of the form (\ref{diagram1}),
with weights given by
\be
 \gamma_i(\delta p,s_0)=j_{\delta p,s_0}+(-1)^{i+\delta}\lfloor\tfrac{i}{2}\rfloor p,\quad i\in\mathbb{N}.
\ee

Now, every Verma module $\Vc_{j_{r,s}}$ with $r\not\in p\mathbb{Z}$ (thus presupposing $p>1$)
appears as a submodule of a Verma module of admissible weight $j_{r_0,s_0}$ for some
$r_0,s_0$ as in (\ref{jadm}). To locate the position of the Verma module $\Vc_{j_{r,s}}$ in the 
corresponding ambient Loewy diagram, we write
\be
 r=\rh_0+\ell p,\qquad s=\sh_0+\ell'p',\qquad 1\leq \rh_0\leq p-1,\qquad 0\leq \sh_0\leq p'-1,
\label{rolsol}
\ee
where $\ell,\ell'\in\mathbb{Z}$ are restricted such that $r,s$ satisfy (\ref{2j1}), that is,
\be
 \ell,\ell'\in\mathbb{N}_0\quad\mathrm{or}\quad
 \ell,\ell'\in(-\mathbb{N}).
\ee
In all cases, we have
\be
 \ell=\lfloor\tfrac{r}{p}\rfloor,\qquad
 \ell'=\lfloor\tfrac{s}{p'}\rfloor.
\label{ellell}
\ee
It then follows that
\be
 \Vc_{j_{\rh_0+\ell p,\sh_0+\ell'p'}}=\begin{cases}
  \Vc_{\al_{\frac{\ell-\ell'}{2}}(\rh_0,\sh_0)}\subset\Vc_{j_{\rh_0,\sh_0}},\ &
    \ell-\ell'\ \,\mathrm{even},
  \\[.3cm]
  \Vc_{\beta_{\frac{\ell-\ell'+1}{2}}(p-\rh_0,\sh_0)}\subset\Vc_{j_{p-\rh_0,\sh_0}},\ &
    \ell-\ell'\ \,\mathrm{odd}.
\end{cases}
\label{VinV2}
\ee
Similarly for $r\in p\mathbb{Z}^\times$, one finds that
\be
 \Vc_{j_{\ell p,\sh_0+\ell'p'}}=\begin{cases}
  \Vc_{\gamma_{|\ell-\ell'-\frac{1}{2}|+\frac{1}{2}}(0,\sh_0)}\subset\Vc_{j_{-p,\sh_0-p'}},\ &
    \ell-\ell'\ \,\mathrm{even},
  \\[.3cm]
  \Vc_{\gamma_{|\ell-\ell'-\frac{1}{2}|+\frac{1}{2}}(p,\sh_0)}\subset\Vc_{j_{p,\sh_0}},\ &
    \ell-\ell'\ \,\mathrm{odd}.
\end{cases}
\label{VinV1}
\ee

\subsection{Singular vectors}
\label{Sec:Sing}

We now turn to the singular vectors generating the submodules of the reducible Verma modules, recalling
that a singular vector is merely a highest-weight vector, as in (\ref{hwv}). Explicit, albeit formal, expressions
were obtained in~\cite{MFF86} and are known as Malikov-Feigin-Fuchs vectors. Here, we elaborate on this result
and recast it in a light and notation suiting our purposes, including the construction of staggered modules.

Let $\ket{j}$ be a highest-weight vector of weight $j$. From the relations
\begin{align}
 [J_0^+,(J_0^-)^x]&=x(J_0^-)^{x-1}\big((1-x)I+2J_0^3\big),
 \\[.15cm]
 [J_1^-,(J_{-1}^+)^x]&=x(J_{-1}^+)^{x-1}\big((k+1-x)I-2J_0^3\big),
\end{align}
it then follows that $(J_0^-)^x\ket{j}$ is a singular vector if and only if
$x=0$ or $x=2j+1$, and that $(J_{-1}^+)^x\ket{j}$ is singular if and only if $x=0$ or $x=k+1-2j$.
Of course, for any of these expressions to be a vector in $\Vc_j$, the corresponding exponent, $x$, must be
a non-negative integer. Following~\cite{MFF86}, however, it is useful to formally extend the 
domain for $x$ to all of $\mathbb{C}$.

In this general setting, we let $F$ denote the set of highest-weight vectors $\ket{j}$, $j\in\mathbb{C}$,
and introduce a pair of linear operators on $F$, defined by
\be
 b:\ket{j}\mapsto(J_0^-)^{2j+1}\ket{j},\qquad b_t:\ket{j}\mapsto(J_{-1}^+)^{t-2j-1}\ket{j}.
\label{bb}
\ee
It follows from
\be
 [J_m^3,(J_0^-)^x]=-xJ_m^-(J_0^-)^{x-1},\qquad
 [J_m^3,(J_{-1}^+)^x]=xJ_{m-1}^+(J_{-1}^+)^{x-1},
\label{J3J}
\ee
with $m=0$, that the images in (\ref{bb}) are indeed in $F$, as
\be
 b\ket{j}\propto\ket{-j-1},\qquad b_t\ket{j}\propto\ket{t-j-1}.
\ee
Since $-(-j-1)-1=j$ and $t-(t-j-1)-1=j$, it also follows that
\be
 b^2=b_t^2=I.
\ee
For any given $j$, we thus have the following chain of Verma modules:
\be
 \ldots
 \begin{array}{c} {}_{b_t}\\[-.1cm] \longleftrightarrow\\[-.1cm] \phantom{{}_{b_t}}\end{array}
 \Vc_{-t-j-1}\begin{array}{c} {}_{b}\\[-.1cm] \longleftrightarrow\\[-.1cm] \phantom{{}_{b}}\end{array}
 \Vc_{j+t}\begin{array}{c} {}_{b_t}\\[-.1cm] \longleftrightarrow\\[-.1cm] \phantom{{}_{b_t}}\end{array}
 \Vc_{-j-1}\begin{array}{c} {}_{b}\\[-.1cm] \longleftrightarrow\\[-.1cm] \phantom{{}_{b}}\end{array}
 \Vc_j\begin{array}{c} {}_{b_t}\\[-.1cm] \longleftrightarrow\\[-.1cm] \phantom{{}_{b_t}}\end{array}
 \Vc_{t-j-1}\begin{array}{c} {}_{b}\\[-.1cm] \longleftrightarrow\\[-.1cm] \phantom{{}_{b}}\end{array}
 \Vc_{j-t}\begin{array}{c} {}_{b_t}\\[-.1cm] \longleftrightarrow\\[-.1cm] \phantom{{}_{b_t}}\end{array}
 \Vc_{2t-j-1}\begin{array}{c} {}_{b}\\[-.1cm] \longleftrightarrow\\[-.1cm] \phantom{{}_{b}}\end{array}
 \ldots
\label{Vbb}
\ee
The highest weights appearing in this chain form the corresponding {\em weight orbit},
\be
 O_j=\{j+nt,\,nt-j-1\,|\,n\in\mathbb{Z}\},
\ee
where we note that $O_{j'}=O_j$ if and only if $j'\in O_j$.

We now fix the relative normalisations of the highest-weight vectors of the Verma modules in (\ref{Vbb}).
For given $\ket{j}$ and $j'\in O_j$, we thus set
\be
 \ket{j'}:=S_{j',j}\ket{j},
\ee
where
\be
 S_{j+nt,j}=(b_tb)^{n},\qquad
 S_{nt-j-1,j}=(b_tb)^{n}b,
\label{SS}
\ee
and note that $(b_tb)^{-1}=b^{-1}b_t^{-1}=bb_t$. It follows that
\be
 S_{j,j}=I,\qquad S_{j,j'}=(S_{j',j})^{-1},\qquad S_{j,j'}=S_{j,j''}S_{j'',j'},\qquad \forall\, j''\in O_j.
\ee
Explicitly, we have
\begin{align}
 \ket{j+nt}&=\begin{cases} (J_{-1}^+)^{(2n-1)t+2j+1}\ldots(J_0^-)^{2t+2j+1}(J_{-1}^+)^{t+2j+1}(J_0^-)^{2j+1}\ket{j},\ &n>0,
  \\[.25cm]
  (J_0^-)^{-2nt-2j-1}\ldots(J_{-1}^+)^{3t-2j-1}(J_0^-)^{2t-2j-1}(J_{-1}^+)^{t-2j-1}\ket{j},\ &n<0,
 \end{cases}
 \\[.4cm]
 \ket{nt-j-1}&=\begin{cases} (J_{-1}^+)^{(2n-1)t-2j-1}\ldots(J_{-1}^+)^{3t-2j-1}(J_0^-)^{2t-2j-1}(J_{-1}^+)^{t-2j-1}\ket{j},\ &n>0,
  \\[.25cm]
  (J_0^-)^{-2nt+2j+1}\ldots(J_0^-)^{2t+2j+1}(J_{-1}^+)^{t+2j+1}(J_0^-)^{2j+1}\ket{j},\ &n\leq0.
 \end{cases}
\end{align}
By commuting the terms in (\ref{SS}) around, we see that
\be
 S_{j',j}=(J_0^-)^{(h_{j'}-h_j)-(j'-j)}(J_{-1}^+)^{h_{j'}-h_j}+\ldots,
\label{SFA}
\ee
where the omitted terms all involve nontrivial (possibly complex) monomials in $\mathfrak{n}_-$-generators different from
$J_0^-$ and $J_{-1}^+$.

As the weights $j'$ and $j$ must differ by an integer for $S_{j',j}\ket{j}\in\Vc_j$,
we also consider the {\em sub-orbit}
\be
 \Ob_j:=\{j'\in O_j\,|\,j'-j\in\mathbb{Z}\}.
\ee
It readily follows that
\be
 j'\in \Ob_j\quad\Longleftrightarrow\quad j\in\Ob_{j'}
\ee
and
\be
 j''\in\Ob_{j'},\ j'\in\Ob_j\quad\Longrightarrow\quad j''\in\Ob_j.
\ee
In the following, we will assume that $2j+1=r-st$ where $t=\frac{p}{p'}$, in which case
\be
 \Ob_j=\{j+np,\,j-r+np\,|\,n\in\mathbb{Z}\}
\ee
and
\be
 j'\in\Ob_j\quad\Longrightarrow\quad h_{j'}-h_j\in\mathbb{Z}.
\ee
Moreover, for $j'\in\Ob_j$, we can write
\be
 S_{j+np,j}=(b_tb)^{np'},\qquad
 S_{j-r+np,j}=(b_tb)^{np'-s}b.
\label{SbbSbb}
\ee
For $r\in p\mathbb{Z}^\times$, we note that every weight in the sub-orbit $\Ob_j$ appears exactly twice, whereas for
$r\notin p\mathbb{Z}$, the weights in $\Ob_j$ are all distinct. In both cases, $\Ob_j$ matches the 
set of weights appearing in the
Loewy diagram of the corresponding ambient Verma module, see (\ref{VinV1}) and (\ref{VinV2}).

Let $j'\in\Ob_j$.
Following (\ref{SFA}), a {\em necessary condition} for $S_{j',j}\ket{j}\in\Vc_j$ is that
\be
 h_{j'}-h_j\geq\max(j'-j,0).
\label{hhjj}
\ee
To appreciate that this is in general not a sufficient condition, we consider $k=-\frac{1}{2}$, $j=-1$ and $j'=2$,
in which case $h_j=0$ and $h_{j'}=4$, thus satisfying (\ref{hhjj}). However, using
\begin{align}
 [J_m^-,(J_{-1}^+)^x]&=x(x-1)J_{m-2}^+(J_{-1}^+)^{x-2}-2xJ_{m-1}^3(J_{-1}^+)^{x-1},
 \\[.2cm]
 [J_m^+,(J_0^-)^x]&=2x(J_0^-)^{x-1}J_m^3-x(x-1)J_m^-(J_0^-)^{x-2},
\end{align}
and (\ref{J3J}), we find that
\begin{align}
 S_{2,-1}&=(J_{-1}^+)^{\frac{7}{2}}(J_0^-)^2(J_{-1}^+)^{\frac{1}{2}}(J_0^-)^{-1}
 \nonumber\\[.2cm]
 &=\big[\tfrac{11}{2}J_{-1}^3J_{-2}^+J_{-1}^+-\tfrac{7}{4}J_{-3}^+J_{-1}^+-\tfrac{1}{2}J_{-2}^3(J_{-1}^+)^2-\tfrac{71}{16}(J_{-2}^+)^2
  -(J_{-1}^3)^2(J_{-1}^+)^2-(J_{-1}^+)^3J_{-1}^-
 \big](J_0^-)^{-1}
 \nonumber\\[.2cm]
 &+(J_{-1}^+)^4J_0^--2J_{-1}^3(J_{-1}^+)^3+\tfrac{13}{2}J_{-2}^+(J_{-1}^+)^2.
\end{align}
As this is expressed in a generalised PBW-type basis,
\be
 \big\{\ldots,(J_{-m_3}^3)^{x_3}\ldots(J_{-m_3'}^3)^{x_3'}
  (J_{-m_+}^+)^{x_+}\ldots(J_{-m_+'}^+)^{x_+'}(J_{-m_-}^-)^{x_-}\ldots(J_{-m_-'}^-)^{x_-'},\ldots\big\},
\ee
and contains terms with exponents not in $\mathbb{N}_0$, we conclude
that $S_{2,-1}\ket{-1}\notin\Vc_{-1}$. This is in accordance with the structure of the Verma module
\psset{unit=1cm}
\be
\begin{pspicture}(-1.5,0.6)(4.1,1.8)
 \rput(-2.1,0.75){$\Vc_0$\,:}
 \rput(-1.06,.75){$_{[0,0]}$}
 \rput(0.12,1.52){$_{[-1,0]}$}
 \rput(0.8,1.5){$\to$}
 \rput(1.6,1.52){$_{[-3,4]}$}
 \rput(2.4,1.5){$\to$}
 \rput(3.2,1.52){$_{[-4,8]}$}
 \rput(4,1.5){$\to$}
 \rput(4.6,1.52){$\cdots$}
 \rput(0.1,0.02){$_{[2,4]}$}
 \rput(0.8,0){$\to$}
 \rput(1.6,0.02){$_{[3,8]}$}
 \rput(2.4,0){$\to$}
 \rput(3.2,0.02){$_{[5,20]}$}
 \rput(4,0){$\to$}
 \rput(4.6,0.02){$\cdots$}
 \rput(-0.6,1.25){$\nearrow$}
 \rput(-0.6,0.25){$\searrow$}
 \rput(0.8,0.73){$\searrow$}
 \rput(0.8,0.73){$\nearrow$}
 \rput(2.4,0.73){$\searrow$}
 \rput(2.4,0.73){$\nearrow$}
 \rput(4,0.73){$\searrow$}
 \rput(4,0.73){$\nearrow$}
\end{pspicture} 
\\[0.7cm]
\ee
where we find the explicit indication of $h_j$ in $[j,h_j]$ helpful. Indeed, $\Vc_{-1}$ is readily seen not to have 
a submodule isomorphic to $\Vc_{2}$. 

We find that necessary \!{\em and} sufficient conditions for $S_{j',j}\ket{j}\in\Vc_j$
are obtained by requiring, in addition to $j'\in\Ob_j$ and (\ref{hhjj}), that
\be
 j+j'\notin\{2j_{r_0,s_0}\,|\,0<r_0<p,\,0\leq s_0<p'\}.
\ee
Note that $j+j'$ is excluded by this supplementary condition exactly if $j$ and $j'$ correspond to a pair of 
nodes located atop one another in the Loewy diagram (\ref{diagram2}) of some ambient Verma module.
In the cases where $S_{j',j}\in U(\mathfrak{n}_-)$, so that $\Vc_{j'}\subseteq\Vc_j$, we see that (\ref{SFA}) offers 
a generalisation of the Fuchs-Astashkevich theorem~\cite{Ast97}, from the Virasoro algebra to $A_1^{(1)}$.

Of particular interest are the two singular vectors generating the submodules of highest weight $\beta_0$ 
and $\beta_1$ 
in (\ref{diagram2}). With $r=r_0$ and $s=s_0$, they are given by the well-known expressions
\begin{align}
 \ket{_{r,s}^{\ 0}}&:=(J_0^-)^{r+st}(J_{-1}^+)^{r+(s-1)t}(J_0^-)^{r+(s-2)t}\ldots
   (J_{-1}^+)^{r-(s-1)t}(J_0^-)^{r-st}\ket{j_{r,s}},
\label{MFF0}
\\[.3cm]
 \ket{_{r,s}^{\ 1}}&:=(J_{-1}^+)^{p-r+(p'-s-1)t}(J_0^-)^{p-r+(p'-s-2)t}(J_{-1}^+)^{p-r+(p'-s-3)t}  
  \ldots\nonumber\\[.2cm]
  &\qquad\qquad\qquad\qquad\qquad\ldots 
  (J_0^-)^{p-r-(p'-s-2)t}(J_{-1}^+)^{p-r-(p'-s-1)t}\ket{j_{r,s}},
\label{MFF1}
\end{align}
and satisfy
\be
\begin{array}{rll}
 &J_0^3\ket{_{r,s}^{\ 0}}=\big(j_{r,s}-r\big)\ket{_{r,s}^{\ 0}},\quad
 &L_0\ket{_{r,s}^{\ 0}}=\big(h_{r,s}+rs\big)\ket{_{r,s}^{\ 0}},
 \\[.3cm]
 &J_0^3\ket{_{r,s}^{\ 1}}=\big(j_{r,s}+p-r\big)\ket{_{r,s}^{\ 1}},\quad
 &L_0\ket{_{r,s}^{\ 1}}=\big(h_{r,s}+(p-r)(p'-s)\big)\ket{_{r,s}^{\ 1}}.
\end{array}
\ee
Similarly, the maximal proper submodule $\Vc_{\gamma_2}$ of $\Vc_{j_{\delta p,s_0}}$\! in (\ref{diagram1})
is generated from the singular vector $\ket{_{0,s_0}^{1-\delta}}$.

\subsection{Admissible characters}

The irreducible highest-weight module of weight $j_{r,s}$ is obtained from the Verma module $\Vc_{j_{r,s}}$ 
by quotienting out the maximal proper submodule. We denote the corresponding affine character by
\be
 \ch_{r,s}(q,z)=\ch_{j_{r,s}}(q,z).
\label{chrs}
\ee
We recall that an admissible module is such an irreducible highest-weight module, 
with highest weight $j_{r_0,s_0}$ subject to (\ref{jadm}).
From the Loewy diagram (\ref{diagram2}) and (\ref{ab}), the affine character
of the admissible module is given by~\cite{KW88}
\be
 \ch_{r_0,s_0}(q,z)
  =\!\sum_{i\in\mathbb{Z}}\big(\chit_{\al_i}(q,z)-\chit_{\beta_i}(q,z)\big)
 =\frac{\Theta_{\la_{r_0,s_0}^+,pp'}(q,z^{\frac{1}{p'}})
   -\Theta_{\la_{r_0,s_0}^-,pp'}(q,z^{\frac{1}{p'}})}{\eta(q,z)},
\label{chadm}
\ee
where
\be
  \la_{r_0,s_0}^\pm:=\pm r_0p'-ps_0.
\label{lar0s0}
\ee
This character can also be written as
\be
 \ch_{r_0,s_0}(q,z)=\frac{1}{\eta(q,z)}\sum_{l\in\mathbb{Z}}
  q^{pp'(l+\frac{\la_{r_0,s_0}^+}{2pp'})^2} z^{-p(l+\frac{\la_{r_0,s_0}^+}{2pp'})}
  \big(1-q^{-r_0(2l p'-s_0)}z^{r_0}\big).
\ee
We occasionally specify the dependence on $t=p/p'$ by writing $\ch_{r,s}^{p,p'}(q,z)$ for the irreducible character
in (\ref{chrs}).

\subsection{Irreducible characters}
\label{SecIrred}

For fractional level, non-admissible yet irreducible modules are rarely discussed in the literature.
However, as they play an important role in the present work,
their affine characters are classified and given below. 
The explicit expressions follow from the details of the corresponding ambient Loewy diagrams.
For weights $j_{r,s}$ with $r=\rh_0+\ell p$ and $s=\sh_0+\ell'p'$ as in (\ref{rolsol}), we thus find that
\be
 \ch_{r,s}(q,z)=\frac{\Theta_{\la_{\rh_0,\sh_0}^+-(\ell'-\ell)pp',pp';\ell'-\ell}(q,z^{\frac{1}{p'}})
  -\Theta_{\la_{\rh_0,\sh_0}^--|\ell'-\ell|pp',pp';|\ell'-\ell|}(q,z^{\frac{1}{p'}})}{\eta(q,z)},
\label{chhrs}
\ee
where the reduced theta functions $\Theta_{n,m;\nu}(q,z)$ are defined in (\ref{affinetheta}).
For $r\in p\mathbb{Z}^\times$, we likewise find that
\be
 \ch_{\ell p,\sh_0+\ell'p'}(q,z)
 =\frac{q^{\frac{p((\ell-\ell')p'-\sh_0)^2}{4p'}}z^{-\frac{p((\ell-\ell')p'-\sh_0)}{2p'}}}{\eta(q,z)}
  \begin{cases}
    \big(1-q^{(\ell-\ell')p\sh_0}z^{(\ell-\ell')p}\big),\quad &\ell>\ell',
   \\[.5cm]
    \big(1-q^{(\ell-\ell'-1)p(\sh_0-p')}z^{(\ell-\ell'-1)p}\big),\quad &\ell\leq\ell',
 \end{cases}
\label{chhps}
\ee
where $\ell>0,\,\ell'\geq0$ or $\ell,\,\ell'\!<0$.
Since the set of {\em distinct} weights of the form $j_{r,s}$, $r,s\in\mathbb{Z}$, is given by
\be
 \{\,j_{r,s}\,|\,r\in\mathbb{Z},\,0\leq s\leq p'-1\}
 =\{\,j_{r,s}\,|\,1\leq r,\,0\leq s\leq p'-1\}\sqcup\{\,j_{r,s}\,|\,r\leq -p,\,-p'\leq s\leq-1\},
\ee
the classification of the distinct irreducible characters can be simplified. 
With $r_0=1,\ldots,p-1$, $s_0=0,\ldots,p'-1$, and $\ell\in\mathbb{N}_0$, we thus find that the
distinct irreducible characters are given by
\begin{align}
 \ch_{r_0+\ell p,s_0}(q,z)
 &=\frac{\Theta_{r_0p'-ps_0+\ell pp',pp';-\ell}(q,z^{\frac{1}{p'}})
  -\Theta_{-r_0p'-ps_0-\ell pp',pp';\ell}(q,z^{\frac{1}{p'}})}{\eta(q,z)},
\label{irr1}
\\[.2cm]
 \ch_{(\ell+1)p,s_0}(q,z)
 &=\frac{q^{\frac{p((\ell+1)p'-s_0)^2}{4p'}}z^{-\frac{p((\ell+1)p'-s_0)}{2p'}}\big(1-q^{(\ell+1)ps_0}z^{(\ell+1)p}\big)}{\eta(q,z)},
\\[.2cm]
 \ch_{r_0-(\ell+2)p,s_0-p'}(q,z)
 &=\frac{\Theta_{r_0p'-ps_0-(\ell+1)pp',pp';\ell+1}(q,z^{\frac{1}{p'}})
  -\Theta_{-r_0p'-ps_0-(\ell+1)pp',pp';\ell+1}(q,z^{\frac{1}{p'}})}{\eta(q,z)},
\\[.2cm]
 \ch_{-(\ell+1)p,s_0-p'}(q,z)
 &=\frac{q^{\frac{p(\ell p'+s_0)^2}{4p'}}z^{\frac{p(\ell p'+s_0)}{2p'}}\big(1-q^{(\ell+1)p(p'-s_0)}z^{-(\ell+1)p}\big)}{\eta(q,z)}.
\label{irr4}
\end{align}
Unlike the expression for the admissible characters in (\ref{chadm}), to the best of our knowledge, 
these explicit general expressions for the irreducible (including non-admissble) characters are new.

\subsection{Integer levels}

In the coset expression (\ref{cosetAAA}), one of the constituent current algebras, $(A_1^{(1)})_n$, 
has integer level of the form
\be
 n=\frac{p}{p'}-2\in\mathbb{N}\qquad\Longrightarrow\qquad p=n+2\in\mathbb{Z}_{\geq3},\qquad p'=1.
\ee
In our analysis of the coset construction in Section~\ref{Sec:Coset}, focus will be on 
the admissible characters for this integer level. These are all irreducible, 
and since $0\leq\sh_0\leq p'-1$ implies $\sh_0=0$, our interest is in
\begin{align}
 \ch_{\rho,0}^{n+2,1}(q,z)
  &=\frac{\Theta_{\rho,n+2}(q,z)-\Theta_{-\rho,n+2}(q,z)}{\eta(q,z)}
 \nonumber\\[.2cm] 
  &=\frac{1}{\eta(q,z)}\sum_{l\in\mathbb{Z}}
  q^{(n+2)(l+\frac{\rho}{2(n+2)})^2}z^{-(n+2)l-\frac{\rho}{2}}\big(1-q^{-2l \rho}z^{\rho}\big),
   \qquad \rho=1,\ldots,n+1.
\label{ch1}
\end{align}
Expanding these affine characters in powers of $z$, we may write
\be
 \ch_{\rho,0}^{n+2,1}(q,z)=
 \sum_{l\in\mathbb{Z}+\frac{\rho-1}{2}}c_{2l}^{\rho-1}(q)\,q^{\frac{l^2}{n}}z^{-l},
\label{chn}
\ee
thereby defining the
$A_1^{(1)}$\! {\em string functions} $c_m^{\,\ell}(q)$, with their dependence on $n$ suppressed.
For integers $\ell,m$ for which $\ell-m\in2\mathbb{Z}$, these functions are given by~\cite{KP80,KP84,JM84,GQ87,DQ90,HNY90}
\be
 c_m^{\,\ell}(q):=\frac{q^{\frac{(\ell+1)^2}{4(n+2)}-\frac{m^2}{4n}}}{\eta^3(q)}
  \sum_{i,l=0}^\infty(-1)^{i+l}q^{il(n+1)+\frac{1}{2}i(i+1+\ell+m)+\frac{1}{2}l(l+1+\ell-m)}
  \big(1-q^{(i+l+1)(n+1-\ell)-(i-l)m}\big).
\label{sf}
\ee
They satisfy
\be
 c_m^{\,\ell}(q)=c_{-m}^{\,\ell}(q)=c_{2n-m}^{\,\ell}(q)=c_{n-m}^{n-\ell}(q),
\label{ccc}
\ee
allowing us to focus on the so-called fundamental domain, where the integers $\ell,m$ are related as
\be
 0\leq m\leq\ell\leq n,\qquad \ell-m\in2\mathbb{Z}.
\ee
The corresponding number of fundamental string functions is thus given by
\be
 \sum_{i=0}^n\big\lfloor\tfrac{i+2}{2}\big\rfloor-\big\lfloor\tfrac{n+1}{2}\big\rfloor
 =\tfrac{1}{4}(n+1)^2+\begin{cases} 0,\ & n\ \,\mathrm{odd},
 \\[.15cm]
 \tfrac{3}{4},\ & n\ \,\mathrm{even}.
 \end{cases}
\ee
Although not needed in this paper, we also note the relation $c_m^\ell(q)=-c_m^{-\ell-2}(q)$.

For $n=1$, there is only one fundamental string function, namely
\be
 c_{\,0}^{\,0}(q)=\frac{1}{\eta(q)},
\label{c00}
\ee
as follows from (\ref{var2}),
and the two characters are given by the well-known expressions
\be
 \ch_{\rho,0}^{3,1}(q,z)
 =\frac{1}{\eta(q)}\sum_{l\in\mathbb{Z}+\frac{\rho-1}{2}}q^{l^2}z^{-l}
 =\frac{\Theta_{\rho-1,1}(q,z)}{\eta(q)},
  \qquad \rho=1,2.
\ee
For $n=2$, there are three fundamental string functions,
\begin{align}
 c_{\,0}^{\,0}(q)=c_{\,2}^{\,2}(q)&=\frac{1}{2q^{\frac{1}{48}}\eta(q)}
  \Big(\prod_{i=1}^\infty(1+q^{i-\frac{1}{2}})+\prod_{i=1}^\infty(1-q^{i-\frac{1}{2}})\Big),
 \\[.15cm]
 c_{\,1}^{\,1}(q)&=\frac{q^{\frac{1}{24}}}{\eta(q)}\prod_{i=1}^\infty(1+q^i),
 \\[.15cm]
 c_{\,2}^{\,0}(q)=c_{\,0}^{\,2}(q)&=\frac{1}{2q^{\frac{1}{48}}\eta(q)}
  \Big(\prod_{i=1}^\infty(1+q^{i-\frac{1}{2}})-\prod_{i=1}^\infty(1-q^{i-\frac{1}{2}})\Big),
\end{align}
giving rise to the well-known character expressions
\begin{align}
 \ch_{1,0}^{4,1}(q,z)&=\frac{1}{2q^{\frac{1}{16}}}\sum_{l\in\mathbb{Z}}
  \Big(\prod_{i=1}^\infty\frac{1+q^{i-\frac{1}{2}}}{1-q^i}+(-1)^l\prod_{i=1}^\infty\frac{1-q^{i-\frac{1}{2}}}{1-q^i}\Big)q^{\frac{l^2}{2}}z^{-l},
 \\
 \ch_{2,0}^{4,1}(q,z)&=\Big(\prod_{i=1}^\infty\frac{1+q^i}{1-q^i}\Big)\!
   \sum_{l\in\mathbb{Z}+\frac{1}{2}}q^{\frac{l^2}{2}}z^{-l},
 \\
  \ch_{3,0}^{4,1}(q,z)&=\frac{1}{2q^{\frac{1}{16}}}\sum_{l\in\mathbb{Z}}
  \Big(\prod_{i=1}^\infty\frac{1+q^{i-\frac{1}{2}}}{1-q^i}-(-1)^l\prod_{i=1}^\infty\frac{1-q^{i-\frac{1}{2}}}{1-q^i}\Big)q^{\frac{l^2}{2}}z^{-l}.
\end{align}

\section{Affine Kac modules}
\label{Sec:Kac}

\subsection{Wakimoto modules}

Resembling the way
Feigin-Fuchs modules~\cite{FF82,FF89} (see also~\cite{Fel89,BMP91,IK11}) are Fock modules 
over the Virasoro algebra,
Wakimoto modules~\cite{Wak86} are Fock modules over the affine Lie algebra $A_1^{(1)}$\!.
Just as the affine Verma modules described above, 
a Wakimoto module is characterised by its highest weight $j$,
and it is reducible if and only if $j=j_{r,s}$ for some $r,s\in\mathbb{Z}$ such that $rs>0$ or $r>0,s=0$. 
Such a Wakimoto module is here denoted by $\Wc_{r,s}$.

For $r\notin p\mathbb{Z}$, the structure of $\Wc_{r,s}$ is examined in~\cite{BF90}; its Loewy diagram is of the form
\psset{unit=.8cm}
\setlength{\unitlength}{.8cm}
\be
\begin{pspicture}(0,0.6)(9,1.7)
 \rput(-3,0.73){$r\not\in p\mathbb{Z}$\,:}
\pscircle[fillstyle=solid,fillcolor=lightgray,linecolor=black,linewidth=0.01](-1.06,0.73){.12}
\pscircle[fillstyle=solid,fillcolor=white,linecolor=black,linewidth=0.01](0,1.5){.12}
 \rput(0.8,1.48){$\rightarrow$}
\pscircle[fillstyle=solid,fillcolor=lightgray,linecolor=black,linewidth=0.01](1.6,1.5){.12}
 \rput(2.4,1.48){$\leftarrow$}
\pscircle[fillstyle=solid,fillcolor=white,linecolor=black,linewidth=0.01](3.2,1.5){.12}
 \rput(4,1.48){$\rightarrow$}
\pscircle[fillstyle=solid,fillcolor=lightgray,linecolor=black,linewidth=0.01](4.8,1.5){.12}
 \rput(5.6,1.48){$\leftarrow$}
\pscircle[fillstyle=solid,fillcolor=white,linecolor=black,linewidth=0.01](6.4,1.5){.12}
 \rput(7.2,1.48){$\rightarrow$}
\pscircle[fillstyle=solid,fillcolor=lightgray,linecolor=black,linewidth=0.01](8,1.5){.12}
 \rput(8.8,1.48){$\leftarrow$}
\pscircle[fillstyle=solid,fillcolor=black,linecolor=black,linewidth=0.01](0,0){.12}
 \rput(0.8,-0.02){$\leftarrow$}
\pscircle[fillstyle=solid,fillcolor=lightgray,linecolor=black,linewidth=0.01](1.6,0){.12}
 \rput(2.4,-0.02){$\rightarrow$}
\pscircle[fillstyle=solid,fillcolor=black,linecolor=black,linewidth=0.01](3.2,0){.12}
 \rput(4,-0.02){$\leftarrow$}
\pscircle[fillstyle=solid,fillcolor=lightgray,linecolor=black,linewidth=0.01](4.8,0){.12}
 \rput(5.6,-0.02){$\rightarrow$}
\pscircle[fillstyle=solid,fillcolor=black,linecolor=black,linewidth=0.01](6.4,0){.12}
 \rput(7.2,-0.02){$\leftarrow$}
\pscircle[fillstyle=solid,fillcolor=lightgray,linecolor=black,linewidth=0.01](8,0){.12}
 \rput(8.8,-0.02){$\rightarrow$}
 \rput(-0.55,1.22){$\swarrow$}
 \rput(-0.55,0.28){$\searrow$}
 \rput(0.8,0.73){$\searrow$}
 \rput(0.8,0.73){$\swarrow$}
 \rput(2.4,0.73){$\searrow$}
 \rput(2.4,0.73){$\swarrow$}
 \rput(4,0.73){$\searrow$}
 \rput(4,0.73){$\swarrow$}
 \rput(5.6,0.73){$\searrow$}
 \rput(5.6,0.73){$\swarrow$}
 \rput(7.2,0.73){$\searrow$}
 \rput(7.2,0.73){$\swarrow$}
 \rput(8.8,0.73){$\searrow$}
 \rput(8.8,0.73){$\swarrow$}
 \rput(9.8,0){$\ldots$}
 \rput(9.8,1.5){$\ldots$}
\end{pspicture} 
\\[0.5cm]
\label{diagramWbraid}
\ee
The corresponding subsingular vectors are here shaded according to their type. Those represented by black dots are singular
and generate the socle,
$\Cc_{r,s}^{\begin{pspicture}(0,0)(0,0) \pscircle[fillstyle=solid,fillcolor=black,linecolor=black,linewidth=0.01](0.15,0.15){.1} \end{pspicture}}$, 
of the Wakimoto module (the socle being the maximal completely reducible submodule).
The grey dots represent subsingular vectors that are singular in the quotient module 
$\Wc_{r,s}/\Cc_{r,s}^{\begin{pspicture}(0,0)(0,0) 
\pscircle[fillstyle=solid,fillcolor=black,linecolor=black,linewidth=0.01](0.15,0.15){.1} \end{pspicture}}$; 
they generate the socle, 
$\Cc_{r,s}^{\begin{pspicture}(0,0)(0,0)
\pscircle[fillstyle=solid,fillcolor=lightgray,linecolor=black,linewidth=0.01](0.15,0.15){.1} \end{pspicture}}$, 
of the quotient module. Finally, the white dots represent subsingular vectors that are singular in the quotient module
$(\Wc_{r,s}/\Cc_{r,s}^{\begin{pspicture}(0,0)(0,0) 
\pscircle[fillstyle=solid,fillcolor=black,linecolor=black,linewidth=0.01](0.15,0.15){.1} 
\end{pspicture}})/\Cc_{r,s}^{\begin{pspicture}(0,0)(0,0) 
\pscircle[fillstyle=solid,fillcolor=lightgray,linecolor=black,linewidth=0.01](0.15,0.15){.1} \end{pspicture}}$; 
they generate the head of the Wakimoto module (the head being the maximal completely reducible quotient).

For $r\in p\mathbb{Z}^\times$, the structure of $\Wc_{r,s}$ is not examined in~\cite{BF90}, but we find that, 
depending on $r,s,p,p'$, its Loewy diagram is of one of the following two types:
\be
\begin{pspicture}(-0.3,0)(9,0.8)
 \rput(-3.1,0.05){$r\in p\mathbb{Z}^\times$:}
 \rput(-0.7,0){$\Bigg\{$}
\pscircle[fillstyle=solid,fillcolor=lightgray,linecolor=black,linewidth=0.01](0,0.62){.12}
 \rput(0.8,0.6){$\rightarrow$}
\pscircle[fillstyle=solid,fillcolor=black,linecolor=black,linewidth=0.01](1.6,0.62){.12}
 \rput(2.4,0.6){$\leftarrow$}
\pscircle[fillstyle=solid,fillcolor=lightgray,linecolor=black,linewidth=0.01](3.2,0.62){.12}
 \rput(4,0.6){$\rightarrow$}
\pscircle[fillstyle=solid,fillcolor=black,linecolor=black,linewidth=0.01](4.8,0.62){.12}
 \rput(5.6,0.6){$\leftarrow$}
\pscircle[fillstyle=solid,fillcolor=lightgray,linecolor=black,linewidth=0.01](6.4,0.62){.12}
 \rput(7.2,0.6){$\rightarrow$}
\pscircle[fillstyle=solid,fillcolor=black,linecolor=black,linewidth=0.01](8,0.62){.12}
 \rput(8.8,0.6){$\leftarrow$}
\pscircle[fillstyle=solid,fillcolor=black,linecolor=black,linewidth=0.01](0,-0.62){.12}
 \rput(0.8,-0.64){$\leftarrow$}
\pscircle[fillstyle=solid,fillcolor=lightgray,linecolor=black,linewidth=0.01](1.6,-0.62){.12}
 \rput(2.4,-0.64){$\rightarrow$}
\pscircle[fillstyle=solid,fillcolor=black,linecolor=black,linewidth=0.01](3.2,-0.62){.12}
 \rput(4,-0.64){$\leftarrow$}
\pscircle[fillstyle=solid,fillcolor=lightgray,linecolor=black,linewidth=0.01](4.8,-0.62){.12}
 \rput(5.6,-0.64){$\rightarrow$}
\pscircle[fillstyle=solid,fillcolor=black,linecolor=black,linewidth=0.01](6.4,-0.62){.12}
 \rput(7.2,-0.64){$\leftarrow$}
\pscircle[fillstyle=solid,fillcolor=lightgray,linecolor=black,linewidth=0.01](8,-0.62){.12}
 \rput(8.8,-0.64){$\rightarrow$}
 \rput(9.8,0.62){$\ldots$}
 \rput(9.8,-0.62){$\ldots$}
\end{pspicture} 
\vspace{0.5cm}
\label{diagramWstrings}
\ee
The subsingular vectors have been shaded as above, although the head of the module is now generated by the vectors
corresponding to grey dots.

As our primary focus is on certain finitely generated submodules, the details of the full Wakimoto modules are omitted.
Instead, the structure of the submodules is described in Section~\ref{Sec:Fin} below.

\subsection{Finitely generated submodules}
\label{Sec:Fin}

Just as the Virasoro Kac modules~\cite{Ras1012,BGT1102,MDRR1503} are finitely generated submodules of Feigin-Fuchs modules,
we now introduce {\em affine Kac modules} as certain finitely generated submodules of Wakimoto modules.
As is clear from their Loewy diagrams below, all affine Kac modules are indecomposable. 
Moreover, every affine Kac module has finitely many composition factors.
\\[.25cm]
\noindent
{\bf Definition.}\quad
For $r,s\in\mathbb{Z}$ such that $rs>0$,
the {\em affine Kac module} $\Ac_{r,s}$ is the submodule of the Wakimoto module $\Wc_{r,s}$, generated by the subsingular vectors of
grade strictly less than $rs$. For $r\in\mathbb{N}$, the affine Kac module $\Ac_{r,0}$ is the highest-weight 
submodule of the Wakimoto module $\Wc_{r,0}$, generated by the highest-weight vector of weight $j_{r,0}$.
The set of affine Kac modules is denoted by $S_{\mathrm{aff}}$.
\\[.25cm]
For $r=\rh_0+\ell p$ and $s=\sh_0+\ell'p'$ as in (\ref{rolsol}), the Loewy diagram of the affine Kac module 
$\Ac_{r,s}$ is given by
\psset{unit=0.9cm}
\setlength{\unitlength}{0.9cm} 
\be
\begin{pspicture}(-1.1,0.6)(9,2.4)
\pscircle[fillstyle=solid,fillcolor=lightgray,linecolor=black,linewidth=0.01](-1.1,0.73){.12}
\pscircle[fillstyle=solid,fillcolor=white,linecolor=black,linewidth=0.01](0,1.5){.12}
 \rput(0.8,1.48){$\rightarrow$}
\pscircle[fillstyle=solid,fillcolor=lightgray,linecolor=black,linewidth=0.01](1.6,1.5){.12}
 \rput(2.4,1.48){$\leftarrow$}
 \rput(4.8,1.48){$\leftarrow$}
\pscircle[fillstyle=solid,fillcolor=white,linecolor=black,linewidth=0.01](5.6,1.5){.12}
 \rput(6.4,1.48){$\rightarrow$}
\pscircle[fillstyle=solid,fillcolor=lightgray,linecolor=black,linewidth=0.01](7.2,1.5){.12}
 \rput(3.6,0){$\ldots$}
 \rput(3.6,1.5){$\ldots$}
\pscircle[fillstyle=solid,fillcolor=black,linecolor=black,linewidth=0.01](0,0){.12}
 \rput(0.8,-0.02){$\leftarrow$}
\pscircle[fillstyle=solid,fillcolor=lightgray,linecolor=black,linewidth=0.01](1.6,0){.12}
 \rput(2.4,-0.02){$\rightarrow$}
 \rput(4.8,-0.02){$\rightarrow$}
\pscircle[fillstyle=solid,fillcolor=black,linecolor=black,linewidth=0.01](5.6,0){.12}
 \rput(6.4,-0.02){$\leftarrow$}
\pscircle[fillstyle=solid,fillcolor=lightgray,linecolor=black,linewidth=0.01](7.2,0){.12}
 \rput(8,-0.02){$\rightarrow$}
\pscircle[fillstyle=solid,fillcolor=black,linecolor=black,linewidth=0.01](8.8,0){.12}
 \rput(-1.65,0.73){$j_{r,s}$}
 \rput(0,-0.45){$v_{2m}$}
 \rput(1.6,-0.45){$v_{2m-1}$}
 \rput(5.62,-0.45){$v_2$}
 \rput(7.22,-0.45){$v_1$}
 \rput(8.82,-0.45){$v_0$}
 \rput(0,1.9){$u_{2m}$}
 \rput(1.6,1.9){$u_{2m-1}$}
 \rput(5.62,1.9){$u_2$}
 \rput(7.22,1.9){$u_1$}
 \rput(-0.6,1.22){$\swarrow$}
 \rput(-0.6,0.28){$\searrow$}
 \rput(0.8,0.73){$\searrow$}
 \rput(0.8,0.73){$\swarrow$}
 \rput(2.4,0.73){$\searrow$}
 \rput(2.4,0.73){$\swarrow$}
 \rput(4.8,0.73){$\searrow$}
 \rput(4.8,0.73){$\swarrow$}
 \rput(6.4,0.73){$\searrow$}
 \rput(6.4,0.73){$\swarrow$}
 \rput(8,0.73){$\searrow$}
\end{pspicture} 
\vspace{1.3cm}
\label{diagramAv}
\ee
where
\be
 m=\min\!\big(|\ell+\tfrac{1}{2}|,|\ell'+\tfrac{1}{2}|\big)-\tfrac{1}{2}
\ee
and
\begin{align}
 \ell,\ell'\geq0:\qquad &
 \left\{\!\!\!\!\!\!\!\begin{array}{rll}
 &u_{2i}=j_{-r+2ip,s},\quad &u_{2i-1}=j_{r-2(\ell+1-i)p,s},
 \\[.25cm]
 &v_{2i}=j_{-r+2(\ell+\ell'+1-i)p,s},\quad &v_{2i-1}=j_{r+2(\ell'+1-i)p,s},
 \end{array}\right.
\\[.4cm]
 \ell,\ell'<0:\qquad &
 \left\{\!\!\!\!\!\!\!\begin{array}{rll}
 &u_{2i}=j_{-r-2ip,s},\quad &u_{2i-1}=j_{r-2(\ell+i)p,s},
 \\[.25cm]
 &v_{2i}=j_{-r+2(\ell+\ell'+1+i)p,s},\quad &v_{2i-1}=j_{r+2(\ell'+i)p,s}.
 \end{array}\right.
\end{align}
It follows that $v_0=j_{-r+2(\ell+\ell'+1)p,s}$ and
\be
 j_{r,s}=\left\{\!\!\!\!\!\!\!\begin{array}{rll}
 &u_{2m+1},\ &|\ell|\leq|\ell'|,
 \\[.25cm]
 &v_{2m+1},\ &|\ell|\geq|\ell'|,
 \end{array}\right.
\ee
stressing that $u_{2m+1}=v_{2m+1}$ for $\ell=\ell'$.
In accordance with the definition of $\Ac_{r,s}$, we also note that $u_0=j_{-r,s}$.

For $r\in p\mathbb{Z}^\times$, in which case $r=\ell p$ and $s=\sh_0+\ell'p'$ with $\ell>0,\,\ell'\geq0$ or $\ell,\ell'<0$, 
the Loewy diagram of the affine Kac module $\Ac_{r,s}$ is one of the following two diagrams:
\psset{unit=0.9cm}
\setlength{\unitlength}{0.9cm} 
\be
\begin{pspicture}(-1.1,0.75)(9,2.2)
 \rput(-1.5,1.53){I\,:}
 \rput(-1.5,0.03){II\,:}
\pscircle[fillstyle=solid,fillcolor=lightgray,linecolor=black,linewidth=0.01](0,1.5){.12}
 \rput(0.8,1.48){$\rightarrow$}
\pscircle[fillstyle=solid,fillcolor=black,linecolor=black,linewidth=0.01](1.6,1.5){.12}
 \rput(2.4,1.48){$\leftarrow$}
\pscircle[fillstyle=solid,fillcolor=lightgray,linecolor=black,linewidth=0.01](3.2,1.5){.12}
 \rput(4,1.48){$\rightarrow$}
 \rput(5.2,1.5){$\ldots$}
 \rput(6.4,1.48){$\leftarrow$}
\pscircle[fillstyle=solid,fillcolor=lightgray,linecolor=black,linewidth=0.01](7.2,1.5){.12}
 \rput(8,1.48){$\rightarrow$}
\pscircle[fillstyle=solid,fillcolor=black,linecolor=black,linewidth=0.01](8.8,1.5){.12}
\pscircle[fillstyle=solid,fillcolor=black,linecolor=black,linewidth=0.01](0,0){.12}
 \rput(0.8,-0.02){$\leftarrow$}
\pscircle[fillstyle=solid,fillcolor=lightgray,linecolor=black,linewidth=0.01](1.6,0){.12}
 \rput(2.4,-0.02){$\rightarrow$}
\pscircle[fillstyle=solid,fillcolor=black,linecolor=black,linewidth=0.01](3.2,0){.12}
 \rput(4,-0.02){$\rightarrow$}
 \rput(5.2,0){$\ldots$}
 \rput(6.4,-0.02){$\leftarrow$}
\pscircle[fillstyle=solid,fillcolor=lightgray,linecolor=black,linewidth=0.01](7.2,0){.12}
 \rput(8,-0.02){$\rightarrow$}
\pscircle[fillstyle=solid,fillcolor=black,linecolor=black,linewidth=0.01](8.8,0){.12}
 \rput(0,-0.45){$j_{r,s}$}
 \rput(1.6,-0.45){$w_{2m}$}
 \rput(3.2,-0.45){$w_{2m-1}$}
 \rput(7.22,-0.45){$w_2$}
 \rput(8.82,-0.45){$w_1$}
 \rput(0,1.9){$j_{r,s}$}
 \rput(1.6,1.9){$w_{2m}$}
 \rput(3.2,1.9){$w_{2m-1}$}
 \rput(7.22,1.9){$w_1$}
 \rput(8.82,1.9){$w_0$}
\end{pspicture} 
\vspace{1.3cm}
\label{diagramAs}
\ee
with
\be
 m=\min\!\big(|\ell|-1,|\ell'+\tfrac{1}{2}|-\tfrac{1}{2}\big).
\ee
Specifically, the diagram type and weights of the subsingular vectors are given by
\be
\begin{array}{rcll}
 \ell'\geq\ell>0:&\quad \mathrm{type\ I}, &\quad
 w_{2i}=j_{r+2(\ell'-i)p,s}, &\quad w_{2i-1}=j_{-r+2ip,s},
\\[.4cm]
 \ell>\ell'\geq0:&\quad \mathrm{type\ II}, &\quad
  w_{2i}=j_{-r+2ip,s}, &\quad w_{2i-1}=j_{r+2(\ell'+1-i)p,s},
\\[.4cm]
 \ell'<\ell<0:&\quad \mathrm{type\ I}, &\quad
 w_{2i}=j_{r+2(\ell'+1+i)p,s}, &\quad w_{2i-1}=j_{-r-2ip,s},
\\[.4cm]
 \ell\leq\ell'<0:&\quad \mathrm{type\ II}, &\quad
 w_{2i}=j_{-r-2ip,s}, &\quad w_{2i-1}=j_{r+2(\ell'+i)p,s}.
\end{array}
\ee
It follows that $j_{r,s}=w_{2m+1}$ and, in accordance with the definition of $\Ac_{r,s}$, that
$w_{-1}=j_{-r,s}$ for type I and $w_0=j_{-r,s}$ for type II.

\subsection{Characters and classes}

The affine character of $\Ac_{r,s}$ is given by the affine Kac character (\ref{Kact}).
For fractional level, it can be written as
\be
 \chit_{r,s}(q,z)=\frac{q^{\frac{(rp'-sp)^2}{4pp'}}z^{-\frac{rp'-sp}{2p'}}}{\eta(q,z)}(1-q^{rs}z^r)
\label{aK}
\ee
and decomposes into a finite sum of irreducible affine characters.
Explicitly, for $r=\rh_0+\ell p$ and $s=\sh_0+\ell'p'$ as in (\ref{rolsol}), we thus find that
\be
 \chit_{r,s}(q,z)=\ch_{r,s}(q,z)+\sum_{i=|\ell-\ell'|+1}^{|\ell+\ell'+1|-1}
  \big[\ch_{w_i^+}(q,z)+\ch_{w_i^-}(q,z)\big]+\ch_{-r+2(\ell+\ell'+1)p,s}(q,z),
\ee
where
\be
 w_i^\pm:=j_{(-1)^{i+\ell+\ell'}\rh_0\pm ip,\,
   \sh_0-\frac{1}{2}(1-(-1)^{i+\ell+\ell'})p'},\qquad i\in\mathbb{N}.
\ee
For $r\in p\mathbb{Z}^\times$, we likewise find that
\be
 \chit_{\ell p,\sh_0+\ell'p'}(q,z)=\begin{cases}
  \displaystyle{\sum_{i=0}^{\min(2\ell-1,2\ell')}
     \ch_{(-1)^i(\ell+\ell'-i)p,\,\sh_0-\frac{1}{2}(1-(-1)^i)p'}(q,z)},
  \quad &\ell>0,\,\ell'\geq0,
 \\[.5cm]
 \displaystyle{\sum_{i=0}^{\min(-2\ell-1,-2\ell'-2)}
  \ch_{(-1)^i(\ell+\ell'+2+i)p,\,\sh_0-\frac{1}{2}(1-(-1)^i)p'}(q,z)},  \quad &\ell,\ell'<0.
\end{cases}
\ee
It follows that
\be
 \chit_{r,s}(q,z)=\ch_{r,s}(q,z)\quad\Longleftrightarrow\quad 
  \big[r\in p\mathbb{N},\,0\leq s\leq p'-1\big]
  \quad \mathrm{or}\quad \big[r\in(-p\mathbb{N}),\,-p'\leq s\leq -1\big],
\ee
so the set of {\em irreducible} affine Kac modules, 
\be
 S_{\mathrm{irr}}:=\{\Ac_{r,s}\in S_{\mathrm{aff}}\,|\,\Ac_{r,s}\cong\Ic_{r,s}\},
\ee
is given by the disjoint union
\be
 S_{\mathrm{irr}}=\{\Ac_{r,s}\in S_{\mathrm{aff}}\,|\,r\in p\mathbb{N},\,0\leq s\leq p'-1\}
 \,\sqcup\,\{\Ac_{r,s}\in S_{\mathrm{aff}}\,|\,r\in(-p\mathbb{N}),\,-p'\leq s\leq -1\}.
\ee
It is also of interest to know whether a given affine Kac module $\Ac_{r,s}$ is isomorphic to the corresponding quotient module 
$\Qc_{r,s}$. From the Loewy diagrams, we see that $\Ac_{r,s}\cong\Qc_{r,s}$ if and only if
\be
 -p\leq r\leq p\quad\;\mathrm{or}\quad -p'\leq s\leq p'-1,
\ee
so
\be
 S_{\mathrm{quo}}:=\{\Ac_{r,s}\in S_{\mathrm{aff}}\,|\,\Ac_{r,s}\cong\Qc_{r,s}\}
\label{Squo}
\ee
is given by the disjoint union
\be
 S_{\mathrm{quo}}=S_{\mathrm{quo}}^>\,\sqcup\,S_{\mathrm{quo}}^<,
\ee
where
\begin{align}
 S_{\mathrm{quo}}^>&=\{\Ac_{r,s}\in S_{\mathrm{aff}}\,|\,1\leq r,\,0\leq s\leq p'-1\}\,
  \sqcup\,\{\Ac_{r,s}\in S_{\mathrm{aff}}\,|\,1\leq r\leq p,\,p'\leq s\},
 \\[.2cm]
 S_{\mathrm{quo}}^<&=\{\Ac_{r,s}\in S_{\mathrm{aff}}\,|\,r\leq-1,\,-p'\leq s\leq -1\}\,
  \sqcup\,\{\Ac_{r,s}\in S_{\mathrm{aff}}\,|\,-p\leq r\leq -1,\,s\leq -p'-1\}. 
\end{align}
It follows that $S_{\mathrm{irr}}\subseteq S_{\mathrm{quo}}$ and that the affine character of $\Ac_{r,s}\in S_{\mathrm{quo}}$
decomposes into a sum of at most two irreducible affine characters.

As already stressed, the affine Kac characters for a given level are all distinct.
The corresponding affine weights $j_{r,s}$ are neatly organised in an (extended) {\em affine
Kac table}. This is illustrated in~Figure~\ref{FigAffineKac} for $k=-\frac{4}{3}$ (for which
$(p,p')=(2,3)$) and for $k=-\frac{1}{2}$ (for which $(p,p')=(3,2)$).
\psset{unit=.79cm}
\begin{figure}
\begin{center}
\begin{pspicture}(15,13)(5,0.3)
\rput(4.5,9){$k=-\frac{4}{3}$}
\rput(4.5,8){$(p,p')=(2,3)$}
\psframe[linewidth=0pt,fillstyle=solid,fillcolor=lightestblue](8,6)(15,13)
\psframe[linewidth=0pt,fillstyle=solid,fillcolor=lightlightblue](9,6)(10,13)
\psframe[linewidth=0pt,fillstyle=solid,fillcolor=lightlightblue](11,6)(12,13)
\psframe[linewidth=0pt,fillstyle=solid,fillcolor=lightlightblue](13,6)(14,13)
\psframe[linewidth=0pt,fillstyle=solid,fillcolor=lightlightblue](8,6)(15,7)
\psframe[linewidth=0pt,fillstyle=solid,fillcolor=lightlightblue](8,9)(15,10)
\psframe[linewidth=0pt,fillstyle=solid,fillcolor=lightlightblue](8,12)(15,13)
\psframe[linewidth=0pt,fillstyle=solid,fillcolor=midblue](9,6)(10,7)
\psframe[linewidth=0pt,fillstyle=solid,fillcolor=midblue](11,6)(12,7)
\psframe[linewidth=0pt,fillstyle=solid,fillcolor=midblue](13,6)(14,7)
\psframe[linewidth=0pt,fillstyle=solid,fillcolor=midblue](9,9)(10,10)
\psframe[linewidth=0pt,fillstyle=solid,fillcolor=midblue](11,9)(12,10)
\psframe[linewidth=0pt,fillstyle=solid,fillcolor=midblue](13,9)(14,10)
\psframe[linewidth=0pt,fillstyle=solid,fillcolor=midblue](9,12)(10,13)
\psframe[linewidth=0pt,fillstyle=solid,fillcolor=midblue](11,12)(12,13)
\psframe[linewidth=0pt,fillstyle=solid,fillcolor=midblue](13,12)(14,13)
\psframe[linewidth=1.4pt,fillstyle=solid,fillcolor=lightpurple](8,6)(9,9)
\pswedge[fillstyle=solid,fillcolor=red,linecolor=red](10,7){.22}{180}{270}
\pswedge[fillstyle=solid,fillcolor=red,linecolor=red](10,8){.22}{180}{270}
\pswedge[fillstyle=solid,fillcolor=red,linecolor=red](10,9){.22}{180}{270}
\pswedge[fillstyle=solid,fillcolor=red,linecolor=red](12,7){.22}{180}{270}
\pswedge[fillstyle=solid,fillcolor=red,linecolor=red](12,8){.22}{180}{270}
\pswedge[fillstyle=solid,fillcolor=red,linecolor=red](12,9){.22}{180}{270}
\pswedge[fillstyle=solid,fillcolor=red,linecolor=red](14,7){.22}{180}{270}
\pswedge[fillstyle=solid,fillcolor=red,linecolor=red](14,8){.22}{180}{270}
\pswedge[fillstyle=solid,fillcolor=red,linecolor=red](14,9){.22}{180}{270}
\psgrid[gridlabels=0pt,subgriddiv=1](8,6)(15,13)
\rput(8.5,6.5){$0$}\rput(9.5,6.5){$\frac{1}{2}$}\rput(10.5,6.5){$1$}\rput(11.5,6.5){$\frac{3}{2}$}\rput(12.5,6.5){$2$}\rput(13.5,6.5){$\frac{5}{2}$}\rput(14.5,6.5){$\hdots$}
\rput(8.5,7.5){$-\frac{1}{3}$}\rput(9.5,7.5){$\frac{1}{6}$}\rput(10.5,7.5){$\frac{2}{3}$}\rput(11.5,7.5){$\frac{7}{6}$}\rput(12.5,7.5){$\frac{5}{3}$}\rput(13.5,7.5){$\frac{13}{6}$}\rput(14.5,7.5){$\hdots$}
\rput(8.5,8.5){$-\frac{2}{3}$}\rput(9.5,8.5){$-\frac{1}{6}$}\rput(10.5,8.5){$\frac{1}{3}$}\rput(11.5,8.5){$\frac{5}{6}$}\rput(12.5,8.5){$\frac{4}{3}$}\rput(13.5,8.5){$\frac{11}{6}$}\rput(14.5,8.5){$\hdots$}
\rput(8.5,9.5){$-1$}\rput(9.5,9.5){$-\frac{1}{2}$}\rput(10.5,9.5){$0$}\rput(11.5,9.5){$\frac{1}{2}$}\rput(12.5,9.5){$1$}\rput(13.5,9.5){$\frac{3}{2}$}\rput(14.5,9.5){$\hdots$}
\rput(8.5,10.5){$-\frac{4}{3}$}\rput(9.5,10.5){$-\frac{5}{6}$}\rput(10.5,10.5){$-\frac{1}{3}$}\rput(11.5,10.5){$\frac{1}{6}$}\rput(12.5,10.5){$\frac{2}{3}$}\rput(13.5,10.5){$\frac{7}{6}$}\rput(14.5,10.5){$\hdots$}
\rput(8.5,11.5){$-\frac{5}{3}$}\rput(9.5,11.5){$-\frac{7}{6}$}\rput(10.5,11.5){$-\frac{2}{3}$}\rput(11.5,11.5){$-\frac{1}{6}$}\rput(12.5,11.5){$\frac{1}{3}$}\rput(13.5,11.5){$\frac{5}{6}$}\rput(14.5,11.5){$\hdots$}
\rput(8.5,12.65){$\vdots$}\rput(9.5,12.65){$\vdots$}\rput(10.5,12.65){$\vdots$}\rput(11.5,12.65){$\vdots$}\rput(12.5,12.65){$\vdots$}\rput(13.5,12.65){$\vdots$}\rput(14.5,12.5){$\vvdots$}
\psframe[linewidth=0pt,fillstyle=solid,fillcolor=lightestblue](0,0)(7,6)
\psframe[linewidth=0pt,fillstyle=solid,fillcolor=lightlightblue](1,0)(2,6)
\psframe[linewidth=0pt,fillstyle=solid,fillcolor=lightlightblue](3,0)(4,6)
\psframe[linewidth=0pt,fillstyle=solid,fillcolor=lightlightblue](5,0)(6,6)
\psframe[linewidth=0pt,fillstyle=solid,fillcolor=lightlightblue](0,0)(7,1)
\psframe[linewidth=0pt,fillstyle=solid,fillcolor=lightlightblue](0,3)(7,4)
\psframe[linewidth=0pt,fillstyle=solid,fillcolor=midblue](1,0)(2,1)
\psframe[linewidth=0pt,fillstyle=solid,fillcolor=midblue](3,0)(4,1)
\psframe[linewidth=0pt,fillstyle=solid,fillcolor=midblue](5,0)(6,1)
\psframe[linewidth=0pt,fillstyle=solid,fillcolor=midblue](1,3)(2,4)
\psframe[linewidth=0pt,fillstyle=solid,fillcolor=midblue](3,3)(4,4)
\psframe[linewidth=0pt,fillstyle=solid,fillcolor=midblue](5,3)(6,4)
\pswedge[fillstyle=solid,fillcolor=red,linecolor=red](6,6){.22}{180}{270}
\pswedge[fillstyle=solid,fillcolor=red,linecolor=red](6,5){.22}{180}{270}
\pswedge[fillstyle=solid,fillcolor=red,linecolor=red](6,4){.22}{180}{270}
\pswedge[fillstyle=solid,fillcolor=red,linecolor=red](4,6){.22}{180}{270}
\pswedge[fillstyle=solid,fillcolor=red,linecolor=red](4,5){.22}{180}{270}
\pswedge[fillstyle=solid,fillcolor=red,linecolor=red](4,4){.22}{180}{270}
\pswedge[fillstyle=solid,fillcolor=red,linecolor=red](2,6){.22}{180}{270}
\pswedge[fillstyle=solid,fillcolor=red,linecolor=red](2,5){.22}{180}{270}
\pswedge[fillstyle=solid,fillcolor=red,linecolor=red](2,4){.22}{180}{270}
\psgrid[gridlabels=0pt,subgriddiv=1](0,0)(7,6)
\rput(0.5,5.5){$\hdots$}\rput(1.5,5.5){$-\frac{19}{6}$}\rput(2.5,5.5){$-\frac{8}{3}$}\rput(3.5,5.5){$-\frac{13}{6}$}\rput(4.5,5.5){$-\frac{5}{3}$}\rput(5.5,5.5){$-\frac{7}{6}$}\rput(6.5,5.5){$-\frac{2}{3}$}
\rput(0.5,4.5){$\hdots$}\rput(1.5,4.5){$-\frac{17}{6}$}\rput(2.5,4.5){$-\frac{7}{3}$}\rput(3.5,4.5){$-\frac{11}{6}$}\rput(4.5,4.5){$-\frac{4}{3}$}\rput(5.5,4.5){$-\frac{5}{6}$}\rput(6.5,4.5){$-\frac{1}{3}$}
\rput(0.5,3.5){$\hdots$}\rput(1.5,3.5){$-\frac{5}{2}$}\rput(2.5,3.5){$-2$}\rput(3.5,3.5){$-\frac{3}{2}$}\rput(4.5,3.5){$-1$}\rput(5.5,3.5){$-\frac{1}{2}$}\rput(6.5,3.5){$0$}
\rput(0.5,2.5){$\hdots$}\rput(1.5,2.5){$-\frac{13}{6}$}\rput(2.5,2.5){$-\frac{5}{3}$}\rput(3.5,2.5){$-\frac{7}{6}$}\rput(4.5,2.5){$-\frac{2}{3}$}\rput(5.5,2.5){$-\frac{1}{6}$}\rput(6.5,2.5){$\frac{1}{3}$}
\rput(0.5,1.5){$\hdots$}\rput(1.5,1.5){$-\frac{11}{6}$}\rput(2.5,1.5){$-\frac{4}{3}$}\rput(3.5,1.5){$-\frac{5}{6}$}\rput(4.5,1.5){$-\frac{1}{3}$}\rput(5.5,1.5){$\frac{1}{6}$}\rput(6.5,1.5){$\frac{2}{3}$}
\rput(0.5,0.5){$\vvdots$}\rput(1.5,0.65){$\vdots$}\rput(2.5,0.65){$\vdots$}\rput(3.5,0.65){$\vdots$}\rput(4.5,0.65){$\vdots$}\rput(5.5,0.65){$\vdots$}\rput(6.5,0.65){$\vdots$}
{\color{gray} % r-part
\rput(0.5,6.5){$\hdots$}
\rput(1.5,6.5){$-6$}
\rput(2.5,6.5){$-5$}
\rput(3.5,6.5){$-4$}
\rput(4.5,6.5){$-3$}
\rput(5.5,6.5){$-2$}
\rput(6.5,6.5){$-1$}
\rput(8.5,5.5){$1$}
\rput(9.5,5.5){$2$}
\rput(10.5,5.5){$3$}
\rput(11.5,5.5){$4$}
\rput(12.5,5.5){$5$}
\rput(13.5,5.5){$6$}
\rput(14.5,5.5){$r$}
}
{\color{blue} % s-part
\rput(7.5,0.65){$\vdots$}
\rput(7.4,1.5){$-5$}
\rput(7.4,2.5){$-4$}
\rput(7.4,3.5){$-3$}
\rput(7.4,4.5){$-2$}
\rput(7.4,5.5){$-1$}
\rput(7.4,6.5){$0$}
\rput(7.4,7.5){$1$}
\rput(7.4,8.5){$2$}
\rput(7.4,9.5){$3$}
\rput(7.4,10.5){$4$}
\rput(7.4,11.5){$5$}
\rput(7.4,12.5){$s$}
}
\end{pspicture}
%%%%%%%%%%%%%%%%%%%%%%%%%%%%%%%%%
%\mbox{}\vspace{1cm}\mbox{}
\\[-.8cm]
%%%%%%%%%%%%%%%%%%%%%%%%%%%%%%%%%
%
\begin{pspicture}(15,13)(-5,0.5)
\rput(4.5,9){$k=-\frac{1}{2}$}
\rput(4.5,8){$(p,p')=(3,2)$}
\psframe[linewidth=0pt,fillstyle=solid,fillcolor=lightestblue](8,6)(15,13)
\psframe[linewidth=0pt,fillstyle=solid,fillcolor=lightlightblue](10,6)(11,13)
\psframe[linewidth=0pt,fillstyle=solid,fillcolor=lightlightblue](13,6)(14,13)
\psframe[linewidth=0pt,fillstyle=solid,fillcolor=lightlightblue](8,6)(15,7)
\psframe[linewidth=0pt,fillstyle=solid,fillcolor=lightlightblue](8,8)(15,9)
\psframe[linewidth=0pt,fillstyle=solid,fillcolor=lightlightblue](8,10)(15,11)
\psframe[linewidth=0pt,fillstyle=solid,fillcolor=lightlightblue](8,12)(15,13)
\psframe[linewidth=0pt,fillstyle=solid,fillcolor=midblue](10,6)(11,7)
\psframe[linewidth=0pt,fillstyle=solid,fillcolor=midblue](13,6)(14,7)
\psframe[linewidth=0pt,fillstyle=solid,fillcolor=midblue](10,8)(11,9)
\psframe[linewidth=0pt,fillstyle=solid,fillcolor=midblue](13,8)(14,9)
\psframe[linewidth=0pt,fillstyle=solid,fillcolor=midblue](10,10)(11,11)
\psframe[linewidth=0pt,fillstyle=solid,fillcolor=midblue](13,10)(14,11)
\psframe[linewidth=0pt,fillstyle=solid,fillcolor=midblue](10,12)(11,13)
\psframe[linewidth=0pt,fillstyle=solid,fillcolor=midblue](13,12)(14,13)
\psframe[linewidth=1.4pt,fillstyle=solid,fillcolor=lightpurple](8,6)(10,8)
\pswedge[fillstyle=solid,fillcolor=red,linecolor=red](11,7){.22}{180}{270}
\pswedge[fillstyle=solid,fillcolor=red,linecolor=red](11,8){.22}{180}{270}
\pswedge[fillstyle=solid,fillcolor=red,linecolor=red](14,7){.22}{180}{270}
\pswedge[fillstyle=solid,fillcolor=red,linecolor=red](14,8){.22}{180}{270}
\psgrid[gridlabels=0pt,subgriddiv=1](8,6)(15,13)
\rput(8.5,6.5){$0$}\rput(9.5,6.5){$\frac{1}{2}$}\rput(10.5,6.5){$1$}\rput(11.5,6.5){$\frac{3}{2}$}\rput(12.5,6.5){$2$}\rput(13.5,6.5){$\frac{5}{2}$}\rput(14.5,6.5){$\hdots$}
\rput(8.5,7.5){$-\frac{3}{4}$}\rput(9.5,7.5){$-\frac{1}{4}$}\rput(10.5,7.5){$\frac{1}{4}$}\rput(11.5,7.5){$\frac{3}{4}$}\rput(12.5,7.5){$\frac{5}{4}$}\rput(13.5,7.5){$\frac{7}{4}$}\rput(14.5,7.5){$\hdots$}
\rput(8.5,8.5){$-\frac{3}{2}$}\rput(9.5,8.5){$-1$}\rput(10.5,8.5){$-\frac{1}{2}$}\rput(11.5,8.5){$0$}\rput(12.5,8.5){$\frac{1}{2}$}\rput(13.5,8.5){$1$}\rput(14.5,8.5){$\hdots$}
\rput(8.5,9.5){$-\frac{9}{4}$}\rput(9.5,9.5){$-\frac{7}{4}$}\rput(10.5,9.5){$-\frac{5}{4}$}\rput(11.5,9.5){$-\frac{3}{4}$}\rput(12.5,9.5){$-\frac{1}{4}$}\rput(13.5,9.5){$\frac{1}{4}$}\rput(14.5,9.5){$\hdots$}
\rput(8.5,10.5){$-3$}\rput(9.5,10.5){$-\frac{5}{2}$}\rput(10.5,10.5){$-2$}\rput(11.5,10.5){$-\frac{3}{2}$}\rput(12.5,10.5){$-1$}\rput(13.5,10.5){$-\frac{1}{2}$}\rput(14.5,10.5){$\hdots$}
\rput(8.5,11.5){$-\frac{15}{4}$}\rput(9.5,11.5){$-\frac{13}{4}$}\rput(10.5,11.5){$-\frac{11}{4}$}\rput(11.5,11.5){$-\frac{9}{4}$}\rput(12.5,11.5){$-\frac{7}{4}$}\rput(13.5,11.5){$-\frac{5}{4}$}\rput(14.5,11.5){$\hdots$}
\rput(8.5,12.65){$\vdots$}\rput(9.5,12.65){$\vdots$}\rput(10.5,12.65){$\vdots$}\rput(11.5,12.65){$\vdots$}\rput(12.5,12.65){$\vdots$}\rput(13.5,12.65){$\vdots$}\rput(14.5,12.5){$\vvdots$}
\psframe[linewidth=0pt,fillstyle=solid,fillcolor=lightestblue](0,0)(7,6)
\psframe[linewidth=0pt,fillstyle=solid,fillcolor=lightlightblue](1,0)(2,6)
\psframe[linewidth=0pt,fillstyle=solid,fillcolor=lightlightblue](4,0)(5,6)
\psframe[linewidth=0pt,fillstyle=solid,fillcolor=lightlightblue](0,0)(7,1)
\psframe[linewidth=0pt,fillstyle=solid,fillcolor=lightlightblue](0,2)(7,3)
\psframe[linewidth=0pt,fillstyle=solid,fillcolor=lightlightblue](0,4)(7,5)
\psframe[linewidth=0pt,fillstyle=solid,fillcolor=midblue](1,0)(2,1)
\psframe[linewidth=0pt,fillstyle=solid,fillcolor=midblue](4,0)(5,1)
\psframe[linewidth=0pt,fillstyle=solid,fillcolor=midblue](1,2)(2,3)
\psframe[linewidth=0pt,fillstyle=solid,fillcolor=midblue](4,2)(5,3)
\psframe[linewidth=0pt,fillstyle=solid,fillcolor=midblue](1,4)(2,5)
\psframe[linewidth=0pt,fillstyle=solid,fillcolor=midblue](4,4)(5,5)
\pswedge[fillstyle=solid,fillcolor=red,linecolor=red](5,6){.22}{180}{270}
\pswedge[fillstyle=solid,fillcolor=red,linecolor=red](5,5){.22}{180}{270}
\pswedge[fillstyle=solid,fillcolor=red,linecolor=red](2,6){.22}{180}{270}
\pswedge[fillstyle=solid,fillcolor=red,linecolor=red](2,5){.22}{180}{270}
\psgrid[gridlabels=0pt,subgriddiv=1](0,0)(7,6)
\rput(0.5,5.5){$\hdots$}\rput(1.5,5.5){$-\frac{11}{4}$}\rput(2.5,5.5){$-\frac{9}{4}$}\rput(3.5,5.5){$-\frac{7}{4}$}\rput(4.5,5.5){$-\frac{5}{4}$}\rput(5.5,5.5){$-\frac{3}{4}$}\rput(6.5,5.5){$-\frac{1}{4}$}
\rput(0.5,4.5){$\hdots$}\rput(1.5,4.5){$-2$}\rput(2.5,4.5){$-\frac{3}{2}$}\rput(3.5,4.5){$-1$}\rput(4.5,4.5){$-\frac{1}{2}$}\rput(5.5,4.5){$0$}\rput(6.5,4.5){$\frac{1}{2}$}
\rput(0.5,3.5){$\hdots$}\rput(1.5,3.5){$-\frac{5}{4}$}\rput(2.5,3.5){$-\frac{3}{4}$}\rput(3.5,3.5){$-\frac{1}{4}$}\rput(4.5,3.5){$\frac{1}{4}$}\rput(5.5,3.5){$\frac{3}{4}$}\rput(6.5,3.5){$\frac{5}{4}$}
\rput(0.5,2.5){$\hdots$}\rput(1.5,2.5){$-\frac{1}{2}$}\rput(2.5,2.5){$0$}\rput(3.5,2.5){$\frac{1}{2}$}\rput(4.5,2.5){$1$}\rput(5.5,2.5){$\frac{3}{2}$}\rput(6.5,2.5){$2$}
\rput(0.5,1.5){$\hdots$}\rput(1.5,1.5){$\frac{1}{4}$}\rput(2.5,1.5){$\frac{3}{4}$}\rput(3.5,1.5){$\frac{5}{4}$}\rput(4.5,1.5){$\frac{7}{4}$}\rput(5.5,1.5){$\frac{9}{4}$}\rput(6.5,1.5){$\frac{11}{4}$}
\rput(0.5,0.5){$\vvdots$}\rput(1.5,0.65){$\vdots$}\rput(2.5,0.65){$\vdots$}\rput(3.5,0.65){$\vdots$}\rput(4.5,0.65){$\vdots$}\rput(5.5,0.65){$\vdots$}\rput(6.5,0.65){$\vdots$}
{\color{gray} % r-part
\rput(0.5,6.5){$\hdots$}
\rput(1.5,6.5){$-6$}
\rput(2.5,6.5){$-5$}
\rput(3.5,6.5){$-4$}
\rput(4.5,6.5){$-3$}
\rput(5.5,6.5){$-2$}
\rput(6.5,6.5){$-1$}
\rput(8.5,5.5){$1$}
\rput(9.5,5.5){$2$}
\rput(10.5,5.5){$3$}
\rput(11.5,5.5){$4$}
\rput(12.5,5.5){$5$}
\rput(13.5,5.5){$6$}
\rput(14.5,5.5){$r$}
}
{\color{blue} % s-part
\rput(7.5,0.65){$\vdots$}
\rput(7.4,1.5){$-5$}
\rput(7.4,2.5){$-4$}
\rput(7.4,3.5){$-3$}
\rput(7.4,4.5){$-2$}
\rput(7.4,5.5){$-1$}
\rput(7.4,6.5){$0$}
\rput(7.4,7.5){$1$}
\rput(7.4,8.5){$2$}
\rput(7.4,9.5){$3$}
\rput(7.4,10.5){$4$}
\rput(7.4,11.5){$5$}
\rput(7.4,12.5){$s$}
}
\end{pspicture}
\end{center}
\caption{Extended affine Kac tables of weights $j_{r,s}$ for $k=-\frac{4}{3}$ 
(for which $p=2$ and $p'=3$) and for $k=-\frac{1}{2}$ (for which $p=3$ and $p'=2$). 
The entries relate to distinct affine Kac modules even if the
weights coincide. An irreducible highest-weight module exists for each weight appearing in a given
affine Kac table. The affine Kac modules which happen to be irreducible are marked
with a shaded quadrant in the top-right corner; the associated weights do {\em not} exhaust the distinct values
of the affine weights. The periodicity $j_{r,s}=j_{r+\ell p,s+\ell p'}$ is made manifest by the shading of
rows and columns. In a given affine Kac table, the {\em admissible} weights comprise the finite 
subtable with the thick frame.
\label{FigAffineKac}}
\end{figure}
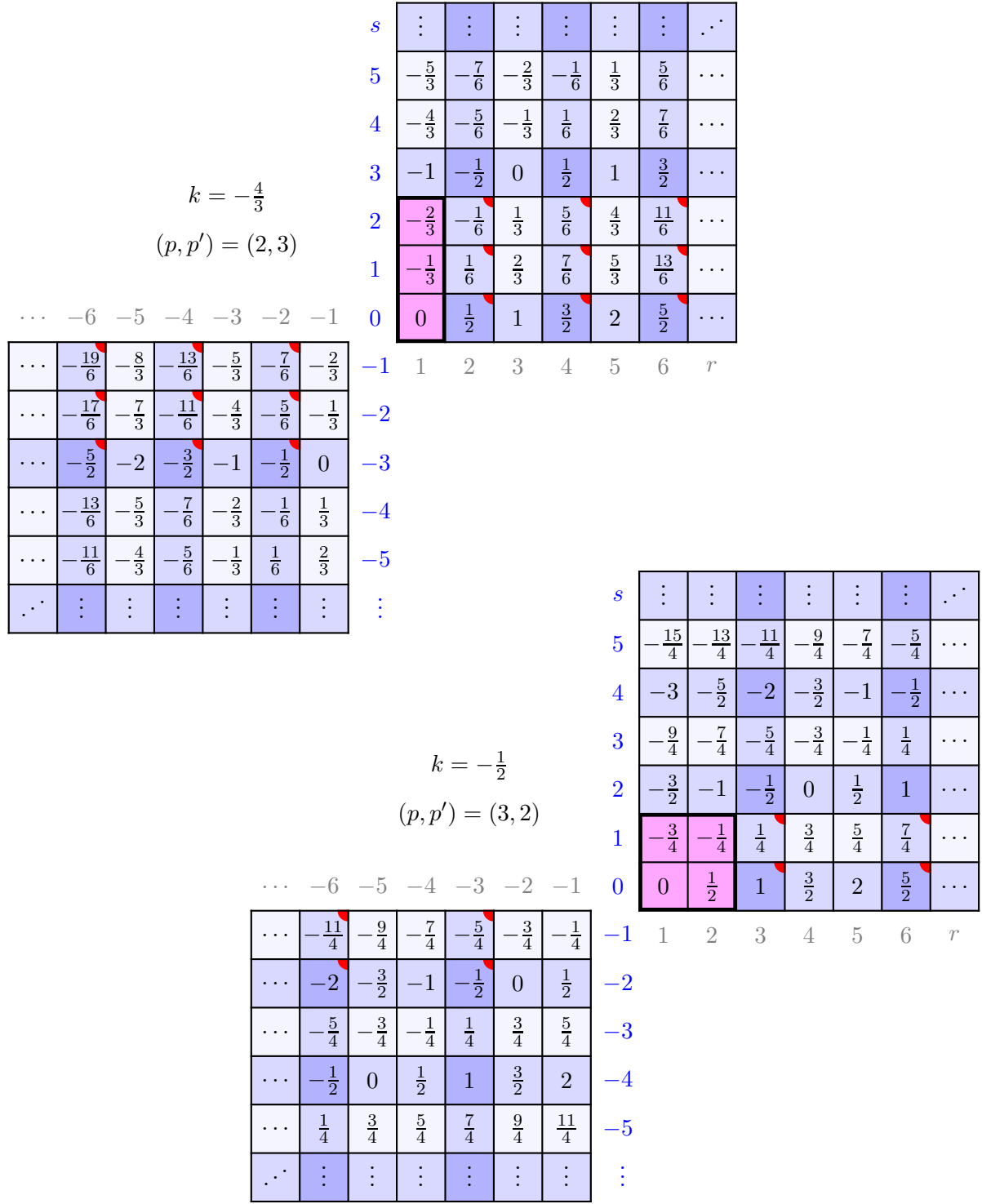

\newpage

\section{Staggered modules}
\label{Sec:Stag}

\subsection{Basic structure}

We are interested in indecomposable $A_1^{(1)}$-modules whose Loewy diagrams are of the form
\psset{unit=.8cm}
\setlength{\unitlength}{.8cm}
\be
\begin{pspicture}(0,1.1)(1.3,2.25)
 \pscircle[fillstyle=solid,fillcolor=lightgray,linecolor=black,linewidth=0.01](0,2.2){.1} 
 \pscircle[fillstyle=solid,fillcolor=black,linecolor=black,linewidth=0.01](0,1.1){.1} 
 \pscircle[fillstyle=solid,fillcolor=white,linecolor=black,linewidth=0.01](1.2,1.1){.1} 
 \pscircle[fillstyle=solid,fillcolor=lightgray,linecolor=black,linewidth=0.01](1.2,0){.1} 
 \rput(0.6,0.55){$\nwarrow$}
 \rput(0.6,1.65){$\nwarrow$}
 \rput(0,1.65){$\downarrow$}
 \rput(1.2,0.55){$\downarrow$}
 \rput(0.6,1.1){$\leftarrow$}
\end{pspicture}
\qquad\qquad
\mathrm{or}
\qquad\qquad
\begin{pspicture}(0,1.1)(1.3,2.25)
 \pscircle[fillstyle=solid,fillcolor=black,linecolor=black,linewidth=0.01](0,1.1){.1} 
 \pscircle[fillstyle=solid,fillcolor=white,linecolor=black,linewidth=0.01](1.2,1.1){.1} 
 \pscircle[fillstyle=solid,fillcolor=lightgray,linecolor=black,linewidth=0.01](1.2,0){.1} 
 \rput(0.6,0.55){$\nwarrow$}
 \rput(1.2,0.55){$\downarrow$}
 \rput(0.6,1.1){$\leftarrow$}
\end{pspicture}
\\[0.8cm]
\label{LS}
\ee
where the black and white nodes represent isomorphic irreducible subquotients.
Such {\em staggered modules} $\Sc$ are described by non-split short exact sequences of the form
\be
 0\,\to\,\Mc^L\,\to\,\Sc\,\to\,\Mc^R\,\to\,0,
\label{MSM}
\ee
where the quotient module $\Mc^R$ is composed of the two (white and grey) nodes on the right in the corresponding diagram.
Accordingly, the affine character of such a staggered module decomposes as
\be
 \chit[\Sc](q,z)=\chit[\Mc^L](q,z)+\chit[\Mc^R](q,z).
\ee
Alternatively, we can characterise the staggered modules in (\ref{LS}) as non-split short exact sequences of the form
\be
 0\,\to\,\bar{\Ac}^B\,\to\,\Sc\,\to\,\bar{\Ac}^T\,\to\,0,
\label{ASA}
\ee
where the submodule $\bar{\Ac}^B$ is the contravariant module to $\Mc^R$, 
composed of the two (black and grey) nodes at the bottom in the corresponding diagram.

With reference to the first diagram in (\ref{LS}), the indecomposable highest-weight modules $\Mc^L$ and $\Mc^R$ in
the corresponding {\em quadrangular} 
staggered module are described by non-split short exact sequences of the form
\be
\begin{array}{rcccccccl}
 0\!&\!\to\!&\!\Ic\!&\!\to\!&\!\Mc^L\!&\!\to\!&\!\Ic_{quo}\!&\!\to\!&\!0,
\\[.3cm]
 0\!&\!\to\!&\!\Ic_{sub}\!&\!\to\!&\!\Mc^R\!&\!\to\!&\!\Ic\!&\!\to\!&\!0,
\end{array}
\label{MLMR}
\ee
where $\Ic$, $\Ic_{quo}$ and $\Ic_{sub}$ are irreducible modules. 
The affine character of $\Sc$ thus decomposes into a sum of four irreducible characters:
\be
 \chit[\Sc](q,z)=\chit[\Ic_{sub}](q,z)+2\chit[\Ic](q,z)+\chit[\Ic_{quo}](q,z).
\ee
With reference to the second diagram in (\ref{LS}), for {\em triangular} $\Sc$, we have
\be
 \Mc^L\cong\Ic,\qquad
 0\,\to\,\Ic_{sub}\,\to\,\Mc^R\,\to\,\Ic\,\to\,0
\ee
and
\be
 \chit[\Sc](q,z)=\chit[\Ic_{sub}](q,z)+2\chit[\Ic](q,z).
\ee

\subsection{Logarithmic vectors and coupling}

Let the highest-weight module $\Mc^L$ in (\ref{MSM}) be generated from the highest-weight vector $\ket{j,h_j}$, and let
the submodule $\Ic$ of $\Mc^L$ in (\ref{MLMR}) be generated from a singular vector $\ket{Q,N}_j$
of charge $Q$ and at grade $N$.
This means that
\be
 U(\mathfrak{n}_+)\ket{j,h_j}=0,\qquad
 J_0^3\ket{j,h_j}=j\ket{j,h_j},\qquad
 L_0\ket{j,h_j}=h_j\ket{j,h_j},
\ee
and that
\be
 \ket{Q,N}_j=S\ket{j,h_j}
\ee
for some $S\in U(\mathfrak{n}_-)$ such that
\be
 U(\mathfrak{n}_+)\ket{Q,N}_j=0,\qquad 
 J_0^3\ket{Q,N}_j=(j+Q)\ket{Q,N}_j,\qquad
 L_0\ket{Q,N}_j=(h_j+N)\ket{Q,N}_j.
\ee
Albeit superfluous, we find the explicit indication of the conformal weight $h_j$ in $\ket{j,h_j}$ helpful.
A convention applicable to any staggered module of the form (\ref{LS}) is to set
\be
 S=S_{j+Q,j},
\ee
given in (\ref{SbbSbb}),
although simpler expressions may exist since $\ket{Q,N}_j$ is in general a singular vector in a {\em quotient} of $\Vc_j$.
For $\Sc$ triangular, the operator $S$ is merely a nonzero scalar multiple of the identity, where we recall that $S_{j,j}=I$.

Given $\ket{Q,N}_j$, let $\ket{R}$ denote a matching {\em logarithmic vector} rooted in the quotient $\Mc^R$ in (\ref{MSM}). 
Although $\ket{R}$ is not uniquely determined, not even up to normalisation, it satisfies
\be
 \big(J_0^3-(j+Q)\big)\ket{R}=\eta\ket{Q,N}_j,\qquad
 \big(L_0-(h_j+N)\big)\ket{R}=\mu\ket{Q,N}_j,
\label{etamujhj}
\ee
for some $\eta,\mu\in\mathbb{C}$. It follows that
\be
 \big(L_0-\tfrac{1}{t}J_0^3(J_0^3+1)\big)\ket{R}=\big(\mu-\tfrac{\eta}{t}\big(2(j+Q)+1\big)\big)\ket{Q,N}_j,
\label{off}
\ee
in accordance with $\frac{\partial}{\partial x}\big(x(x+1)\big)=2x+1$.
The presence of the horizontal arrow in the diagram (\ref{LS}) implies that
{\em at most one} of the parameters $\eta,\mu$ can be zero. 

For every $P\in U(\mathfrak{n}_-)$, there exist $P_-',P_-'',P_-''',P_+',P_+''\in U(\mathfrak{n}_-)$
such that
\be
 J_1^-P=P_-'+P_-''J_0^3+P_-'''K+PJ_1^-,\qquad
 J_0^+P=P_+'+P_+''J_0^3+PJ_0^+.
\ee
The vanishing conditions
\be
 J_1^-S\ket{j,h_j}=J_1^-\ket{Q,N}_j=0,\qquad J_0^+S\ket{j,h_j}=J_0^+\ket{Q,N}_j=0
\ee
thus imply that
\be
 S_-'+jS_-''+kS_-'''=0,\qquad S_+'+jS_+''=0.
\ee
Moreover, we see that
\be
 J_1^-\ket{R}\in[\Mc^L]_{Q-1,N-1},\qquad J_0^+\ket{R}\in[\Mc^L]_{Q+1,N}
\ee
and $S^\dagger\ket{R}\in[\Mc^L]_{0,0}$, so that
\be
 S^\dagger\ket{R}=\beta\ket{j,h_j}
\ee
for some $\beta\in\mathbb{C}$. It readily follows that
\be
 SS^\dagger\ket{R}=\beta\ket{Q,N}_j,
\label{SSR}
\ee
showing that $\beta$ depends quadratically on the normalisation of $S$.
The {\em logarithmic coupling constant} $\beta$ plays a role akin to the similar coupling constants appearing in the description of 
staggered Virasoro modules, see~\cite{Roh96,GK96,MR08,KR09,GJSV13}. 

Since $\Mc^R$ in (\ref{MSM}) is a quotient of a Verma module, the staggered module $\Sc$
is likewise a quotient of an indecomposable module, obtained by setting to zero a submodule-generating
singular vector of the form
\be
 P\ket{R}+\ket{L},\qquad P\in U(\mathfrak{n}_-),\qquad \ket{L}\in\Mc^L,
\label{PRL}
\ee
where $P\ket{R}=0$ in $\Mc^R$.
From
\be
 J_1^-P\ket{R}=\eta P_-''\ket{Q,N}_j+PJ_1^-\ket{R}\in\Mc^L,\qquad
 J_0^+P\ket{R}=\eta P_+''\ket{Q,N}_j+PJ_0^+\ket{R}\in\Mc^L,
\ee
it follows that
\be
 \eta P_-''\ket{Q,N}_j+PJ_1^-\ket{R}+J_1^-\ket{L}=0,\qquad
 \eta P_+''\ket{Q,N}_j+PJ_0^+\ket{R}+J_0^+\ket{L}=0.
\ee
As the examples considered in Sections~\ref{Sec:StagEx} and~\ref{Sec:j10} demonstrate,
these conditions will in general affect the possible values of $\beta$.

\subsection{Conjectures}
\label{Sec:Conjectures}

The staggered modules described in the following are all either quadrangular or triangular, 
but we do not claim that our list of such modules is exhaustive.
In all the examples we have examined explicitly, the generator $L_0$ acts non-diagonalisably, while $J_0^3$ need not.
The examples for $k=-\frac{4}{3}$ discussed in Sections~\ref{Sec:StagEx} and~\ref{Sec:j10} have been selected to illustrate
both possibilities for $J_0^3$.
The staggered modules differ from the reducible yet indecomposable $A_1^{(1)}$-modules discussed in 
\cite{Gab01,LMRS04,Ras0508,Rid09}.
\\[.15cm]
\noindent
{\bf Conjecture 1.}\quad
For every $3$-tuple $(a,s_0,\ell)\in\mathbb{Z}^3$, where
\be
 1\leq a\leq p-1,\qquad 
 0\leq s_0\leq p'-1,\qquad
 \ell\geq1,
\ee
there exists a pair of quadrangular staggered modules, $\Sc_{\ell p,s_0}^{a,0;+}$ and $\Sc_{\ell p,s_0}^{a,0;-}$, such that
the following short exact sequences are non-split:
\be
\begin{array}{rcccccccl}
 0\!&\!\to\!&\!\Qc_{\ell p-a,s_0}\!&\!\to\!&\!\Sc_{\ell p,s_0}^{a,0;+}\!&\!\to\!&\!\Qc_{\ell p+a,s_0}\!&\!\to\!&\!0,
\\[.3cm]
 0\!&\!\to\!&\!\Qc_{-\ell p+a,s_0-p'}\!&\!\to\!&\!\Sc_{\ell p,s_0}^{a,0;-}\!&\!\to\!&\!\Qc_{-\ell p-a,s_0-p'}\!&\!\to\!&\!0.
\end{array}
\ee
\noindent
{\bf Conjecture 2.}\quad
For every $3$-tuple $(r_0,b,\ell)\in\mathbb{Z}^3$, where
\be
 1\leq r_0\leq p-1,\qquad
 1\leq b\leq p'-1,\qquad
 \ell\geq1,
\ee
there exists a pair of quadrangular staggered modules, $\Sc_{r_0,\ell p'}^{0,b;+}$ and $\Sc_{r_0,\ell p'}^{0,b;-}$, such that
the following short exact sequences are non-split:
\be
 0\,\to\,\Qc_{\pm r_0,\pm(\ell p'-b)}\,\to\,\Sc_{r_0,\ell p'}^{0,b;\pm}\,\to\,\Qc_{\mp r_0,\mp(\ell p'+b)}\,\to\,0.
\ee
\noindent
{\bf Conjecture 3.}\quad
For every $2$-tuple $(b,\ell)\in\mathbb{Z}^2$, where
\be
 1\leq b\leq p'-1,\qquad
 \ell\geq1,
\ee
there exists a pair of staggered modules, $\Sc_{p,\ell p'}^{0,b;+}$ and $\Sc_{p,\ell p'}^{0,b;-}$, such that
the following short exact sequences are non-split:
\be
 0\,\to\,\Qc_{\pm p,\pm(\ell p'-b)}\,\to\,\Sc_{p,\ell p'}^{0,b;\pm}\,\to\,\Qc_{\mp p,\mp(\ell p'+b)}\,\to\,0.
\ee
For $\ell=1$, the staggered modules $\Sc_{p,\ell p'}^{0,b;\pm}$ are triangular, whereas for $\ell>1$, they are quadrangular.
\\[.25cm]
By construction, the affine characters of the staggered modules proposed above are given by
\begin{align}
 \chit[\Sc_{\ell p,s_0}^{a,0;+}](q,z)&=\ch_{\ell p-a,s_0}(q,z)+2\,\ch_{\ell p+a,s_0}(q,z)+\ch_{(\ell+2)p-a,s_0}(q,z),
\\[.15cm]
 \chit[\Sc_{\ell p,s_0}^{a,0;-}](q,z)&=\ch_{-\ell p+a,s_0-p'}(q,z)+2\,\ch_{-\ell p-a,s_0-p'}(q,z)+\ch_{-(\ell+2)p+a,s_0-p'}(q,z),
\\[.15cm]
 \chit[\Sc_{r_0,\ell p'}^{0,b;\pm}](q,z)&=\ch_{\pm r_0,\pm(\ell p'-b)}(q,z)+2\,\ch_{\mp r_0,\mp(\ell p'+b)}(q,z)
  +\ch_{\pm r_0,\pm((\ell+2)p'-b)}(q,z),
\\[.15cm]
 \chit[\Sc_{p,\ell p'}^{0,b;\pm}](q,z)&=(1-\delta_{\ell,1})\ch_{\pm p,\pm(\ell p'-b)}(q,z)+2\,\ch_{\mp p,\mp(\ell p'+b)}(q,z)
  +\ch_{\pm p,\pm((\ell+2)p'-b)}(q,z).
\end{align}
We also note that all the quotient modules $\Qc_{r,s}$ appearing in the short exact sequences 
above are elements of $S_{\mathrm{quo}}$, as they are isomorphic to the similarly labelled affine 
Kac modules $\Ac_{r,s}$, see (\ref{Squo}).

\subsection{Example I}
\label{Sec:StagEx}

To illustrate and provide evidence for the conjectures of Section~\ref{Sec:Conjectures}, 
we consider two examples of staggered modules for $k=-\frac{4}{3}$ (in which case $t=\frac{2}{3}$, $p=2$ and $p'=3$).
The first example is discussed in the following; the second in Section~\ref{Sec:j10}.
Both examples are related to the Verma module $\Vc_{-\frac{2}{3}}$ whose Loewy diagram is given by
\psset{unit=1cm}
\be
\begin{pspicture}(-1.6,0.6)(2.5,1.9)
 \rput(-2.4,0.68){$\Vc_{-\frac{2}{3}}$\,:}
 \rput(-1.06,.75){$_{[-\frac{2}{3},-\frac{1}{3}]}$}
 \rput(0.12,1.52){$_{[-\frac{5}{3},\frac{5}{3}]}$}
 \rput(0.8,1.5){$\to$}
 \rput(1.6,1.52){$_{[-\frac{8}{3},\frac{20}{3}]}$}
 \rput(2.4,1.5){$\to$}
 \rput(3,1.52){$\cdots$}
 \rput(0.1,0.02){$_{[\frac{1}{3},\frac{2}{3}]}$}
 \rput(0.8,0){$\to$}
 \rput(1.6,0.02){$_{[\frac{4}{3},\frac{14}{3}]}$}
 \rput(2.4,0){$\to$}
 \rput(3,0.02){$\cdots$}
 \rput(-0.6,1.25){$\nearrow$}
 \rput(-0.6,0.25){$\searrow$}
 \rput(0.8,0.73){$\searrow$}
 \rput(0.8,0.73){$\nearrow$}
 \rput(2.4,0.73){$\searrow$}
 \rput(2.4,0.73){$\nearrow$}
\end{pspicture} 
\\[0.85cm]
\label{VVV}
\ee
Following (\ref{MFF0}) and (\ref{MFF1}), the maximal proper submodule of
$\Vc_{-\frac{2}{3}}$ is generated from the two singular vectors
\begin{align}
 \ket{1,1}_{-\frac{2}{3}}&=J_{-1}^+\ket{-\tfrac{2}{3},-\tfrac{1}{3}},
 \\[.2cm]
 \ket{-1,2}_{-\frac{2}{3}}
  &=(J_0^-)^{\frac{7}{3}}(J_{-1}^+)^{\frac{5}{3}}(J_0^-)(J_{-1}^+)^{\frac{1}{3}}(J_0^-)^{-\frac{1}{3}}\ket{-\tfrac{2}{3},-\tfrac{1}{3}}
 \nonumber\\[.2cm]
 &=-\tfrac{4}{81}\Big(2J_{-2}^-+6J_{-1}^-J_{-1}^3+3J_0^-J_{-2}^3+9J_0^-(J_{-1}^3)^2+9(J_0^-)^2J_{-2}^+
 \nonumber\\[.2cm]
 &\qquad\qquad-\tfrac{9}{4}\big[2J_{-1}^-J_0^-+18(J_0^-)^2J_{-1}^3+9(J_0^-)^3J_{-1}^+\big]J_{-1}^+\Big)\ket{-\tfrac{2}{3},-\tfrac{1}{3}}.
\label{-12sing}
\end{align}

Now, Conjecture 1 asserts that there exists a quadrangular staggered module, denoted by $\Sc_{2,2}^{1,0;-}$, whose 
Loewy diagram is of the form
\psset{unit=1cm}
\be
\begin{pspicture}(0,1.1)(1.4,2.7)
 \rput(-0.2,2.2){{\scriptsize $[-\frac{2}{3},-\frac{1}{3}]$}}
 \rput(-0.2,1.1){{\scriptsize $[-\frac{5}{3},\frac{5}{3}]$}}
 \rput(1.4,1.1){{\scriptsize $[-\frac{5}{3},\frac{5}{3}]$}}
 \rput(1.4,0){{\scriptsize $[-\frac{8}{3},\frac{20}{3}]$}}
 \rput(0.6,0.55){$\nwarrow$}
 \rput(0.6,1.65){$\nwarrow$}
 \rput(-0.2,1.65){$\downarrow$}
 \rput(1.4,0.55){$\downarrow$}
 \rput(0.6,1.1){$\leftarrow$}
\end{pspicture}
\\[1.2cm]
\label{S1}
\ee
corresponding to the short exact sequence
\be
 0\,\to\,\Qc_{-1,-1}\,\to\,\Sc_{2,2}^{1,0;-}\,\to\,\Qc_{-3,-1}\,\to\,0,
\label{Q1}
\ee
where $\Qc_{-1,-1}=\Vc_{-\frac{2}{3}}/\Vc_{\frac{1}{3}}$ and $\Qc_{-3,-1}=\Vc_{-\frac{5}{3}}/\Vc_{\frac{4}{3}}$.
As in (\ref{VVV}), the four pairs of numbers $[j,h_j]$ in the diagram (\ref{S1}) are the affine and conformal weights of 
the associated irreducible sub-quotients. In the submodule $\Qc_{-1,-1}$,
the maximal proper submodule (in fact, the only proper submodule) is generated from the singular vector
\be
 \ket{-1,2}_{-\frac{2}{3}}
  =S\ket{-\tfrac{2}{3},-\tfrac{1}{3}},\qquad
  S:=-\tfrac{4}{81}\big(2J_{-2}^-+6J_{-1}^-J_{-1}^3+3J_0^-J_{-2}^3+9J_0^-(J_{-1}^3)^2+9(J_0^-)^2J_{-2}^+\big),
\label{m12j}
\ee
obtained from (\ref{-12sing}) by setting $J_{-1}^+\ket{-\frac{2}{3},-\frac{1}{3}}\equiv0$.
Let $\ket{R}$ denote a corresponding logarithmic vector (\ref{etamujhj}),
\be
 (J_0^3+\tfrac{5}{3})\ket{R}=\eta\ket{-1,2}_{-\frac{2}{3}}\qquad
 (L_0-\tfrac{5}{3})\ket{R}=\mu\ket{-1,2}_{-\frac{2}{3}},
\label{etamu1}
\ee
in which case the relation (\ref{off}) specialises to
\be
 \big(L_0-\tfrac{3}{2}J_0^3(J_0^3+1)\big)\ket{R}=\big(\mu+\tfrac{7}{2}\eta\big)\ket{-1,2}_{-\frac{2}{3}}.
\label{off1}
\ee
From
\be
 J_1^-\ket{R}\in[\Qc_{-1,-1}]_{-2,1},\qquad 
 J_0^+\ket{R}\in[\Qc_{-1,-1}]_{0,2},
\ee
it follows that
\begin{align}
 J_1^-\ket{R}&=\big(\delta_1J_{-1}^-J_0^-+\delta_2(J_0^-)^2J_{-1}^3\big)\ket{-\tfrac{2}{3},-\tfrac{1}{3}},
\\[.2cm]
 J_0^+\ket{R}&=\big(\epsilon_1J_0^-J_{-2}^++\epsilon_2(J_{-1}^3)^2
   +\epsilon_3J_{-2}^3\big)\ket{-\tfrac{2}{3},-\tfrac{1}{3}},
\end{align}
for some $\delta_1,\delta_2,\epsilon_1,\epsilon_2,\epsilon_3\in\mathbb{C}$.
Applying (\ref{JJ}) to $\ket{R}$ then implies the relation
\be
 (6\delta_1-2\delta_2+6\epsilon_1-3\epsilon_2)J_{-1}^+\ket{-\tfrac{2}{3},-\tfrac{1}{3}}=0.
\label{rel1}
\ee
However, $J_{-1}^+\ket{-\tfrac{2}{3},-\tfrac{1}{3}}\equiv0$ in $\Qc_{-1,-1}$, so this does not impose any constraints 
on the parameters $\delta_1,\ldots,\epsilon_3$.
Using (\ref{dagger}), (\ref{L0}) and (\ref{J1+})-(\ref{J23}), we also find that
\begin{align}
 \big(L_0-\tfrac{3}{2}J_0^3(J_0^3+1)\big)\ket{R}
  &=-\tfrac{9}{8}(6\delta_1-2\delta_2+6\epsilon_1-3\epsilon_2)\ket{-1,2}_{-\frac{2}{3}},
 \\[.2cm]
 S^\dagger\ket{R}
  &=\tfrac{560}{2187}(6\delta_1-2\delta_2+6\epsilon_1-3\epsilon_2)\ket{-\tfrac{2}{3},-\tfrac{1}{3}}.
\end{align}

In $\Vc_{-\frac{5}{3}}$, the submodule isomorphic to $\Vc_{\frac{4}{3}}$ is generated from the
singular vector
\be
 \ket{3,3}_{-\frac{5}{3}}=(J_{-1}^+)^3\ket{-\tfrac{5}{3},\tfrac{5}{3}}.
\ee
For the quotient in (\ref{Q1}) to be isomorphic to $\Vc_{-\frac{5}{3}}/\Vc_{\frac{4}{3}}$ instead of
some other quotient of $\Vc_{-\frac{5}{3}}$, a singular vector of the form (\ref{PRL}),
\be
 \ket{2,5}_{-\frac{2}{3}}^{\mathrm{log}}
  =(J_{-1}^+)^3\ket{R}+\big(a J_{-3}^+J_{-2}^++bJ_{-1}^3(J_{-2}^+)^2\big)\ket{-\tfrac{2}{3},-\tfrac{1}{3}},\qquad
  a,b\in\mathbb{C},
\ee
must therefore vanish in the staggered module. Accordingly, imposing
\be
 J_1^-\ket{2,5}_{-\frac{2}{3}}^{\mathrm{log}}=J_0^+\ket{2,5}_{-\frac{2}{3}}^{\mathrm{log}}=0
\ee
is seen to fix
\be
 a=-6\delta_1+6\delta_2,\qquad b=-12\delta_2
\ee
and imply that
\be
 \eta=0.
\ee
The vanishing of $\eta$ means that $J_0^3$ acts {\em diagonalisably} on the staggered module.
This is contrasted by $L_0$ whose off-diagonal action in (\ref{off1}) we may normalise by setting $\mu=1$, 
thereby obtaining
\be
 6\delta_1-2\delta_2+6\epsilon_1-3\epsilon_2=-\tfrac{8}{9},
\ee
hence
\be
 S^\dagger\ket{R}=\beta\ket{-\tfrac{2}{3},-\tfrac{1}{3}},\qquad
 \beta=-\tfrac{4480}{19683}.
\ee
As expected, the logarithmic coupling constant $\beta$ is independent of the specific values for 
$\delta_1,\delta_2,\epsilon_1,\epsilon_2,\epsilon_3$.
It does depend, however, on the
convention for $\ket{-1,2}_{-\frac{2}{3}}$ in (\ref{m12j})
and $\mu$ in (\ref{etamu1}). For the {\em existence} of the staggered module $\Sc_{2,2}^{1,0;-}$, 
the key requirement is that there {\em exist} such parameters respecting all the
self-consistency conditions imposed by the 
structure of the module.

\subsection{Example II}
\label{Sec:j10}

For $k=-\frac{4}{3}$,
Conjecture 2 asserts that there exists a quadrangular staggered module, denoted by $\Sc_{1,3}^{0,1;+}$, whose 
Loewy diagram is of the form
\psset{unit=1cm}
\be
\begin{pspicture}(0,1.1)(1.4,2.7)
 \rput(-0.2,2.2){{\scriptsize $[-\frac{2}{3},-\frac{1}{3}]$}}
 \rput(-0.2,1.1){{\scriptsize $[\frac{1}{3},\frac{2}{3}]$}}
 \rput(1.4,1.1){{\scriptsize $[\frac{1}{3},\frac{2}{3}]$}}
 \rput(1.4,0){{\scriptsize $[-\frac{8}{3},\frac{20}{3}]$}}
 \rput(0.6,0.55){$\nwarrow$}
 \rput(0.6,1.65){$\nwarrow$}
 \rput(-0.2,1.65){$\downarrow$}
 \rput(1.4,0.55){$\downarrow$}
 \rput(0.6,1.1){$\leftarrow$}
\end{pspicture}
\\[1.2cm]
\label{S2}
\ee
corresponding to the short exact sequence
\be
 0\,\to\,\Qc_{1,2}\,\to\,\Sc_{1,3}^{0,1;+}\,\to\,\Qc_{-1,-4}\,\to\,0,
\label{Q2}
\ee
where $\Qc_{1,2}=\Vc_{-\frac{2}{3}}/\Vc_{-\frac{5}{3}}$ and $\Qc_{-1,-4}=\Vc_{\frac{1}{3}}/\Vc_{\frac{4}{3}}$.
In the submodule $\Qc_{1,2}$, the maximal proper submodule is generated from the singular vector
\be
 \ket{1,1}_{-\frac{2}{3}}
  =S\ket{-\tfrac{2}{3},-\tfrac{1}{3}},\qquad
  S:=J_{-1}^+.
\label{m12j2}
\ee
As before, we let $\ket{R}$ denote a matching logarithmic vector such that
\be
 \big(J_0^3-\tfrac{1}{3}\big)\ket{R}=\eta\ket{1,1}_{-\frac{2}{3}},\qquad
 \big(L_0-\tfrac{2}{3}\big)\ket{R}=\mu\ket{1,1}_{-\frac{2}{3}},
\label{etamu2}
\ee
hence
\be
 \big(L_0-\tfrac{3}{2}J_0^3(J_0^3+1)\big)\ket{R}=\big(\mu-\tfrac{5}{2}\eta\big)\ket{1,1}_{-\frac{2}{3}},
\label{off2}
\ee
for some $\eta,\mu\in\mathbb{C}$, at most one of which can be zero.
Moreover,
\be
 J_1^-\ket{R}\in[\Qc_{1,2}]_{0,0},\qquad 
 J_0^+\ket{R}\in[\Qc_{1,2}]_{2,1}=\{0\},
\ee
implying that
\be
 J_1^-\ket{R}=\delta\ket{-\tfrac{2}{3},-\tfrac{1}{3}},\qquad J_0^+\ket{R}=0,
\ee
for some $\delta\in\mathbb{C}$. Using (\ref{L0}) and (\ref{dagger}), it then follows that
\be
 \big(L_0-\tfrac{3}{2}J_0^3(J_0^3+1)\big)\ket{R}=\tfrac{3}{2}\delta\ket{1,1}_{-\frac{2}{3}},
 \qquad
 S^\dagger\ket{R}=\delta\ket{-\tfrac{2}{3},-\tfrac{1}{3}}.
\ee

Following the discussion in Section~\ref{Sec:Sing},
the submodule of $\Vc_{\frac{1}{3}}$ that is isomorphic to $\Vc_{\frac{4}{3}}$ is generated from the singular vector
\be
 \ket{1,4}_{\frac{1}{3}}
 =S_{\frac{4}{3},\frac{1}{3}}\ket{\tfrac{1}{3},\tfrac{2}{3}}
  =(J_{-1}^+)^3(J_0^-)^{\frac{7}{3}}(J_{-1}^+)^{\frac{5}{3}}(J_0^-)(J_{-1}^+)^{\frac{1}{3}}(J_0^-)^{-\frac{1}{3}}
    (J_{-1}^+)^{-1}\ket{\tfrac{1}{3},\tfrac{2}{3}}.
\ee
This can be rewritten as
\be
 \ket{1,4}_{\frac{1}{3}}=P\ket{\tfrac{1}{3},\tfrac{2}{3}},\qquad P\in U(\mathfrak{n}_-),
\ee
where
\begin{align}
 P=&-\tfrac{4}{81}\Big(420J_{-4}^+-1260J_{-1}^3J_{-3}^+-630J_{-2}^3J_{-2}^+-420J_{-3}^3J_{-1}^+-700J_{-2}^-(J_{-1}^+)^2
  +1890J_{-2}^3J_{-1}^3J_{-1}^+
 \nonumber\\[.2cm]
 &-1260J_{-1}^-J_{-2}^+J_{-1}^++1680J_{-1}^-J_{-1}^3(J_{-1}^+)^2-1170J_0^-J_{-3}^+J_{-1}^++705J_0^-J_{-2}^3(J_{-1}^+)^2
  +1890(J_{-1}^3)^2J_{-2}^+
 \nonumber\\[.2cm]
 &-540J_0^-(J_{-2}^+)^2-1890(J_{-1}^3)^3J_{-1}^++2970J_0^-J_{-1}^3J_{-2}^+J_{-1}^+
  -1935J_0^-(J_{-1}^3)^2(J_{-1}^+)^2
 \nonumber\\[.2cm]
 &+495(J_0^-)^2J_{-2}^+(J_{-1}^+)^2+360J_{-1}^-J_0^-(J_{-1}^+)^3-405(J_0^-)^2J_{-1}^3(J_{-1}^+)^3-\tfrac{81}{4}(J_0^-)^3(J_{-1}^+)^4\Big).
\end{align}
A singular vector of the form
\be
 \ket{2,5}_{-\frac{2}{3}}^{\mathrm{log}}
  =P\ket{R}+\ket{L},\qquad \ket{L}\in[\Qc_{1,2}]_{2,5},
\ee
must therefore vanish in the staggered module.
Accordingly, we find that imposing
\be
 J_1^-\ket{2,5}_{-\frac{2}{3}}^{\mathrm{log}}=J_0^+\ket{2,5}_{-\frac{2}{3}}^{\mathrm{log}}=0
\ee
fixes $\ket{L}$ and implies that
\be
 \eta=\mu,\qquad \delta=-\mu.
\ee
It follows that both $J_0^3$ and $L_0$ act non-diagonalisably on the staggered module and that
setting $\mu=1$ yields
\be
 S^\dagger\ket{R}=\beta\ket{-\tfrac{2}{3},-\tfrac{1}{3}},\qquad
 \beta=-1.
\ee

\section{Coset construction}
\label{Sec:Coset}

\subsection{Branching rules}

Let $\h$ denote a subalgebra of the affine Lie algebra $\g$, and $L^\h_n$ and $L^\g_n$ the generators of 
the corresponding Segal-Sugawara constructions. In~\cite{GKO85,GKO86},
Goddard, Kent and Olive found that the generators
\be
 L_n^{\g/\h}:=L_n^\g-L_n^\h,\qquad n\in\mathbb{Z},
\ee
fom a Virasoro algebra of central charge $c^{\g/\h}=c^\g-c^\h$, and
that they commute with the generators of $\h$,
\be
 [L_n^{\g/\h},\h]=0, \qquad n\in\mathbb{Z}.
\ee
This suggests that a $\g$-module $\mathcal{M}_\la^\g$ decomposes in terms of $\h$-modules,
\be
 \Mc_\la^{\g}\simeq\bigoplus_{\mu}B_\la^{\mu}\otimes\Mc_{\mu}^\h,
\ee
in such a way that the `coefficients' $B_\la^{\mu}$ may be viewed as modules over the coset Virasoro
algebra generated by $L_n^{\g/\h}$. In terms of characters, we have the corresponding
{\em branching rule}
\be
 \chit[\mathcal{M}_\la^\g](q,z)=\sum_{\mu}\chit[B_\la^{\mu}](q)\,\chit[\mathcal{M}_{\mu}^\h](q,z'),
\label{chiMBM}
\ee
where $z'$ is a function of $z$, determined by the embedding $\h\subseteq\g$, while
the Virasoro characters $\chit[B_\la^{\mu}](q)$ are known as {\em branching functions}.

Here, we are interested in the case (\ref{cosetAAA}) where
$\h=(A_1^{(1)})_{k+n}$ is the diagonal subalgebra of $\g=(A_1^{(1)})_k\oplus(A_1^{(1)})_n$ for
$t=k+2\in\mathbb{Q}_+$ and $n\in\mathbb{N}$.
The branching rules then take the form of generalised Kac-Peterson multiplication formulas (\ref{KP}),
\be
 \chit[\Mc_{k;\lambda}](q,z)\,\chit[\Mc_{n;\lambda'}](q,z)=\sum_\mu\chit[\Mc_{k,n;\mu}^{\mathrm{Vir}}](q)
  \,\chit[\Mc_{k+n;\mu}](q,z),
\label{KPa}
\ee
where $\Mc_{k;\lambda}$ is an $(A_1^{(1)})_k$-module labeled by $\lambda$, while the central charge of the
coset Virasoro algebra is given by
\be
 c^{k;n}=c_k+c_n-c_{k+n}
 =\frac{3n}{n+2}-\frac{6n}{t(t+n)}.
\label{ckn}
\ee
In the decomposition (\ref{KPa}) of a product of {\em admissible} characters, 
the branching functions, $\chit[\Mc_{k,n;\mu}^{\mathrm{Vir}}](q)$, are 
known~\cite{ACT91,BMSW9702,SW02} and given in terms of the string functions (\ref{sf}).
Our objective is to determine the branching functions that arise when considering affine Kac modules 
and staggered modules. Our results agree with and extend those found in~\cite{PR13} by computing the 
so-called logarithmic limit~\cite{Ras0405,Ras0406,PS1207} of the branching rules for admissible characters.

\subsection{Branching of affine Kac characters}

We recall that the affine Kac character $\chit_{r,s}^{p,p'}(q,z)$ is defined for $r>0$ and $s\geq0$ and for $r,s<0$,
while the integer-level admissible character $\ch_{\rho,0}^{n+2,1}(q,z)$ is defined for $\rho=1,\ldots,n+1$.
Following (\ref{KPa}), the goal here is to express $\chit_{r,s}^{p,p'}(q,z)\,\ch_{\rho,0}^{n+2,1}(q,z)$ as a sum 
of $(A_1^{(1)})_{k+n}$-characters with $q$-dependent coefficients.
As an ansatz, we take these characters to be affine Kac characters. We thus write
\be
 \chit_{r,s}^{p,p'}(q,z)\,\ch_{\rho,0}^{n+2,1}(q,z)
  =\sum_{\sigma}\chit_{r,s;\rho;\sigma}^{p,p';n}(q)\,
    \chit_{\sigma,s}^{p+np',p'}(q,z)
\label{cosetKac}
\ee
and seek to determine the branching functions $\chit_{r,s;\rho;\sigma}^{p,p';n}(q)$. 
For consistency, the summation variable $\sigma$ must
satisfy $\sigma\in\mathbb{N}$ if $r>0$, and $\sigma\in-\mathbb{N}$ if $r<0$. As the analysis in Appendix~\ref{Sec:Alt} 
reveals, $\sigma$ is also subject to the parity constraint
\be
 \sigma\equiv r+ns+\rho-1 \mod 2,
\label{sigma2}
\ee
while
\be
 \chit_{r,s;\rho;\sigma}^{p,p';n}(q)=\chit_{-r,-s;\rho;-\sigma}^{p,p';n}(q),\qquad r,s,\sigma<0.
\label{msigma}
\ee
Because of the relation (\ref{msigma}), all {\em distinct} such branching functions appear for $r,\sigma>0$ and $s\geq0$.

Explicit expressions for the branching functions were found in~\cite{PR13} by evaluating 
the logarithmic limit~\cite{Ras0405,Ras0406,PS1207} of the branching relations (\ref{KPa}) for admissible characters.
For $r,\sigma>0$ and $s\geq0$, it was found that
\be
 \chit_{r,s;\rho;\sigma}^{p,p';n}(q)=q^{\frac{(r(p+np')-\sigma p)^2}{4np(p+np')}}
  \big[c_{r-\sigma}^{\,\ell}(q)-q^{\frac{r\sigma}{n}}c_{r+\sigma}^{\,\ell}(q)\big],\qquad \ell=\begin{cases}
  \rho-1,\ & s\ \,\mathrm{even},\\[.15cm] n+1-\rho,\ & s\ \,\mathrm{odd}.\end{cases}
\label{br}
\ee
We now prove this result algebraically. 
With (\ref{br}) and using (\ref{aK}), the branching rule (\ref{cosetKac}) is seen to be equivalent to
\be
 (1-q^{rs}z^r)\,\ch_{\rho,0}^{n+2,1}(q,z)=\sum_\sigma q^{\frac{(r+ns-\sigma)^2}{4n}}z^{\frac{r+ns-\sigma}{2}}
  (1-q^{\sigma s}z^\sigma)\big[c_{r-\sigma}^{\,\ell}(q)-q^{\frac{r\sigma}{n}}c_{r+\sigma}^{\,\ell}(q)\big],
\ee
whose right-hand side can be re-expressed as
\begin{align}
    RHS
 &=\sum_\sigma\Big(q^{\frac{(r+ns-\sigma)^2}{4n}}z^{\frac{r+ns-\sigma}{2}}c_{r-\sigma}^{\,\ell}(q)
  +q^{\frac{(r+ns+\sigma)^2}{4n}}z^{\frac{r+ns+\sigma}{2}}c_{r+\sigma}^{\,\ell}(q)\Big)
 \nonumber\\[.15cm]
 &-\sum_\sigma q^{rs}z^r\Big(q^{\frac{(-r+ns-\sigma)^2}{4n}}z^{\frac{-r+ns-\sigma}{2}}c_{r+\sigma}^{\,\ell}(q)
  +q^{\frac{(-r+ns+\sigma)^2}{4n}}z^{\frac{-r+ns+\sigma}{2}}c_{r-\sigma}^{\,\ell}(q)\Big).
\end{align}
Since the summands cancel for $\sigma=0$, we see that
\begin{align}
 RHS&=\sum_{\sigma\,\in\,2\mathbb{Z}+(r+ns+\rho-1)} 
 \Big(q^{\frac{(r+ns-\sigma)^2}{4n}}z^{\frac{r+ns-\sigma}{2}}c_{r-\sigma}^{\,\ell}(q)
  -q^{rs}z^rq^{\frac{(-r+ns+\sigma)^2}{4n}}z^{\frac{-r+ns+\sigma}{2}}c_{r-\sigma}^{\,\ell}(q)\Big)
 \nonumber\\[.15cm]
 &=(1-q^{rs}z^r)\sum_{\sigma\,\in\,2\mathbb{Z}+\rho-1}c_{ns+\sigma}^{\,\ell}(q)\,q^{\frac{\sigma^2}{4n}}z^{-\frac{\sigma}{2}},
\end{align}
where we have used the symmetry properties (\ref{ccc}).
For $s$ even, we furthermore have
\be
 \sum_{\sigma\,\in\,2\mathbb{Z}+\rho-1}c_{ns+\sigma}^{\,\ell}(q)\,q^{\frac{\sigma^2}{4n}}z^{-\frac{\sigma}{2}}
 =\sum_{\sigma\,\in\,2\mathbb{Z}+\rho-1}c_{\sigma}^{\rho-1}(q)\,q^{\frac{\sigma^2}{4n}}z^{-\frac{\sigma}{2}}
 =\ch_{\rho,0}^{n+2,1}(q,z),
\ee
while for $s$ odd, we similarly have
\be
 \sum_{\sigma\,\in\,2\mathbb{Z}+\rho-1}\!\!c_{ns+\sigma}^{\,\ell}(q)\,q^{\frac{\sigma^2}{4n}}z^{-\frac{\sigma}{2}}
 =\!\sum_{\sigma\,\in\,2\mathbb{Z}+\rho-1}\!\!c_{ns+\sigma}^{n-\rho+1}(q)\,q^{\frac{\sigma^2}{4n}}z^{-\frac{\sigma}{2}}
  =\!\sum_{\sigma\,\in\,2\mathbb{Z}+\rho-1}\!\!c_{\sigma}^{\rho-1}(q)\,q^{\frac{\sigma^2}{4n}}z^{-\frac{\sigma}{2}}
 =\ch_{\rho,0}^{n+2,1}(q,z).
\ee
This concludes the proof of (\ref{br}).

The exponent of $q$ in (\ref{br}) suggests that the central charge $c^{k;n}$ (\ref{ckn}) of the coset Virasoro algebra
 may be interpreted as a $(p,p')$-dependent expression $c^{(n)}$ in which the parameters
have been shifted as $(p,p')\to(p,p+np')$. 
Thus requiring that
\be
 c^{t-2;n}=c^{(n)}\big(\frac{t}{t+n}\big),\qquad t=\frac{p}{p'},
\label{ct2n}
\ee
as well as
\be
 c^{(n)}(t)\in\mathbb{C}[t,t^{-1}],\qquad
 c^{(n)}(t^{-1})=c^{(n)}(t),
% c^{(n)}(1)=\frac{3n}{n+2},\qquad
\ee
we recover the familiar expression
\be
 c^{(n)}(t)=\frac{3n}{n+2}-\frac{6(1-t)^2}{nt}.
\label{cnt}
\ee

\subsection{Virasoro Kac characters}
\label{Sec:VirKac}

Virasoro Kac characters first appeared in~\cite{PRZ0607} as the characters of certain modules arising as natural building blocks
in the logarithmic minimal models ${\cal LM}(p,p')$, but whose properties were only
partly known. The precise identification of the underlying modules was made in~\cite{Ras1012,MDRR1503} and will be reviewed
in Section~\ref{Sec:PhiKac}.
Here, it suffices to recall their characters.
For every pair $r,s\in\mathbb{N}$, the corresponding Virasoro Kac character is thus given by
\be
 \chit_{r,s}(q)=\frac{q^{\bar{h}_{r,s}-\frac{\bar{c}}{24}}(1-q^{rs})}{\varphi(q)},
\label{VirKac}
\ee
where
\be
 \bar{h}_{r,s}:=\frac{(r-st)^2-(1-t)^2}{4t},\qquad \bar{c}=\bar{c}(t):=1-\frac{6(1-t)^2}{t}.
\label{hbar}
\ee
To indicate the dependence on $t=p/p'$, we may write $\chit_{r,s}^{p,p'}(q)$.

The Virasoro Kac characters also appear as the branching functions in (\ref{cosetKac}) for $n=1$. Indeed, for
$r,\sigma>0$, $s\geq0$ and $\rho=1,2$, such that $\sigma\equiv r+s+\rho-1$ mod $2$, we have
\be
 \chit_{r,s;\rho;\sigma}^{p,p';1}(q)=q^{\frac{(r(p+p')-\sigma p)^2}{4p(p+p')}}(1-q^{r\sigma})\,c_{\,0}^{\,0}(q)
  =\chit_{r,\sigma}^{p,p+p'}(q),
\ee
where we have used (\ref{c00}). We note the shifted dependence on $p,p'$,
in accordance with $\bar{c}=c^{(1)}$, see (\ref{ct2n}) and (\ref{cnt}).

\subsection{Superconfomal Kac characters}

Superconformal Kac characters first appeared in~\cite{PR13,PRT14} and are given
for $r,s\in\mathbb{N}$ by
\be
 \chih_{r,s}(q)=2(1-\kappa)q^{\hat{h}_{r,s}-\frac{\hat{c}}{24}}
  \displaystyle{\Big(\prod_{i=1}^\infty\frac{1+q^{i-\kappa}}{1-q^i}\Big)(1-q^{\frac{rs}{2}})},\qquad
  \kappa=\begin{cases} 0,\ & r+s\ \,\mathrm{odd},\\[.15cm]  \tfrac{1}{2},\ & r+s\ \,\mathrm{even},
\end{cases}
\label{chih}
\ee
where
\be
 \hat{h}_{r,s}:=\frac{(r-st)^2-(1-t)^2}{8t}+\frac{1}{32}\big(1-(-1)^{r+s}\big),
 \qquad 
 \hat{c}=\hat{c}(t):=\frac{3}{2}-\frac{3(1-t)^2}{t}.
\label{hhat}
\ee
To indicate the dependence on $t=p/p'$, we may write $\chih_{r,s}^{p,p'}(q)$.
In a way very similar to the Virasoro Kac modules, the underlying modules are defined 
for the Neveu-Schwarz sector ($\kappa=0$) in~\cite{CRR15} and for the Ramond sector ($\kappa=\frac{1}{2}$) in~\cite{CR16}.
We refer to~\cite{IK03} for a discussion of Ramond algebra Verma modules and 
characters in the special case $\hat{h}=\frac{\hat{c}}{24}$.
As discussed in the following, the superconformal Kac characters are closely related to the branching functions (\ref{br}) for $n=2$.

A Neveu-Schwarz Kac character naturally splits into two parts, as
\be
 \prod_{i=1}^\infty\frac{1+q^{i-\frac{1}{2}}}{1-q^i}=q^{\frac{1}{16}}\big[c_{\,0}^{\,0}(q)+c_{\,0}^{\,2}(q)\big],
\ee
where
\be
 q^{\frac{1}{16}}c_{\,0}^{\,0}(q)\in\mathbb{N}_0[[q]],\qquad 
 q^{\frac{1}{16}}c_{\,0}^{\,2}(q)\in q^{\frac{1}{2}}\mathbb{N}_0[[q]],
\ee
while $\frac{rs}{2}\in\frac{1}{2}\mathbb{N}_0$.
To describe this, we find it convenient to introduce the operators $P_\delta$, $\delta\in\mathbb{Z}_2$,
where $P_0$ (respectively $P_1$) singles out the contribution from vectors of the same (respectively opposite) parity 
as the highest-weight vector. Extending this action to the generating function, we thus have
\be
 P_\delta\,\Big(\prod_{i=1}^\infty\frac{1+q^{i-\frac{1}{2}}}{1-q^i}\Big)
  =q^{\frac{1}{16}}c_{\,0}^{\,2\delta}(q)
  =\frac{1}{2}\Big(\prod_{i=1}^\infty\frac{1+q^{i-\frac{1}{2}}}{1-q^i}+(-1)^\delta\prod_{i=1}^\infty\frac{1-q^{i-\frac{1}{2}}}{1-q^i}\Big),
   \qquad \delta\in\mathbb{Z}_2.
\ee
In the Ramond sector, the similar split into even and odd parts yields two identical contributions
to the Virasoro character (due to the action of the odd algebra generator $G_0$), as
\be
 P_\delta\,\Big(\prod_{i=1}^\infty\frac{1+q^{i}}{1-q^i}\Big)
  =c_{\,1}^{\,1}(q),\qquad\delta\in\mathbb{Z}_2.
\ee
This also explains the factor of $2$ in (\ref{chih}).

Still with $n=2$, we now let $r,\sigma>0$ and $s\geq0$ such that $\sigma\equiv r+\rho-1$ mod $2$.
For $\rho=1,3$, we then find that
\be
 \chit_{r,s;\rho;\sigma}^{p,p';2}(q)
  =P_\delta\big(\chih_{r,\sigma}^{p,p+2p'}(q)\big),\qquad r+\sigma\ \,\mathrm{even},
  \qquad
  \delta=\begin{cases} 0,\ & \tfrac{1}{2}(r+2s+\rho-\sigma-1)\ \,\mathrm{even},\\[.15cm]
   1,\ & \tfrac{1}{2}(r+2s+\rho-\sigma-1)\ \,\mathrm{odd}, \end{cases}
\label{NS}
\ee
corresponding to the Neveu-Schwarz sector.
For $\rho=2$, we likewise find that
\be
 \chit_{r,s;2;\sigma}^{p,p';2}(q)
  =P_0\big(\chih_{r,\sigma}^{p,p+2p'}(q)\big)
  =P_1\big(\chih_{r,\sigma}^{p,p+2p'}(q)\big)
  =\tfrac{1}{2}\chih_{r,\sigma}^{p,p+2p'}(q),\qquad r+\sigma\ \,\mathrm{odd},
\label{Ram}
\ee
corresponding to the Ramond sector.
In both sectors, we thus have
\be
  \chit_{r,s;\rho;\sigma}^{p,p';2}(q)+\chit_{r,s;4-\rho;\sigma}^{p,p';2}(q)=\chih_{r,\sigma}^{p,p+2p'}(q),\qquad
   \rho=1,2,3.
\ee
Up to a sign depending on the parity of the highest-weight vector,
the associated {\em super-characters} are given by
\be
 \chit_{r,s;1;\sigma}^{p,p';2}(q)-\chit_{r,s;3;\sigma}^{p,p';2}(q)=
  (-1)^{\frac{r+2s-\sigma}{2}}q^{\frac{(r(p+2p')-\sigma p)^2}{8p(p+2p')}}\big(c_{\,0}^{\,0}(q)-c_{\,0}^{\,2}(q)\big)
    \big(1-(-1)^\sigma q^{\frac{r\sigma}{2}}\big)
\label{super}
\ee
in the Neveu-Schwarz sector, whereas they are zero in the Ramond sector.
We note the shifted dependence on $p,p'$ in the superconformal characters (\ref{NS})-(\ref{super}), 
in accordance with $\hat{c}=c^{(2)}$, 
see (\ref{ct2n}) and (\ref{cnt}).

\subsection{Branching of staggered characters}

The branching rules for the staggered modules proposed in Section~\ref{Sec:Conjectures} follow from
the branching rules for affine Kac characters. 
For simplicity, we only consider staggered branching for $n=1$, in which case the branching functions are Virasoro characters.
As discussed in the following, some of these are characters of staggered Virasoro modules.
More generally, we will be interested in characters of the form
\begin{align}
 \chit[R_{\ell p,s}^{a,0}](q)&=\chit_{\ell p-a,s}(q)+\chit_{\ell p+a,s}(q),\qquad\ 1\leq a<p,\qquad \ell,s\geq1,
\label{Ra0}
 \\[.2cm]
 \chit[R_{r,\ell p'}^{0,b}](q)&=\chit_{r,\ell p'-b}(q)+\chit_{r,\ell p'+b}(q),\qquad 1\leq b<p',\qquad \ell,r\geq1,
\label{R0b}
\end{align}
where $R_{\ell p,s}^{a,0}$ and $R_{r,\ell p'}^{0,b}$ with $s\leq p'$ and $r\leq p$
are indeed staggered Virasoro modules known~\cite{RP0706,RP0707} from logarithmic minimal models.
The Loewy diagrams of these modules are discussed in Section~\ref{Sec:PhiStag}.
To emphasise the dependence on $p,p'$, we write $\chit^{p,p'}[R_{r,s}^{a,b}](q)$.

Now, for the staggered modules in Conjecture~$1$, we find the branching rules
\begin{align}
 \chit[\Sc_{\ell p,s_0}^{a,0;+}](q,z)\,\ch_{\rho,0}^{n+2,1}(q,z)
 &={\sum_{\sigma>0}}'\chit^{p,p+p'}[R_{\ell p,\sigma}^{a,0}](q)\,\chit_{\sigma,s_0}^{p+p',p'}(q,z),
 \\[.15cm]
 \chit[\Sc_{\ell p,s_0}^{a,0;-}](q,z)\,\ch_{\rho,0}^{n+2,1}(q,z)
 &={\sum_{\sigma>0}}'\chit^{p,p+p'}[R_{\ell p,\sigma}^{a,0}](q)\,\chit_{-\sigma,s_0-p'}^{p+p',p'}(q,z),
\end{align}
where the primes on $\sum$ indicate that the parity of the corresponding summation variables is constrained 
respectively by
$\sigma\equiv\ell p-a+s_0+\rho-1$ and $\sigma\equiv\ell p-a+p'+s_0+\rho-1$ mod $2$.
To describe the branching of the characters of the
staggered modules in Conjecture~$2$ and~$3$, we let $r=1,\ldots,p$. We then find that
\be
  \chit[\Sc_{r,\ell p'}^{0,b;\pm}](q,z)\,\ch_{\rho,0}^{n+2,1}(q,z)
  ={\sum_{\sigma>0}}'\chit_{r,\sigma}^{p,p+p'}(q)\,\chit^{p+p',p'}[\Sc_{\sigma,\ell p'}^{0,b;\pm}](q,z),
\ee
where the prime on $\sum$ indicates that $\sigma\equiv r+\ell p'-b+\rho-1$ mod $2$, and where we have introduced
\be
 \chit[\Sc_{\sigma,\ell p'}^{0,b;\pm}](q,z)=\chit_{-\sigma,-\ell p'+b}(q,z)+\chit_{\sigma,\ell p'+b}(q,z).
\ee
As the notation suggests, for every $\sigma,\ell>0$ and $b=1,\ldots,p'-1$, 
we believe that there exists a pair of $A_1^{(1)}$-modules, $\Sc_{\sigma,\ell p'}^{0,b;\pm}$, 
generalising the staggered ones in Conjecture~$2$ and~$3$.

\section{Module functor}
\label{Sec:Fun}

\subsection{Mukhi-Panda residues}

It follows from the rewriting of the Jacobi triple product in (\ref{varphiqz})
or from the expression (\ref{1varphiqz}) for the reciprocal of the product,
together with the result (\ref{sumf}), that
\be
 \lim_{z\to1}\frac{1-z}{\eta(q,z)}=\frac{1}{\eta^3(q)}.
\ee
Specialised affine characters $\chit(q,1)$ are therefore ill-defined in general. 
Instead, the {\em residue} of an affine character at $z=1$ may be well-defined.
Indeed, in~\cite{MP90}, it was found that the residue of an admissible character is given
in terms of an irreducible Virasoro character. Concretely, these {\em Mukhi-Panda residues} are given by
\be
 \mathrm{Res}_{z=1}\big[\ch_{r_0,s_0}^{p,p'}(q,z)\big]
  =-\frac{\ch_{r_0,s_0}^{p,p'}(q)}{\eta^2(q)},\qquad 1\leq r_0\leq p-1,\quad 1\leq s_0\leq p'-1,
\label{MP}
\ee
where
\be
 \ch_{r_0,s_0}^{p,p'}(q)=\frac{\Theta_{\lambda_{r_0,s_0}^+,pp'}(q)-\Theta_{\lambda_{r_0,s_0}^-,pp'}(q)}{\eta(q)}
\ee
is an irreducible minimal-model Virasoro character. We note that the minus sign in (\ref{MP}) is a consequence 
of the minus sign on $J_0^3$ in the character (\ref{chiM}). For $s_0=0$, (\ref{thetam}) implies that
\be
  \mathrm{Res}_{z=1}\big[\ch_{r_0,0}^{p,p'}(q,z)\big]
  =-\frac{\Theta_{r_0p',pp'}(q)-\Theta_{-r_0p',pp'}(q)}{\eta^3(q)}
  =0.
\label{r00}
\ee
We will revisit the vanishing of these residues after (\ref{sumQ}).

\subsection{Residues of general affine characters}

The residue formula (\ref{MP}) admits extensions to the various families of characters considered in the preceding sections.
Indeed, we find that the act of taking the residue at $z=1$ of an affine character is equivalent to a linear
map from the space of affine characters to the space of appropriately renormalised Virasoro characters.
To appreciate this, we introduce the operator
\be
 \phi=-\eta^2(q)\mathrm{Res}_{z=1},
\ee
allowing us to re-express (\ref{MP}) as
\be
 \phi\big[\ch_{r_0,s_0}^{p,p'}(q,z)\big]=\ch_{r_0,s_0}^{p,p'}(q).
\ee
Likewise, with $\bar{h}_{r,s}$ and $\bar{c}$ as in (\ref{hbar}),
so that
\be
 h_{r,s}-\tfrac{c}{24}=\bar{h}_{r,s}-\tfrac{\bar{c}}{24}-\tfrac{1}{12},
\ee
we have
\be
 \phi\big[\chit_{j_{r,s}}(q,z)\big]=\phi\big[\chit_{-j_{r,s}-1}(q,z)\big]=\chit_{\bar{h}_{r,s}}(q),
\ee
where
\be
 \chit_{\bar{h}_{r,s}}(q)=\frac{q^{\bar{h}_{r,s}-\frac{\bar{c}}{24}}}{\varphi(q)}
\ee
is the character of the Verma module of highest weight $\bar{h}_{r,s}$ over the Virasoro algebra of central charge $\bar{c}$.
As the affine and Virasoro Verma characters only depend on $r,s$ through $(r-st)$ and $(r-st)^2$, respectively, for fractional
$t=\frac{p}{p'}$, the parameterisations of the weights can always be translated into the quadrant where $r,s>0$.
In addition, the conformal weights are invariant under simultaneous sign changes, as
\be
 \bar{h}_{-r,-s}=\bar{h}_{r,s}.
\ee
We also note the symmetries 
\be
 \bar{h}_{r,s}(\tfrac{1}{t})=\bar{h}_{s,r}(t),\qquad
 \bar{c}(\tfrac{1}{t})=\bar{c}(t),
\ee
justifying the common restriction to $p\leq p'$ in studies of the Virasoro representation theory. 

The situation is only slightly more involved for the Kac modules. We thus find that
\be
 \phi\big[\chit_{r,s}(q,z)\big]=\phi\big[\chit_{-r,-s}(q,z)\big]
 =\chit_{r,s}(q),\qquad r,s\in\mathbb{N},
\ee
where $\chit_{r,s}(q)$ is the Virasoro Kac character (\ref{VirKac}).
For $s=0$, we simply have
\be
 \phi\big[\chit_{r,0}(q,z)\big]=0,\qquad r\in\mathbb{N}.
\ee
This result reflects the fact that the singular vector generating the vanishing submodule in the associated quotient module
$\Qc_{r,0}$ appears at Virasoro level $0$ in the Verma module, 
ensuring that the space of vectors at any grade (Virasoro level) $N$ is finite-dimensional:
\be
 \sum_{Q\in\mathbb{Z}}\dim[\Qc_{r,0}]_{Q,N}<\infty,\qquad N\in\mathbb{N}_0.
\label{sumQ}
\ee
This also explains the result (\ref{r00}), as the irreducible module of highest weight $j_{r,0}$ is obtained
by a further quotient of $\Qc_{r,0}$ (except for $r\in p\mathbb{N}$, in which case $\Qc_{r,0}$ is already irreducible).
In fact, the irreducible modules with only finite-dimensional $L_0$-eigenspaces are exactly the
quasi-integrable ones.

The Mukhi-Panda residue formula (\ref{MP}) also extends to the {\em full family} of irreducible affine characters for fractional level.
We thus find that the $\phi$-images of the distinct irreducible characters (\ref{irr1})-(\ref{irr4}) are given by
\begin{align}
 \phi\big[\ch_{r_0+\ell p,s_0}(q,z)\big]
 &=\frac{\Theta_{-r_0p'+ps_0-\ell pp',pp';\ell}(q)
  -\Theta_{-r_0p'-ps_0-\ell pp',pp';\ell}(q)}{\eta(q)},
\\[.2cm]
 \phi\big[\ch_{(\ell+1)p,s_0}(q,z)\big]
 &=\frac{q^{\frac{p((\ell+1)p'-s_0)^2}{4p'}}\big(1-q^{(\ell+1)ps_0}\big)}{\eta(q)},
\\[.2cm]
 \phi\big[\ch_{r_0-(\ell+2)p,s_0-p'}(q,z)\big]
 &=\frac{\Theta_{r_0p'-ps_0-(\ell+1)pp',pp';\ell+1}(q)
  -\Theta_{-r_0p'-ps_0-(\ell+1)pp',pp';\ell+1}(q)}{\eta(q)},
\\[.2cm]
 \phi\big[\ch_{-(\ell+1)p,s_0-p'}(q,z)\big]
 &=\frac{q^{\frac{p(\ell p'+s_0)^2}{4p'}}\big(1-q^{(\ell+1)p(p'-s_0)}\big)}{\eta(q)}.
\end{align}
For $s_0=0$, these expressions specialise to
\be
 \phi\big[\ch_{r_0+\ell p,0}(q,z)\big]=\phi\big[\ch_{(\ell+1)p,0}(q,z)\big]=0
\ee
and
\be
 \phi\big[\ch_{r_0-(\ell+2)p,-p'}(q,z)\big]=\ch_{p-r_0+(\ell+1)p,p'}(q),\qquad
 \phi\big[\ch_{-(\ell+1)p,-p'}(q,z)\big]=\ch_{(\ell+1)p,p'}(q),
\label{Vir0}
\ee
whereas for $s_0\neq0$,
\begin{align}
 \phi\big[\ch_{r_0+\ell p,s_0}(q,z)\big]
 &=\ch_{r_0+\ell p,s_0}(q),
\label{Vir1}
\\[.2cm]
 \phi\big[\ch_{(\ell+1)p,s_0}(q,z)\big]
 &=\ch_{(\ell+1)p,s_0}(q),
\label{Vir2}
\\[.2cm]
 \phi\big[\ch_{r_0-(\ell+2)p,s_0-p'}(q,z)\big]
 &=\ch_{p-r_0+(\ell+1)p,p'-s_0}(q),
\label{Vir3}
\\[.2cm]
 \phi\big[\ch_{-(\ell+1)p,s_0-p'}(q,z)\big]
 &=\ch_{(\ell+1)p,p'-s_0}(q).
\label{Vir4}
\end{align}
Compactly, the action of $\phi$ on an irreducible affine character is thus given by
\be
 \phi\big[\ch_{\rho,\sigma}(q,z)\big]=\ch_{\rho,\sigma}(q),
\ee
with the Virasoro characters satisfying
\be
 \ch_{-\rho,-\sigma}(q)=\ch_{\rho,\sigma}(q),\qquad
 \ch_{\rho,0}(q)=\begin{cases} 0,\ &\rho>0,\\[.15cm] \ch_{p-\rho,p'}(q),\ &\rho\leq0. \end{cases}
\ee

As the notation suggests, the $\phi$-images (\ref{Vir0})-(\ref{Vir4}) are recognised as the irreducible Virasoro 
characters~\cite{DiFSZ87,RP0707} appearing in the logarithmic minimal model 
$\mathcal{LM}(p,p')$~\cite{PRZ0607}. In the comparison with~\cite{RP0707}, in particular, we note that the 
notation employed there is related to ours by $K_{m,n;\ell}(q)=\Theta_{-n,m/2;\ell}(q)/\eta(q)$.
We also note that most of the Virasoro characters appear twice, as
the expressions (\ref{Vir3}) and (\ref{Vir4}) respectively follow from (\ref{Vir1}) and (\ref{Vir2}) by relabelling 
$r_0$ and $s_0$ by $p-r_0$ and $p'-s_0$. On the set of irreducible affine characters, $\phi$ is therefore 
not injective, but it does map surjectively
onto the set of irreducible Virasoro characters appearing in ${\cal LM}(p,p')$.

The $\phi$-images of the staggered modules proposed in Section~\ref{Sec:Conjectures}
are themselves characters of staggered modules. Concretely, we find that
\be
 \phi\big[\chit[\Sc_{\ell p,s_0}^{a,0;-}](q,z)\big]=\chit[R_{\ell p,p'-s_0}^{a,0}](q),
 \qquad
 \phi\big[\chit[\Sc_{\ell p,s_0}^{a,0;+}](q,z)\big]=\begin{cases} \chit[R_{\ell p,s_0}^{a,0}](q),\ &s_0\neq0,\\[.15cm] 
  0,\ &s_0=0,\end{cases}
\label{phiS}
\ee
and
\be
 \phi\big[\chit[\Sc_{r_0,\ell p'}^{0,b;\pm}](q,z)\big]=\chit[R_{r_0,\ell p'}^{0,b}](q),
 \qquad
 \phi\big[\chit[\Sc_{p,\ell p'}^{0,b;\pm}](q,z)\big]=\chit[R_{p,\ell p'}^{0,b}](q),
\label{phiS2}
\ee
where the characters of the Virasoro modules $R_{r,s}^{a,b}$ are given 
in (\ref{Ra0})-(\ref{R0b}).

Reducibility of a character is not necessarily preserved by $\phi$. Indeed, for $r=1,\ldots,p$ and $\ell\in\mathbb{N}$,
the affine Kac module $\Ac_{r,\ell p'}$ is reducible, while $\phi\big[\chit_{r,\ell p'}(q,z)\big]=\chit_{r,\ell p'}(q)$ 
is an irreducible Virasoro character. On the other hand, an irreducible character cannot be mapped to a reducible one.
Likewise, the shape of the staggered modules underlying the character mappings in (\ref{phiS}) may change from 
quadrangular to triangular, but not the other way around. 
Indeed, this happens for the quadrangular staggered module $\Sc_{p,0}^{a,0;-}$.
These observations will be revisited from a `module perspective' in the following.

\subsection{Functor characteristics}
\label{Sec:Gen}

Here, we outline how the map $\phi$ between the sets of $A_1^{(1)}$\! and Virasoro characters 
elevates to a functor,
\be
 \Phi:\ \Cc_{A_1^{(1)}}\to\,\Cc_{\mathrm{Vir}},
\ee
between categories of $A_1^{(1)}$-modules and Virasoro modules.
The affine Lie algebra $A_1^{(1)}$ is at level $k$ with $k+2=p/p'$ as in (\ref{tpp}), while the Virasoro algebra
has central charge $\bar{c}$ as in (\ref{hbar}), parameterised by the same $p,p'$.
For concreteness, we only consider the action of $\Phi$ on irreducible, affine Kac and staggered 
$A_1^{(1)}$-modules. Let $\Mc$ be such a module. As will become clear in the following, 
we then have the following commutative diagram
\newpage
\psset{unit=1cm}
\be
 \begin{pspicture}(0.2,1.4)(3,3.4)
 \rput(0,2.5){$\Mc$}
 \rput(0,0.5){$\bar{\Mc}$}
 \psline[arrowscale=1.5]{->}(0,2.1)(0,0.85)
 \psline[arrowscale=1.5]{->}(3,2.1)(3,0.85)
 \psline[arrowscale=1.5]{->}(0.5,2.5)(2.3,2.5)
 \psline[arrowscale=1.5]{->}(0.5,0.5)(2.3,0.5)
\rput(3,2.5){$\chit(q,z)$}
\rput(3,0.5){$\chit(q)$}
\rput(-0.4,1.5){$\Phi$}
\rput(3.4,1.5){$\phi$}
\rput(1.5,0.1){$\chit$}
\rput(1.5,2.85){$\chit$}
 \end{pspicture}
\\[1.4cm]
\label{comm}
\ee
where it is understood that the (affine) character of the zero module is $0$.

We use Kac-type labels of the form $(r,s)$ to characterise
the weights of subsingular vectors. For such a vector in an $A_1^{(1)}$-module, the labels thus
indicate that the vector is of affine weight $j_{r,s}$. In a Virasoro module, the same labels
indicate that the subsingular vector is of conformal weight $\bar{h}_{r,s}$. 
Because of the symmetries $j_{r+lp,s+lp'}=j_{r,s}$, 
$\bar{h}_{r+lp,s+lp'}=\bar{h}_{r,s}$, $l\in\mathbb{Z}$, and $\bar{h}_{-r,-s}=\bar{h}_{r,s}$, 
we have the equivalences  $(r+lp,s+lp')\equiv(r,s)$ for both types of labels and $(-r,-s)\equiv(r,s)$
for Virasoro labels only.

Now, on the irreducible $A_1^{(1)}$-module $\Ic$ of highest weight $j$, the action of $\Phi$ is simply defined by
\be
 \Phi:\ \Ic\,\mapsto\,\begin{cases} \,\bar{\Ic},\ &j\notin\tfrac{1}{2}\mathbb{N}_0,\\[.15cm] 
  \,0,\ &j\in\tfrac{1}{2}\mathbb{N}_0, \end{cases}
\label{PhiIrr}
\ee
where $\bar{\Ic}$ is the irreducible Virasoro module of highest weight
\be
 \bar{h}=h_j+\frac{2+\bar{c}-c}{24}=\frac{(2j+1)^2-(1-t)^2}{4t}.
\ee
In terms of Loewy diagrams, this corresponds to
\psset{unit=1cm}
\be
 \begin{pspicture}(0,0)(0,0) 
 \pscircle[fillstyle=solid,fillcolor=black,linecolor=black,linewidth=0.01](0.1,0.1){.1} 
 \end{pspicture}
\quad\ \begin{array}{c}\Phi\\[-.2cm] \longmapsto\\[-.2cm] \phantom{\Phi}\end{array}\quad\!\!
 \begin{cases}
 \begin{pspicture}(-0.06,0)(0.25,0) 
 \pscircle[fillstyle=solid,fillcolor=black,linecolor=black,linewidth=0.01](0.1,0.1){.1}
 \end{pspicture},
 \ &j\notin\tfrac{1}{2}\mathbb{N}_0,
 \\[.4cm]
 {\color{red} \times},\ &j\in\tfrac{1}{2}\mathbb{N}_0,
 \end{cases}
\label{PhiIrred}
\ee
where an ${\color{red} \times}$ indicates the {\em absence} of a subsingular vector. In this case, the $\Phi$-image 
for $j\in\tfrac{1}{2}\mathbb{N}_0$ is therefore nothing but the (Virasoro) zero module, 
as already indicated in (\ref{PhiIrr}).

On more general modules, $\Phi$ acts by eliminating certain subquotients. 
This procedure is carried out explicitly for 
affine Kac modules and staggered modules in Sections~\ref{Sec:PhiKac} and~\ref{Sec:PhiStag}.
Although these $A_1^{(1)}$-modules are all indecomposable, their Virasoro images need not be. 
The affine Kac modules also have the key property that the quasi-integrable subquotients 
(the irreducible subquotients with only finite-dimensional 
$L_0$-eigenspaces, see (\ref{sumQ})) form a submodule or can be used to form a quotient of the affine 
module\footnote{We recall that every such subquotient is an irreducible highest-weight module of affine weight $j_{r,0}$ for some 
$r\in\mathbb{N}$.}.
Combined with the strong similarities between the $A_1^{(1)}$\! and Virasoro
representation theories, this allows us to define $\Phi$ by detailing its action on 
the corresponding Loewy diagrams: In a diagram associated with a given $A_1^{(1)}$-module,
the functor removes all the quasi-integrable subquotients, and 
reinterprets the Kac-type labels of any remaining subsingular vectors as labels of subsingular vectors in the Loewy 
diagram of a module over the Virasoro algebra with central charge $\bar{c}$.
As a consequence, $\Phi$ is an {\em exact functor} (meaning that it transforms short exact sequences into short exact sequences).

\subsection{Kac modules}
\label{Sec:PhiKac}

The Virasoro Kac module $\Kc_{r,s}$ is defined for $r,s\in\mathbb{N}$ in much the same 
way~\cite{Ras1012,MDRR1503} as the affine Kac modules in Section~\ref{Sec:Fin}. It is thus the submodule 
of the Feigin-Fuchs module of weight $\bar{h}_{r,s}$, generated by the subsingular vectors of grade strictly less 
than $rs$. As discussed in the following, these Virasoro
modules are the (nonzero) images of $\Phi$ acting on affine Kac modules.

For $s\notin p'\mathbb{Z}$, the affine Kac module $\Ac_{r,s}$ does not contain any quasi-integrable subquotients. 
The functor $\Phi$ therefore preserves the structure of the Loewy diagram, 
that being of braided type (\ref{diagramAv}) for $r\notin p\mathbb{Z}$ and of chain type (\ref{diagramAs}) for 
$r\in p\mathbb{Z}^\times$. Indeed, the ensuing Virasoro module is the Virasoro Kac module $\Kc_{|r|,|s|}$.

For $s\in p'\mathbb{Z}$, on the other hand, the affine Kac module $\Ac_{r,s}$ {\em does}, in general, 
contain quasi-integrable subquotients. In these cases, the functor $\Phi$ reduces 
the diagram to one corresponding to a quotient (for $s\geq0$) or submodule (for $s<0$) of the original module. 
To appreciate this and to identify the ensuing Virasoro module, we write $s=\ell'p'$, with $\ell'\in\mathbb{Z}$, and
divide the affine Kac modules into classes.

First, let $r\notin p\mathbb{Z}$, in which case $r=\rh_0+\ell p$, as before. 
For $s\geq0$, we then have
\psset{unit=0.58cm}
\setlength{\unitlength}{0.58cm} 
\be
\begin{pspicture}(-1.1,0.6)(9,2)
\pscircle[fillstyle=solid,fillcolor=lightgray,linecolor=black,linewidth=0.01](-1.1,0.73){.12}
\pscircle[fillstyle=solid,fillcolor=white,linecolor=black,linewidth=0.01](0,1.5){.12}
 \rput(0.8,1.48){$\rightarrow$}
\pscircle[fillstyle=solid,fillcolor=lightgray,linecolor=black,linewidth=0.01](1.6,1.5){.12}
 \rput(2.4,1.48){$\leftarrow$}
 \rput(4.8,1.48){$\leftarrow$}
\pscircle[fillstyle=solid,fillcolor=white,linecolor=black,linewidth=0.01](5.6,1.5){.12}
 \rput(6.4,1.48){$\rightarrow$}
\pscircle[fillstyle=solid,fillcolor=lightgray,linecolor=black,linewidth=0.01](7.2,1.5){.12}
 \rput(3.6,0){$\ldots$}
 \rput(3.6,1.5){$\ldots$}
\pscircle[fillstyle=solid,fillcolor=black,linecolor=black,linewidth=0.01](0,0){.12}
 \rput(0.8,-0.02){$\leftarrow$}
\pscircle[fillstyle=solid,fillcolor=lightgray,linecolor=black,linewidth=0.01](1.6,0){.12}
 \rput(2.4,-0.02){$\rightarrow$}
 \rput(4.8,-0.02){$\rightarrow$}
\pscircle[fillstyle=solid,fillcolor=black,linecolor=black,linewidth=0.01](5.6,0){.12}
 \rput(6.4,-0.02){$\leftarrow$}
\pscircle[fillstyle=solid,fillcolor=lightgray,linecolor=black,linewidth=0.01](7.2,0){.12}
 \rput(8,-0.02){$\rightarrow$}
\pscircle[fillstyle=solid,fillcolor=black,linecolor=black,linewidth=0.01](8.8,0){.12}
 \rput(-0.6,1.22){$\swarrow$}
 \rput(-0.6,0.28){$\searrow$}
 \rput(0.8,0.73){$\searrow$}
 \rput(0.8,0.73){$\swarrow$}
 \rput(2.4,0.73){$\searrow$}
 \rput(2.4,0.73){$\swarrow$}
 \rput(4.8,0.73){$\searrow$}
 \rput(4.8,0.73){$\swarrow$}
 \rput(6.4,0.73){$\searrow$}
 \rput(6.4,0.73){$\swarrow$}
 \rput(8,0.73){$\searrow$}
\end{pspicture}
\ \ \begin{array}{c}\Phi\\[-.2cm] \longmapsto\\[-.2cm] \phantom{\Phi}\end{array}\
\begin{cases}
\begin{pspicture}(-1.3,0)(12,1.8)
\pscircle[fillstyle=solid,fillcolor=black,linecolor=black,linewidth=0.01](-1.1,0.73){.12}
\pscircle[fillstyle=solid,fillcolor=lightgray,linecolor=black,linewidth=0.01](0,1.5){.12}
 \rput(0.8,1.48){$\rightarrow$}
\pscircle[fillstyle=solid,fillcolor=black,linecolor=black,linewidth=0.01](1.6,1.5){.12}
 \rput(2.4,1.48){$\leftarrow$}
 \rput(4.8,1.48){$\leftarrow$}
\pscircle[fillstyle=solid,fillcolor=lightgray,linecolor=black,linewidth=0.01](5.6,1.5){.12}
 \rput(6.4,1.48){$\rightarrow$}
\pscircle[fillstyle=solid,fillcolor=black,linecolor=black,linewidth=0.01](7.2,1.5){.12}
 \rput(3.6,0){${\color{red} \ldots}$}
 \rput(3.6,1.5){$\ldots$}
\rput(0,0){${\color{red} \times}$}
\rput(1.6,0){${\color{red} \times}$}
\rput(5.6,0){${\color{red} \times}$}
\rput(7.2,0){${\color{red} \times}$}
\rput(8.8,0){${\color{red} \times}$}
 \rput(-0.6,1.22){$\swarrow$}
\rput(11.5,0.75){$(\ell'>\ell\geq0)$}
\end{pspicture}
\\[.4cm]
\begin{pspicture}(-1.3,0)(12,2)
\pscircle[fillstyle=solid,fillcolor=lightgray,linecolor=black,linewidth=0.01](0,1.5){.12}
 \rput(0.8,1.48){$\rightarrow$}
\pscircle[fillstyle=solid,fillcolor=black,linecolor=black,linewidth=0.01](1.6,1.5){.12}
 \rput(2.4,1.48){$\leftarrow$}
 \rput(4.8,1.48){$\leftarrow$}
\pscircle[fillstyle=solid,fillcolor=lightgray,linecolor=black,linewidth=0.01](5.6,1.5){.12}
 \rput(6.4,1.48){$\rightarrow$}
\pscircle[fillstyle=solid,fillcolor=black,linecolor=black,linewidth=0.01](7.2,1.5){.12}
 \rput(3.6,0){${\color{red} \ldots}$}
 \rput(3.6,1.5){$\ldots$}
\rput(-1.1,0.73){${\color{red} \times}$}
\rput(0,0){${\color{red} \times}$}
\rput(1.6,0){${\color{red} \times}$}
\rput(5.6,0){${\color{red} \times}$}
\rput(7.2,0){${\color{red} \times}$}
\rput(8.8,0){${\color{red} \times}$}
\rput(11.5,0.75){$(\ell\geq\ell'\geq0)$}
\end{pspicture}
\end{cases}
\vspace{0.1cm}
\label{Phi1}
\ee
whereas for $s<0$, we have
\psset{unit=0.58cm}
\setlength{\unitlength}{0.58cm} 
\be
\begin{pspicture}(-1.1,0.6)(9,2)
\pscircle[fillstyle=solid,fillcolor=lightgray,linecolor=black,linewidth=0.01](-1.1,0.73){.12}
\pscircle[fillstyle=solid,fillcolor=white,linecolor=black,linewidth=0.01](0,1.5){.12}
 \rput(0.8,1.48){$\rightarrow$}
\pscircle[fillstyle=solid,fillcolor=lightgray,linecolor=black,linewidth=0.01](1.6,1.5){.12}
 \rput(2.4,1.48){$\leftarrow$}
 \rput(4.8,1.48){$\leftarrow$}
\pscircle[fillstyle=solid,fillcolor=white,linecolor=black,linewidth=0.01](5.6,1.5){.12}
 \rput(6.4,1.48){$\rightarrow$}
\pscircle[fillstyle=solid,fillcolor=lightgray,linecolor=black,linewidth=0.01](7.2,1.5){.12}
 \rput(3.6,0){$\ldots$}
 \rput(3.6,1.5){$\ldots$}
\pscircle[fillstyle=solid,fillcolor=black,linecolor=black,linewidth=0.01](0,0){.12}
 \rput(0.8,-0.02){$\leftarrow$}
\pscircle[fillstyle=solid,fillcolor=lightgray,linecolor=black,linewidth=0.01](1.6,0){.12}
 \rput(2.4,-0.02){$\rightarrow$}
 \rput(4.8,-0.02){$\rightarrow$}
\pscircle[fillstyle=solid,fillcolor=black,linecolor=black,linewidth=0.01](5.6,0){.12}
 \rput(6.4,-0.02){$\leftarrow$}
\pscircle[fillstyle=solid,fillcolor=lightgray,linecolor=black,linewidth=0.01](7.2,0){.12}
 \rput(8,-0.02){$\rightarrow$}
\pscircle[fillstyle=solid,fillcolor=black,linecolor=black,linewidth=0.01](8.8,0){.12}
 \rput(-0.6,1.22){$\swarrow$}
 \rput(-0.6,0.28){$\searrow$}
 \rput(0.8,0.73){$\searrow$}
 \rput(0.8,0.73){$\swarrow$}
 \rput(2.4,0.73){$\searrow$}
 \rput(2.4,0.73){$\swarrow$}
 \rput(4.8,0.73){$\searrow$}
 \rput(4.8,0.73){$\swarrow$}
 \rput(6.4,0.73){$\searrow$}
 \rput(6.4,0.73){$\swarrow$}
 \rput(8,0.73){$\searrow$}
\end{pspicture}
\ \ \begin{array}{c}\Phi\\[-.2cm] \longmapsto\\[-.2cm] \phantom{\Phi}\end{array}\
\begin{cases}
\begin{pspicture}(-1.3,0)(12,1.8)
\pscircle[fillstyle=solid,fillcolor=lightgray,linecolor=black,linewidth=0.01](-1.1,0.73){.12}
 \rput(3.6,1.5){${\color{red} \ldots}$}
 \rput(3.6,0){$\ldots$}
\rput(0,1.5){${\color{red} \times}$}
\rput(1.6,1.5){${\color{red} \times}$}
\rput(5.6,1.5){${\color{red} \times}$}
\rput(7.2,1.5){${\color{red} \times}$}
\pscircle[fillstyle=solid,fillcolor=black,linecolor=black,linewidth=0.01](0,0){.12}
 \rput(0.8,-0.02){$\leftarrow$}
\pscircle[fillstyle=solid,fillcolor=lightgray,linecolor=black,linewidth=0.01](1.6,0){.12}
 \rput(2.4,-0.02){$\rightarrow$}
 \rput(4.8,-0.02){$\rightarrow$}
\pscircle[fillstyle=solid,fillcolor=black,linecolor=black,linewidth=0.01](5.6,0){.12}
 \rput(6.4,-0.02){$\leftarrow$}
\pscircle[fillstyle=solid,fillcolor=lightgray,linecolor=black,linewidth=0.01](7.2,0){.12}
 \rput(8,-0.02){$\rightarrow$}
\pscircle[fillstyle=solid,fillcolor=black,linecolor=black,linewidth=0.01](8.8,0){.12}
 \rput(-0.6,0.28){$\searrow$}
\rput(11.5,0.75){$(\ell<\ell'<0)$}
\end{pspicture}
\\[.4cm]
\begin{pspicture}(-1.3,0)(12,2)
 \rput(3.6,1.5){${\color{red} \ldots}$}
 \rput(3.6,0){$\ldots$}
\rput(-1.1,0.73){${\color{red} \times}$}
\rput(0,1.5){${\color{red} \times}$}
\rput(1.6,1.5){${\color{red} \times}$}
\rput(5.6,1.5){${\color{red} \times}$}
\rput(7.2,1.5){${\color{red} \times}$}
\pscircle[fillstyle=solid,fillcolor=black,linecolor=black,linewidth=0.01](0,0){.12}
 \rput(0.8,-0.02){$\leftarrow$}
\pscircle[fillstyle=solid,fillcolor=lightgray,linecolor=black,linewidth=0.01](1.6,0){.12}
 \rput(2.4,-0.02){$\rightarrow$}
 \rput(4.8,-0.02){$\rightarrow$}
\pscircle[fillstyle=solid,fillcolor=black,linecolor=black,linewidth=0.01](5.6,0){.12}
 \rput(6.4,-0.02){$\leftarrow$}
\pscircle[fillstyle=solid,fillcolor=lightgray,linecolor=black,linewidth=0.01](7.2,0){.12}
 \rput(8,-0.02){$\rightarrow$}
\pscircle[fillstyle=solid,fillcolor=black,linecolor=black,linewidth=0.01](8.8,0){.12}
\rput(11.5,0.75){$(\ell'\leq\ell<0)$}
\end{pspicture}
\end{cases}
\vspace{0.1cm}
\ee
Note that the shadings of the subsingular vectors have changed in the images in (\ref{Phi1}) to reflect the changed nature 
(socle depth) of the vectors.
In diagrams where the leftmost node is absent in the image, the labels of the leftmost present node 
(corresponding to the lowest $L_0$-eigenvalue in the ensuing Virasoro module) are $(-r,-s)$.
However, the Virasoro equivalence $(-r,-s)\equiv(r,s)$ ensures that the
weight of this node is, after all, $\bar{h}_{r,s}$. In fact, the image for $s\neq0$ is the Loewy diagram
of the Virasoro Kac module $\Kc_{|r|,|s|}$. For $s=0$, we have $m=0$ in (\ref{diagramAv}) and a 
vanishing $\Phi$-image corresponding to the (Virasoro) zero module (here represented by the two 
leftmost absences in the lower case of (\ref{Phi1})).

We now let $r\in p\mathbb{Z}^\times$, in which case $r=\ell p$ with $\ell\neq0$.
For $s\geq0$, we then have
\psset{unit=0.58cm}
\setlength{\unitlength}{0.58cm} 
\be
\begin{array}{rl}
\begin{pspicture}(0,-0.15)(9,0.4)
\pscircle[fillstyle=solid,fillcolor=lightgray,linecolor=black,linewidth=0.01](0,0){.12}
 \rput(0.8,-0.02){$\rightarrow$}
\pscircle[fillstyle=solid,fillcolor=black,linecolor=black,linewidth=0.01](1.6,0){.12}
 \rput(2.4,-0.02){$\leftarrow$}
\pscircle[fillstyle=solid,fillcolor=lightgray,linecolor=black,linewidth=0.01](3.2,0){.12}
 \rput(4,-0.02){$\rightarrow$}
 \rput(5.2,0){$\ldots$}
 \rput(6.4,-0.02){$\leftarrow$}
\pscircle[fillstyle=solid,fillcolor=lightgray,linecolor=black,linewidth=0.01](7.2,0){.12}
 \rput(8,-0.02){$\rightarrow$}
\pscircle[fillstyle=solid,fillcolor=black,linecolor=black,linewidth=0.01](8.8,0){.12}
\end{pspicture}
\quad \begin{array}{c}\Phi\\[-.2cm] \longmapsto\\[-.2cm] \phantom{\Phi}\end{array}\quad\
\begin{pspicture}(0,-0.15)(9,0.4)
\pscircle[fillstyle=solid,fillcolor=black,linecolor=black,linewidth=0.01](0,0){.12}
\pscircle[fillstyle=solid,fillcolor=black,linecolor=black,linewidth=0.01](3.2,0){.12}
 \rput(5.2,0){$\ldots$}
\pscircle[fillstyle=solid,fillcolor=black,linecolor=black,linewidth=0.01](7.2,0){.12}
\rput(1.6,0){${\color{red} \times}$}
\rput(8.8,0){${\color{red} \times}$}
\end{pspicture}
\qquad &(\ell'\geq\ell>0)
\\[.5cm]
\begin{pspicture}(0,-0.15)(9,0.4)
\pscircle[fillstyle=solid,fillcolor=black,linecolor=black,linewidth=0.01](0,0){.12}
 \rput(0.8,-0.02){$\leftarrow$}
\pscircle[fillstyle=solid,fillcolor=lightgray,linecolor=black,linewidth=0.01](1.6,0){.12}
 \rput(2.4,-0.02){$\rightarrow$}
\pscircle[fillstyle=solid,fillcolor=black,linecolor=black,linewidth=0.01](3.2,0){.12}
 \rput(4,-0.02){$\leftarrow$}
 \rput(5.2,0){$\ldots$}
 \rput(6.4,-0.02){$\leftarrow$}
\pscircle[fillstyle=solid,fillcolor=lightgray,linecolor=black,linewidth=0.01](7.2,0){.12}
 \rput(8,-0.02){$\rightarrow$}
\pscircle[fillstyle=solid,fillcolor=black,linecolor=black,linewidth=0.01](8.8,0){.12}
\end{pspicture}
\quad \begin{array}{c}\Phi\\[-.2cm] \longmapsto\\[-.2cm] \phantom{\Phi}\end{array}\quad\
\begin{pspicture}(0,-0.15)(9,0.4)
\pscircle[fillstyle=solid,fillcolor=black,linecolor=black,linewidth=0.01](1.6,0){.12}
 \rput(5.2,0){$\ldots$}
\pscircle[fillstyle=solid,fillcolor=black,linecolor=black,linewidth=0.01](7.2,0){.12}
\rput(0,0){${\color{red} \times}$}
\rput(3.2,0){${\color{red} \times}$}
\rput(8.8,0){${\color{red} \times}$}
\end{pspicture}
\qquad &(\ell>\ell'\geq0)
\end{array}
\\[-.3cm]
\label{Phi3}
\ee
whereas for $s<0$, we have
\psset{unit=0.58cm}
\setlength{\unitlength}{0.58cm} 
\be
\begin{array}{rl}
\begin{pspicture}(0,-0.15)(9,0.4)
\pscircle[fillstyle=solid,fillcolor=lightgray,linecolor=black,linewidth=0.01](0,0){.12}
 \rput(0.8,-0.02){$\rightarrow$}
\pscircle[fillstyle=solid,fillcolor=black,linecolor=black,linewidth=0.01](1.6,0){.12}
 \rput(2.4,-0.02){$\leftarrow$}
\pscircle[fillstyle=solid,fillcolor=lightgray,linecolor=black,linewidth=0.01](3.2,0){.12}
 \rput(4,-0.02){$\rightarrow$}
 \rput(5.2,0){$\ldots$}
 \rput(6.4,-0.02){$\leftarrow$}
\pscircle[fillstyle=solid,fillcolor=lightgray,linecolor=black,linewidth=0.01](7.2,0){.12}
 \rput(8,-0.02){$\rightarrow$}
\pscircle[fillstyle=solid,fillcolor=black,linecolor=black,linewidth=0.01](8.8,0){.12}
\end{pspicture}
\quad \begin{array}{c}\Phi\\[-.2cm] \longmapsto\\[-.2cm] \phantom{\Phi}\end{array}\quad\
\begin{pspicture}(0,-0.15)(9,0.4)
\pscircle[fillstyle=solid,fillcolor=black,linecolor=black,linewidth=0.01](1.6,0){.12}
 \rput(5.2,0){$\ldots$}
\pscircle[fillstyle=solid,fillcolor=black,linecolor=black,linewidth=0.01](8.8,0){.12}
\rput(0,0){${\color{red} \times}$}
\rput(3.2,0){${\color{red} \times}$}
\rput(7.2,0){${\color{red} \times}$}
\end{pspicture}
\qquad &(\ell'<\ell<0)
\\[.5cm]
\begin{pspicture}(0,-0.15)(9,0.4)
\pscircle[fillstyle=solid,fillcolor=black,linecolor=black,linewidth=0.01](0,0){.12}
 \rput(0.8,-0.02){$\leftarrow$}
\pscircle[fillstyle=solid,fillcolor=lightgray,linecolor=black,linewidth=0.01](1.6,0){.12}
 \rput(2.4,-0.02){$\rightarrow$}
\pscircle[fillstyle=solid,fillcolor=black,linecolor=black,linewidth=0.01](3.2,0){.12}
 \rput(4,-0.02){$\leftarrow$}
 \rput(5.2,0){$\ldots$}
 \rput(6.4,-0.02){$\leftarrow$}
\pscircle[fillstyle=solid,fillcolor=lightgray,linecolor=black,linewidth=0.01](7.2,0){.12}
 \rput(8,-0.02){$\rightarrow$}
\pscircle[fillstyle=solid,fillcolor=black,linecolor=black,linewidth=0.01](8.8,0){.12}
\end{pspicture}
\quad \begin{array}{c}\Phi\\[-.2cm] \longmapsto\\[-.2cm] \phantom{\Phi}\end{array}\quad\
\begin{pspicture}(0,-0.15)(9,0.4)
\pscircle[fillstyle=solid,fillcolor=black,linecolor=black,linewidth=0.01](0,0){.12}
\pscircle[fillstyle=solid,fillcolor=black,linecolor=black,linewidth=0.01](3.2,0){.12}
 \rput(5.2,0){$\ldots$}
\pscircle[fillstyle=solid,fillcolor=black,linecolor=black,linewidth=0.01](8.8,0){.12}
\rput(1.6,0){${\color{red} \times}$}
\rput(7.2,0){${\color{red} \times}$}
\end{pspicture}
\qquad &(\ell\leq\ell'<0)
\end{array}
\\[-.3cm]
\ee
The ensuing Loewy diagrams are of the so-called {\em islands type} indicating completely reducible 
Virasoro modules. Nevertheless, for $s\neq0$, the image of the affine Kac module $\Ac_{r,s}$ is the 
Virasoro Kac module $\Kc_{|r|,|s|}$. Indeed, it is well known~\cite{RP0707,MDRR1503} that $\Kc_{r,s}$ 
is completely reducible if $r\in p\mathbb{N}$ and $s\in p'\mathbb{N}$.
The islands diagrams thus arise when applying $\Phi$ to $\Ac_{r,s}$ 
where $r\in p\mathbb{Z}^\times$ and $s\in p'\mathbb{Z}^\times$. For $s=0$, the affine Kac module is irreducible.
Following from (\ref{PhiIrred}), its $\Phi$-image is the zero
module (represented by the leftmost absence in the lower case of (\ref{Phi3})).

In summary, we have
\be
 \Phi:\ \Ac_{r,s}\ \mapsto\ \begin{cases} \Kc_{|r|,|s|},\ &rs>0,\\[.2cm] 0,\ &s=0.\end{cases}
\ee

\subsection{Staggered modules}
\label{Sec:PhiStag}

\psset{unit=.6cm}
\setlength{\unitlength}{.6cm}
On the staggered modules in Conjecture 1 in Section~\ref{Sec:Conjectures}, $\Phi$ acts as
\be
 \hspace{-1cm}
 \Phi:\ \Sc_{\ell p,s_0}^{a,0;+}\ \mapsto\ \begin{cases} R_{\ell p,s_0}^{a,0},\ &s_0\neq0,\\[.15cm] 0,\ &s_0=0,\end{cases} 
\qquad\quad
\begin{pspicture}(0,1)(1.3,2.3)
 \pscircle[fillstyle=solid,fillcolor=lightgray,linecolor=black,linewidth=0.01](0,2.2){.1} 
 \pscircle[fillstyle=solid,fillcolor=black,linecolor=black,linewidth=0.01](0,1.1){.1} 
 \pscircle[fillstyle=solid,fillcolor=white,linecolor=black,linewidth=0.01](1.2,1.1){.1} 
 \pscircle[fillstyle=solid,fillcolor=lightgray,linecolor=black,linewidth=0.01](1.2,0){.1} 
 \rput(0.6,0.55){$\nwarrow$}
 \rput(0.6,1.65){$\nwarrow$}
 \rput(0,1.65){$\downarrow$}
 \rput(1.2,0.55){$\downarrow$}
 \rput(0.6,1.1){$\leftarrow$}
\end{pspicture}
\quad \begin{array}{c}\Phi\\[-.2cm] \longmapsto\\[-.2cm] \phantom{\Phi}\end{array}\quad\
\begin{cases}
\ \
\begin{pspicture}(0,1.3)(3.8,2.5)
 \pscircle[fillstyle=solid,fillcolor=lightgray,linecolor=black,linewidth=0.01](0,2.2){.1} 
 \pscircle[fillstyle=solid,fillcolor=black,linecolor=black,linewidth=0.01](0,1.1){.1} 
 \pscircle[fillstyle=solid,fillcolor=white,linecolor=black,linewidth=0.01](1.2,1.1){.1} 
 \pscircle[fillstyle=solid,fillcolor=lightgray,linecolor=black,linewidth=0.01](1.2,0){.1} 
 \rput(0.6,0.55){$\nwarrow$}
 \rput(0.6,1.65){$\nwarrow$}
 \rput(0,1.65){$\downarrow$}
 \rput(1.2,0.55){$\downarrow$}
 \rput(0.6,1.1){$\leftarrow$}
 \rput(3.8,1.16){$(s_0\neq0)$}
\end{pspicture}
\\
\\[0.5cm]
\ \
\begin{pspicture}(0,0)(3.8,2.3)
 \rput(0,2.2){${\color{red} \times}$}
 \rput(0,1.1){${\color{red} \times}$}
 \rput(1.2,1.1){${\color{red} \times}$}
 \rput(1.2,0){${\color{red} \times}$}
 \rput(3.8,1.16){$(s_0=0)$}
\end{pspicture}
\end{cases}
%\\[.2cm]
\ee
and
\be
 \hspace{-2cm}
 \Phi:\ \Sc_{\ell p,s_0}^{a,0;-}\ \mapsto\ R_{\ell p,p'-s_0}^{a,0},
\qquad\quad
\begin{pspicture}(0,1)(1.3,2.3)
 \pscircle[fillstyle=solid,fillcolor=lightgray,linecolor=black,linewidth=0.01](0,2.2){.1} 
 \pscircle[fillstyle=solid,fillcolor=black,linecolor=black,linewidth=0.01](0,1.1){.1} 
 \pscircle[fillstyle=solid,fillcolor=white,linecolor=black,linewidth=0.01](1.2,1.1){.1} 
 \pscircle[fillstyle=solid,fillcolor=lightgray,linecolor=black,linewidth=0.01](1.2,0){.1} 
 \rput(0.6,0.55){$\nwarrow$}
 \rput(0.6,1.65){$\nwarrow$}
 \rput(0,1.65){$\downarrow$}
 \rput(1.2,0.55){$\downarrow$}
 \rput(0.6,1.1){$\leftarrow$}
\end{pspicture}
\quad \begin{array}{c}\Phi\\[-.2cm] \longmapsto\\[-.2cm] \phantom{\Phi}\end{array}\quad\
\begin{cases}
\ \
\begin{pspicture}(0,1.3)(3.8,2.5)
 \pscircle[fillstyle=solid,fillcolor=lightgray,linecolor=black,linewidth=0.01](0,2.2){.1} 
 \pscircle[fillstyle=solid,fillcolor=black,linecolor=black,linewidth=0.01](0,1.1){.1} 
 \pscircle[fillstyle=solid,fillcolor=white,linecolor=black,linewidth=0.01](1.2,1.1){.1} 
 \pscircle[fillstyle=solid,fillcolor=lightgray,linecolor=black,linewidth=0.01](1.2,0){.1} 
 \rput(0.6,0.55){$\nwarrow$}
 \rput(0.6,1.65){$\nwarrow$}
 \rput(0,1.65){$\downarrow$}
 \rput(1.2,0.55){$\downarrow$}
 \rput(0.6,1.1){$\leftarrow$}
 \rput(4.6,1.16){$((s_0,\ell)\neq(0,1))$}
\end{pspicture}
\\
\\[0.5cm]
\ \
\begin{pspicture}(0,0)(3.8,2.3)
 \pscircle[fillstyle=solid,fillcolor=black,linecolor=black,linewidth=0.01](0,1.1){.1} 
 \pscircle[fillstyle=solid,fillcolor=white,linecolor=black,linewidth=0.01](1.2,1.1){.1} 
 \pscircle[fillstyle=solid,fillcolor=lightgray,linecolor=black,linewidth=0.01](1.2,0){.1} 
 \rput(0,2.2){${\color{red} \times}$}
 \rput(0.6,0.55){$\nwarrow$}
 \rput(1.2,0.55){$\downarrow$}
 \rput(0.6,1.1){$\leftarrow$}
 \rput(4.6,1.16){$((s_0,\ell)=(0,1))$}
\end{pspicture}
\end{cases}
\\[.1cm]
\ee
On the staggered modules in Conjecture 2, $\Phi$ acts as
\be
 \hspace{-1cm}
 \Phi:\ \Sc_{r_0,\ell p'}^{0,b;\pm}\ \mapsto\ R_{r_0,\ell p'}^{0,b},
\qquad\quad
\begin{pspicture}(0,1)(1.3,2.3)
 \pscircle[fillstyle=solid,fillcolor=lightgray,linecolor=black,linewidth=0.01](0,2.2){.1} 
 \pscircle[fillstyle=solid,fillcolor=black,linecolor=black,linewidth=0.01](0,1.1){.1} 
 \pscircle[fillstyle=solid,fillcolor=white,linecolor=black,linewidth=0.01](1.2,1.1){.1} 
 \pscircle[fillstyle=solid,fillcolor=lightgray,linecolor=black,linewidth=0.01](1.2,0){.1} 
 \rput(0.6,0.55){$\nwarrow$}
 \rput(0.6,1.65){$\nwarrow$}
 \rput(0,1.65){$\downarrow$}
 \rput(1.2,0.55){$\downarrow$}
 \rput(0.6,1.1){$\leftarrow$}
\end{pspicture}
\quad \begin{array}{c}\Phi\\[-.2cm] \longmapsto\\[-.2cm] \phantom{\Phi}\end{array}\quad\
\begin{pspicture}(0,1)(1.3,2.3)
 \pscircle[fillstyle=solid,fillcolor=lightgray,linecolor=black,linewidth=0.01](0,2.2){.1} 
 \pscircle[fillstyle=solid,fillcolor=black,linecolor=black,linewidth=0.01](0,1.1){.1} 
 \pscircle[fillstyle=solid,fillcolor=white,linecolor=black,linewidth=0.01](1.2,1.1){.1} 
 \pscircle[fillstyle=solid,fillcolor=lightgray,linecolor=black,linewidth=0.01](1.2,0){.1} 
 \rput(0.6,0.55){$\nwarrow$}
 \rput(0.6,1.65){$\nwarrow$}
 \rput(0,1.65){$\downarrow$}
 \rput(1.2,0.55){$\downarrow$}
 \rput(0.6,1.1){$\leftarrow$}
\end{pspicture}
\\[.2cm]
\ee
On the staggered modules in Conjecture 3, $\Phi$ acts as
\be
 \hspace{-1cm}
 \Phi:\ \Sc_{p,\ell p'}^{0,b;\pm}\ \mapsto\ R_{p,\ell p'}^{0,b},
\qquad
\begin{cases}
\ \
\begin{pspicture}(0,1)(1.3,2)
 \pscircle[fillstyle=solid,fillcolor=black,linecolor=black,linewidth=0.01](0,1.1){.1} 
 \pscircle[fillstyle=solid,fillcolor=white,linecolor=black,linewidth=0.01](1.2,1.1){.1} 
 \pscircle[fillstyle=solid,fillcolor=lightgray,linecolor=black,linewidth=0.01](1.2,0){.1} 
 \rput(0.6,0.55){$\nwarrow$}
 \rput(1.2,0.55){$\downarrow$}
 \rput(0.6,1.1){$\leftarrow$}
\end{pspicture}
\quad \begin{array}{c}\Phi\\[-.2cm] \longmapsto\\[-.2cm] \phantom{\Phi}\end{array}\quad\
\begin{pspicture}(0,1)(2.8,2)
 \pscircle[fillstyle=solid,fillcolor=black,linecolor=black,linewidth=0.01](0,1.1){.1} 
 \pscircle[fillstyle=solid,fillcolor=white,linecolor=black,linewidth=0.01](1.2,1.1){.1} 
 \pscircle[fillstyle=solid,fillcolor=lightgray,linecolor=black,linewidth=0.01](1.2,0){.1} 
 \rput(0.6,0.55){$\nwarrow$}
 \rput(1.2,0.55){$\downarrow$}
 \rput(0.6,1.1){$\leftarrow$}
 \rput(3.8,1.16){$(\ell=1)$}
\end{pspicture}
\\
\\[0.1cm]
\ \
\begin{pspicture}(0,1)(1.3,2)
 \pscircle[fillstyle=solid,fillcolor=lightgray,linecolor=black,linewidth=0.01](0,2.2){.1} 
 \pscircle[fillstyle=solid,fillcolor=black,linecolor=black,linewidth=0.01](0,1.1){.1} 
 \pscircle[fillstyle=solid,fillcolor=white,linecolor=black,linewidth=0.01](1.2,1.1){.1} 
 \pscircle[fillstyle=solid,fillcolor=lightgray,linecolor=black,linewidth=0.01](1.2,0){.1} 
 \rput(0.6,0.55){$\nwarrow$}
 \rput(0.6,1.65){$\nwarrow$}
 \rput(0,1.65){$\downarrow$}
 \rput(1.2,0.55){$\downarrow$}
 \rput(0.6,1.1){$\leftarrow$}
\end{pspicture}
\quad \begin{array}{c}\Phi\\[-.2cm] \longmapsto\\[-.2cm] \phantom{\Phi}\end{array}\quad\
\begin{pspicture}(0,1)(3.8,2)
 \pscircle[fillstyle=solid,fillcolor=lightgray,linecolor=black,linewidth=0.01](0,2.2){.1} 
 \pscircle[fillstyle=solid,fillcolor=black,linecolor=black,linewidth=0.01](0,1.1){.1} 
 \pscircle[fillstyle=solid,fillcolor=white,linecolor=black,linewidth=0.01](1.2,1.1){.1} 
 \pscircle[fillstyle=solid,fillcolor=lightgray,linecolor=black,linewidth=0.01](1.2,0){.1} 
 \rput(0.6,0.55){$\nwarrow$}
 \rput(0.6,1.65){$\nwarrow$}
 \rput(0,1.65){$\downarrow$}
 \rput(1.2,0.55){$\downarrow$}
 \rput(0.6,1.1){$\leftarrow$}
 \rput(3.8,1.16){$(\ell>1)$}
\end{pspicture}
\end{cases}
\\[.1cm]
\ee
Combined with (\ref{phiS}) and (\ref{phiS2}), these results confirm the commutativity of the diagram (\ref{comm}).
The exactness of $\Phi$ is also readily confirmed.

\section{Conclusion}
\label{Sec:Concl}

We have revisited the representation theory of the affine Lie algebra $A_1^{(1)}$\! and introduced two new classes of $A_1^{(1)}$-modules:
affine Kac modules and certain staggered modules, although the existence of the latter class has in general only been conjectured. 
We have derived the corresponding branching functions following from the Goddard-Kent-Olive construction, 
and introduced a functor mapping the new modules to similar Virasoro modules.

It would be of great interest to extend our results to other affine Lie algebras, primarily to the affine $s\ell(3)$ algebra $A_2^{(1)}$\!.
This would yield new insight, not only into the somewhat poorly understood representation theory of 
higher-rank affine Lie algebras, but also into the representation theory of the $\mathcal{W}_3$ algebra \cite{Zam85,BMPbook}.
The recent advances \cite{Ras1803} in the representation theory of finite-dimensional simple Lie algebras
are likely to play a role.
We are also interested in developing a systematic approach to logarithmic $A_1^{(1)}$\! models. This includes 
using the Nahm-Gaberdiel-Kausch algorithm \cite{Nahm94,GK96} to compute fusion products of 
affine Kac modules, in particular, noting that such fusion products differ from the ones considered previously 
in the literature \cite{Gab01,Rid11}.
We hope to address these problems elsewhere.

\section*{Acknowledgments}
\vskip.1cm
\noindent
This work is supported by the Australian Research Council under the Discovery Project scheme, 
project number DP160101376. The author thanks Denis Bernard, Kenji Iohara, 
Masoud Kamgarpour, Paul Pearce, Christopher Raymond, David Ridout, Ole Warnaar, and Simon Wood
for helpful discussions and comments.

\appendix

\section{Generalised theta functions}
\label{Sec:GenTheta}

The Jacobi-Riemann theta functions, $\Theta_{n,m}(q,z)$, are defined by
\be
 \Theta_{n,m}(q,z):=\sum_{l\in\mathbb{Z}+\frac{n}{2m}}q^{ml^2}z^{-ml},\qquad 
  n\in\mathbb{Z},\ m\in\mathbb{Z}^\times,
\ee
and satisfy
\be
 \Theta_{n,m}(q,z)=\Theta_{2im\pm n,m}(q,z^{\pm1}),
   \qquad i\in\mathbb{Z}.
\label{thetasym}
\ee
We find it convenient to introduce, for every $n,\nu\in\mathbb{Z}$ and $m\in\mathbb{Z}^\times$, the {\em reduced theta function}
\be
 \Theta_{n,m;\nu}(q,z):=
  \begin{cases}
   \displaystyle{\,\Theta_{n,m}(q,z)-\sum_{l=\nu}^{-1}
     q^{m(l+\frac{n}{2m})^2}z^{-m(l+\frac{n}{2m})}},\ &\nu<0,
  \\[0.5cm]
 \,\Theta_{n,m}(q,z),\ &\nu=0,
  \\[.13cm]
   \displaystyle{\,\Theta_{n,m}(q,z)-\sum_{l=1}^\nu
   q^{m(l+\frac{n}{2m})^2}z^{-m(l+\frac{n}{2m})}},\ &\nu>0,
 \end{cases}
\label{affinetheta}
\ee
where we note the relations
\be
 \Theta_{n,m;-\nu}(q,z)=\Theta_{-n,m;\nu}(q,z^{-1})=\Theta_{n-2(\nu+1)m,m;\nu}(q,z).
\label{mnu}
\ee
We are also interested in the {\em specialised} theta functions obtained by setting $z=1$, denoting them by
\be
 \Theta_{n,m;\nu}(q):=\Theta_{n,m;\nu}(q,1),\qquad \Theta_{n,m}(q):=\Theta_{n,m}(q,1).
\ee
It follows that
\be
 \Theta_{n,m;\nu}(q)=
  \begin{cases}
   \displaystyle{\,\Theta_{n,m}(q)-\sum_{l=\nu}^{-1}
     q^{m(l+\frac{n}{2m})^2}},\ &\nu<0,
  \\[0.5cm]
 \,\Theta_{n,m}(q),\ &\nu=0,
  \\[.13cm]
   \displaystyle{\,\Theta_{n,m}(q)-\sum_{l=1}^\nu
   q^{m(l+\frac{n}{2m})^2}},\ &\nu>0,
 \end{cases}
\label{theta}
\ee
while $\Theta_{n,m}(q)$ is recognised as the familiar theta function
\be
 \Theta_{n,m}(q)=\sum_{l\in\mathbb{Z}+\frac{n}{2m}}q^{ml^2}.
\ee
From (\ref{mnu}) and (\ref{thetasym}), we see that
\be
 \Theta_{n,m;-\nu}(q)=\Theta_{-n,m;\nu}(q)=\Theta_{n-2(\nu+1)m,m;\nu}(q),\qquad
 \Theta_{n,m}(q)=\Theta_{2im\pm n,m}(q),
   \qquad i\in\mathbb{Z}.
\label{thetam}
\ee

In our notation, the Kac-Peterson~\cite{KP84} multiplication formula reads
\be
 \Theta_{n,m}(q,z)\Theta_{n',m'}(q,z)
  =\sum_{\ell=0}^{m+m'-1}\Theta_{nm'-n'm+2\ell mm',mm'(m+m')}(q)\Theta_{n+n'+2\ell m,m+m'}(q,z).
\label{KP}
\ee
Using (\ref{thetasym}) and (\ref{thetam}), one verifies that the right-hand side is invariant 
under interchanging $(n,m)$ and $(n',m')$, as the left-hand side evidently is.
The Jacobi triple product
\be
 \varphi(q,z):=\prod_{i=1}^\infty(1-q^i)(1-q^iz^{-1})(1-q^{i-1}z)
 =(1-z)\prod_{i=1}^\infty(1-q^i)(1-q^iz^{-1})(1-q^iz)
\label{varphiqz}
\ee
can be expressed as
\be
 \varphi(q,z)=\sum_{l\in\mathbb{Z}}(-1)^l q^{\frac{1}{2}l(l-1)}z^l
 =q^{-\frac{1}{8}}z^{\frac{1}{2}}\big[\Theta_{1,2}(q,z)-\Theta_{-1,2}(q,z)\big].
\label{varphiqz2}
\ee
This suggests the introduction of the {\em affine Dedekind eta function}
\be
 \eta(q,z):=q^{\frac{1}{8}}z^{-\frac{1}{2}}\varphi(q,z),
\label{etaqz}
\ee
much akin to the definition of the usual Dedekind eta function, $\eta(q)$, in terms of the Euler product, 
$\varphi(q)$, as
\be
 \eta(q):=q^{\frac{1}{24}}\varphi(q),\qquad 
  \varphi(q):=\prod_{i=1}^\infty(1-q^i).
\ee
We note the properties
\be
 \varphi(q,z)=-z\varphi(q,z^{-1}),\qquad \eta(q,z)=-\eta(q,z^{-1}),
\label{zm1}
\ee
and the well-known expressions
\be
 \varphi(q)=\sum_{l\in\mathbb{Z}}(-1)^lq^{\frac{1}{2}(3l^2-l)},\qquad
 \varphi^3(q)=\sum_{l\in\mathbb{Z}}(4l+1)q^{l(2l+1)}
  =\sum_{l=0}^\infty(-1)^l(2l+1)q^{\frac{1}{2}l(l+1)},
\label{varphi3}
\ee
as well as
\be
 \varphi^2(q)=\sum_{i,j=0}^\infty(-1)^{i+j}q^{2ij+\frac{1}{2}i(i+1)+\frac{1}{2}j(j+1)}\big(1-q^{2(1+i+j)}\big).
\label{var2}
\ee

\section{The reciprocal of the Jacobi triple product}
\label{Sec:rec}

The reciprocal of the Jacobi triple product $\varphi(q,z)$ has been known for more than a century~\cite{TM1898}.
In a notation convenient for our purposes, it is given by
\be
 \frac{1}{\varphi(q,z)}=\frac{1}{\varphi^3(q)(1-z)}\sum_{m\in\mathbb{Z}}f_m(q)z^m,
\label{1varphiqz}
\ee
where
\be
 f_m(q)
 :=\sum_{l=1}^\infty(-1)^l\big(q^{\frac{1}{2}l(l+1+2|m|)}-q^{\frac{1}{2}l(l-1+2|m|)}\big),\qquad m\in\mathbb{Z}.
\label{fm}
\ee
It follows that
\be
 f_{-m}(q)=f_{m}(q),
\label{fmf2m}
\ee
in accordance with (\ref{zm1}). Useful rewritings of $f_m(q)$ include
\be
 f_m(q)
 =\sum_{l=1}^\infty(-1)^lq^{\frac{1}{2}l(l-1+2|m|)}(q^l-1)
 =q^{|m|}+(-1)^mq^{-\frac{1}{2}m^2}(q^{\frac{m}{2}}+q^{-\frac{m}{2}})(F_\infty(q)-F_{|m|}(q)),
\ee
where
\be
 F_n(q):=\sum_{l=0}^n(-1)^l q^{\frac{1}{2}l(l+1)},\qquad n\in\mathbb{N}_0\cup\{\infty\}.
\ee
Adding up the functions $f_m(q)$ defined in (\ref{fm}) yields
\be
 \sum_{m\in\mathbb{Z}}f_m(q)
  =\sum_{m=0}^\infty(2-\delta_{m,0})f_m(q)
  =1,
\label{sumf}
\ee
where we have used (\ref{fmf2m}) and that 
\be
 F_m(q)-F_{m-1}(q)=(-1)^m q^{\frac{1}{2}m(m+1)},\qquad
 \sum_{m=0}^\infty (-1)^m q^{-\frac{1}{2}m^2}(q^{\frac{m}{2}}+q^{-\frac{m}{2}})=1.
\ee

The expression (\ref{1varphiqz}) for the reciprocal of $\varphi(q,z)$ can be proven as follows.
First, with
\be
 R_\ell(q):=\sum_{m\in\mathbb{Z}}(-1)^{\ell-m}f_m(q)q^{\frac{1}{2}(\ell-m)(\ell-m-1)}
\ee
and using (\ref{varphiqz2}), the relation (\ref{1varphiqz}) is seen to be equivalent to
\be
 \varphi^3(q)(1-z)=\sum_{\ell\in\mathbb{Z}}R_\ell(q)z^\ell.
\label{R}
\ee
Since (\ref{fmf2m}) implies
\be
 R_\ell(q)=-R_{1-\ell}(q),
\ee
the relation (\ref{1varphiqz}) is thus equivalent to
\be
 R_\ell(q)=-\delta_{\ell,1}\varphi^3(q),\qquad \ell\in\mathbb{N}.
\label{Rvarphi}
\ee
Using (\ref{fm}), we find
\begin{align}
 R_\ell(q)&=(-1)^{\ell}f_0(q)q^{\frac{1}{2}\ell(\ell-1)}
  +\sum_{m=1}^\infty (-1)^{\ell-m}f_{m}(q)\big(q^{\frac{1}{2}(\ell-m)(\ell-m-1)}
   +q^{\frac{1}{2}(m+\ell)(m+\ell-1)}\big)
\nonumber
\\
 &=(-1)^{\ell}q^{\frac{1}{2}\ell(\ell-1)}\sum_{n=1}^\infty (-1)^n q^{\frac{1}{2}n(n-1)}\Big(
  (q^n-q^{\ell n})\sum_{m=0}^{n-1}q^{-\ell m}+(q^{\ell n}-1)\sum_{m=0}^{n-1}q^{-(\ell-1)m}\Big).
\label{Rlq2}
\end{align}
For $\ell\neq0,1$, this simplifies to
\be
 R_\ell(q)=\frac{(-1)^{\ell}q^{\frac{1}{2}\ell(\ell-1)}(q^{\ell-1}-q^\ell)}{(1-q^{\ell-1})(1-q^\ell)}
  \sum_{n=1}^\infty q^{\frac{1}{2}n(n-1)}(1-q^{-\ell n})(q^n-q^{\ell n})=0, %\qquad \ell\neq0,1,
\ee
where the last equality follows from
\be
 \sum_{n=1}^{\infty}(-1)^n \big(q^{\frac{1}{2}n(n-2\ell+1)}+q^{\frac{1}{2}n(n+2\ell-1)}\big)
  =-1.
\ee
Finally, for $\ell=1$, we find 
\be
 R_1(q)=-\sum_{n=1}^\infty (-1)^n n\big(q^{\frac{1}{2}n(n+1)}-q^{\frac{1}{2}n(n-1)}\big)
  =-\sum_{n=0}^\infty(-1)^n(2n+1)q^{\frac{1}{2}n(n+1)}=-\varphi^3(q),
\ee
thereby completing the proof of (\ref{Rvarphi}), hence of (\ref{1varphiqz}).

\section{Alternative expressions for branching functions}
\label{Sec:Alt}

Here, we re-analyse the branching rule (\ref{cosetKac}) without the use of string functions and derive 
new expressions for the branching functions. To do this,
we first use (\ref{1varphiqz}) to re-express the left-hand side of (\ref{cosetKac}) as
\begin{align}
 \chit_{r,s}^{p,p'}(q,z)\,\ch_{\rho,0}^{n+2,1}(q,z)
 &=\frac{q^{\frac{r^2p'}{4p}+\frac{s^2p}{4p'}+\frac{\rho^2}{4(n+2)}}z^{\frac{sp}{2p'}+\frac{1}{2}}}{\eta(q,z)\eta^3(q)(1-z)}
 \nonumber\\[.2cm]
 &\times
  \sum_{l,m\in\mathbb{Z}}q^{(n+2)l^2}\!f_m(q)z^{m-(n+2)l}
  \big(q^{-\frac{rs}{2}}z^{-\frac{r}{2}}-q^{\frac{rs}{2}}z^{\frac{r}{2}}\big)
  \big(q^{l\rho}z^{-\frac{\rho}{2}}-q^{-l\rho}z^{\frac{\rho}{2}}\big).
\label{chitch}
\end{align}
Expanding the sum in powers of $z$, we find that
\be
 \chit_{r,s}^{p,p'}(q,z)\,\ch_{\rho,0}^{n+2,1}(q,z)=
 \frac{q^{\frac{r^2p'}{4p}+\frac{s^2p}{4p'}+\frac{\rho^2}{4(n+2)}}z^{\frac{sp}{2p'}+\frac{ns}{2}}}{\eta(q,z)\eta^3(q)(1-z)}
  \sum_{i\in\mathbb{Z}+\kappa}H_i(q)z^i,
\ee
where
\be
 \kappa=\begin{cases}0,\ &r+ns+\rho\ \, \mathrm{odd},
  \\[.2cm]
   \frac{1}{2},\ &r+ns+\rho\ \, \mathrm{even},
 \end{cases}
\ee
and
\begin{align}
 H_i(q)&=\sum_{l\in\mathbb{Z}}q^{(n+2)l^2}
  \big[q^{l\rho-\frac{rs}{2}}f_{i+(n+2)l+\frac{ns-1+r+\rho}{2}}(q)
   -q^{-l\rho-\frac{rs}{2}}f_{i+(n+2)l+\frac{ns-1+r-\rho}{2}}(q)
\nonumber\\[.2cm]
  &\qquad\qquad\qquad
   -q^{l\rho+\frac{rs}{2}}f_{i+(n+2)l+\frac{ns-1-r+\rho}{2}}(q)
   +q^{-l\rho+\frac{rs}{2}}f_{i+(n+2)l+\frac{ns-1-r-\rho}{2}}(q)\big].
\end{align}
Writing $H_i^{r,s}(q)$ to indicate explicitly the dependence on $r,s$, it follows from (\ref{fmf2m}) that
\be
 H_i^{-r,-s}(q)=-H_{1-i}^{r,s}(q).
\label{HH}
\ee
Likewise, to indicate whether $r>0$ (hence $s\geq0$) or $r<0$ (hence $s<0$), we may write $H_i^+(q)$ and $H_i^-(q)$, respectively.
We then find that
\begin{align}
 \sum_{i=1}^\ell H_i^+(q)
  &=q^{2\ell s}\sum_{i=1}^\ell H_{1-i}^+(q),
\label{sumH1}
 \\[.15cm]
 \sum_{i=1}^\ell(1+q^s-\delta_{i,1})H_{i-\frac{1}{2}}^+(q)
  &=q^{2\ell s}\sum_{i=1}^\ell(1+q^{-s}-\delta_{i,1})H_{\frac{3}{2}-i}^+(q).
\label{sumH2}
\end{align}
Similar relations for the functions $H^-(q)$ readily follow by application of (\ref{HH}).

Now, with the shorthand notation
\be
 X_\sigma(q)=q^{\frac{\sigma^2 p'}{4(p+np')}-\frac{r^2 p'}{4p}+\frac{s^2 n}{4}-\frac{\rho^2}{4(n+2)}}
  \eta^3(q)\chit_{r,s;\rho;\sigma}^{p,p';n}(q),
\ee
the relation (\ref{cosetKac}) is seen to be equivalent to
\be
 \sum_{i\in\mathbb{Z}+\kappa}H_i(q)z^i
   =\sum_\sigma X_\sigma(q) (1-z)\big(q^{-\frac{\sigma s}{2}}z^{-\frac{\sigma}{2}}
   -q^{\frac{\sigma s}{2}}z^{\frac{\sigma}{2}}\big),
\label{HX}
\ee
based on which we conclude that (\ref{sigma2}) must be respected.
By expanding the right-hand side of (\ref{HX}) in powers of $z$ and using the $H$-relations above, we then find that
\be
 X_\sigma(q)=-q^{-\frac{\sigma s}{2}}\sum_{i=1-\kappa}^{\frac{\sigma}{2}}\frac{1+q^s-\delta_{i,\kappa}}{1+q^s}\,H_i^+(q),\qquad
  \sigma\in\tfrac{1}{2}\mathbb{N},
\ee
with similar expressions in terms of $H^-(q)$ and for $\sigma\in(-\frac{1}{2}\mathbb{N})$.
For $r>0$ and $s\geq0$ (hence $\sigma>0$), we have thus arrived at the explicit expression
\be
 \chit_{r,s;\rho;\sigma}^{p,p';n}(q)
  =-\frac{q^{-\frac{\sigma^2 p'}{4(p+np')}+\frac{r^2 p'}{4p}-\frac{s^2 n}{4}+\frac{\rho^2}{4(n+2)}-\frac{\sigma s}{2}}}{\eta^3(q)}
  \sum_{i=1-\kappa}^{\frac{\sigma}{2}}\frac{1+q^s-\delta_{i,\kappa}}{1+q^s}\,H_i^+(q)
\ee
for the branching functions in (\ref{cosetKac}), with (\ref{HH}) implying the relations (\ref{msigma}).

%%%%%%%%%%%%%%%%%%%%%%%%%%%%%%%%%%%%%%%%

\end{document}